\documentclass[12pt]{article}
\pdfoutput=1
\usepackage[nosort]{cite}

\usepackage{xcolor}

\usepackage{epsfig}
\usepackage{amsfonts}
\usepackage{amscd}
\usepackage{latexsym}
\usepackage{amsmath,amssymb}
\usepackage{verbatim}
\usepackage{setspace}
\usepackage{color}
\usepackage{fancyhdr}
\usepackage{cite}
\usepackage{hyperref}
\usepackage{tikz}
\usepackage{slashed}
\usepackage{multirow}
\usetikzlibrary{calc}
\usetikzlibrary{topaths}
\usetikzlibrary{decorations}
\usetikzlibrary{decorations.pathmorphing}
\usetikzlibrary{arrows,decorations.markings,cd}
\usetikzlibrary{calc,arrows,cd,decorations.markings,snakes}
\tikzset{
->-/.style args={#1rotate#2}{decoration={markings, mark=at position #1 with {\arrow[scale=1.5,rotate = #2 ]{stealth}}}, postaction={decorate}}
}
\usetikzlibrary{shapes.geometric}
\usetikzlibrary{knots}
\usepackage{tikz-3dplot}

\usepackage{draft}
\usepackage{hyperref}
\usepackage{graphicx,color,subfig}
\usepackage{cite}
\usepackage{mciteplus}
\usepackage{skak}
\usepackage{bbm}

\numberwithin{equation}{section}

\def\ee{\mathrm{e}}
\def\mm{\mathrm{m}}
\def\bZ{\mathbb{Z}}

\def\({\left(}
\def\){\right)}

\def\a{\mathbf{A}}
\def\b{\mathbf{B}}
\def\f{\phi}

\begin{document}

\begin{titlepage}
\hfill YITP-SB-2022-14

\title{\bf Higher Gauging \\ and Non-invertible Condensation Defects}

\author{Konstantinos Roumpedakis$^{1}$, Sahand Seifnashri$^{2,3}$, and Shu-Heng Shao$^{2}$}

\address{${}^{1}$Mani L.\ Bhaumik Institute for Theoretical Physics, Department of Physics and Astronomy, University of California, Los Angeles, CA 90095, USA}
\address{${}^{2}$C.\ N.\ Yang Institute for Theoretical Physics, Stony Brook University}
\address{${}^{3}$Simons Center for Geometry and Physics, Stony Brook University}

\abstract{We discuss invertible and non-invertible topological condensation defects  arising from gauging a discrete higher-form symmetry on a higher codimensional manifold in spacetime, which we define as higher gauging.  
A  $q$-form symmetry is called $p$-gaugeable if it can be gauged on a codimension-$p$ manifold in spacetime. 
We focus on 1-gaugeable 1-form symmetries in general 2+1d QFT, and gauge them on a surface in spacetime. 
The universal  fusion rules of the resulting invertible and non-invertible condensation surfaces are determined. 
In the special case of 2+1d TQFT, every (invertible and non-invertible) 0-form global symmetry, including the  $\mathbb{Z}_2$ electromagnetic symmetry of the $\mathbb{Z}_2$ gauge theory,   is realized from higher gauging. 
We further compute the fusion rules between the surfaces, the bulk lines, and lines that only live on the surfaces, determining some of the most basic data for the underlying fusion 2-category. 
We emphasize that the fusion ``coefficients" in these non-invertible fusion rules are generally not numbers, but rather 1+1d TQFTs. Finally, we discuss examples of non-invertible symmetries in  non-topological 2+1d QFTs such as the  free $U(1)$ Maxwell theory and QED. }

\end{titlepage}

\tableofcontents

\section{Introduction}

\subsection{How common are the non-invertible  symmetries?}

In relativistic quantum field theory (QFT) in $D$ spacetime dimensions, the modern characterization of a global symmetry is in terms of the topological operator/defect that generates the symmetry transformation \cite{Gaiotto:2014kfa}.\footnote{For the most part of this paper, we  only consider relativistic quantum systems in Euclidean signature, in which case the  distinction between an operator that acts on the Hilbert space and a defect that extends in time  is not very sharp. We will therefore use these two terms interchangeably. In non-relativistic systems, however, global symmetries are not necessarily generated by topological operators. Furthermore, the distinction between an operator and a defect is important. For instance, see \cite{Gorantla:2022eem} for a recent discussion that emphasizes this distinction in certain non-relativistic systems.  \label{defectvsoperator}}  
In the simplest example of a $U(1)$ global symmetry with a conserved Noether current $\partial_\mu J^\mu(x)=0$, the topological operator is the unitary operator $U_\theta(M^{(D-1)})  = \exp\left( i \theta\oint_{M^{(D-1)}} dn_\mu J^\mu\right)$ supported on a codimension-1 manifold $M^{(D-1)}$ in spacetime (which is commonly taken to be the whole space at a fixed time).

In recent years, the notion of global symmetries has been generalized in several different directions. 
Two exciting developments are the \textit{higher-form global symmetries} \cite{Gaiotto:2014kfa} and the \textit{non-invertible global symmetries} \cite{Bhardwaj:2017xup,Chang:2018iay}. 
In this paper we will discuss the connection between these two seemingly unrelated generalized global symmetries.

We start with a lightening introduction to non-invertible symmetries.  
Given that every ordinary and higher-form global symmetry is associated with a topological operator, it is natural to ask if the converse is true. 
The answer is no: There are topological operators that do not obey a group multiplication law. 
More specifically, a topological operator $U$ is called \textit{non-invertible} if there does \textit{not} exist an inverse topological operator $U^{-1}$ such that under parallel fusion, $U\times U^{-1} = U^{-1} \times U =1$.

Non-invertible symmetries were discussed extensively  in the context of 1+1d CFT \cite{Verlinde:1988sn,Petkova:2000ip,Fuchs:2002cm}  (see also \cite{Bachas:2004sy,Fuchs:2007tx,Bachas:2009mc} and more recently \cite{Ji:2019ugf,Lin:2019hks,Gaiotto:2020fdr,Gaiotto:2020iye,Komargodski:2020mxz,Gaiotto:2020dhf,Lin:2021udi,Thorngren:2021yso,Burbano:2021loy}). 
They have also been realized on the lattice \cite{Grimm:1992ni,Feiguin:2006ydp,Hauru:2015abi,Aasen:2016dop,Buican:2017rxc,Aasen:2020jwb,Inamura:2021szw,Koide:2021zxj,Huang:2021nvb,Vanhove:2021zop,Liu:2022qwn} and are sometimes discussed under the name of algebraic higher symmetries \cite{Ji:2019jhk,Kong:2020cie}. 
The simplest example is the Kramers-Wannier duality, generated by a topological line in the 1+1d Ising model \cite{Frohlich:2004ef,Frohlich:2006ch,Frohlich:2009gb}.
Recently, non-invertible  defects have  been found in many familiar gauge theories in 2+1d and 3+1d, both in the continuum \cite{Choi:2021kmx,Kaidi:2021xfk} and on the lattice \cite{Nguyen:2021yld,Koide:2021zxj,Choi:2021kmx}.

It has been advocated that these non-invertible topological operators should be viewed as a generalization of the ordinary global symmetry \cite{Bhardwaj:2017xup,Chang:2018iay}.  
Following this proposal, we will refer to non-invertible topological operator of codimension-$(q+1)$ as a \textit{non-invertible $q$-form symmetry}. 
Below we review some of the arguments behind this proposal.
\begin{itemize}
\item Like the ordinary and higher-form global symmetries, some of these non-invertible symmetries can be gauged in a generalized way.   
This has been discussed extensively for non-invertible 0-form symmetries in 1+1d \cite{Frohlich:2009gb,Carqueville:2012dk,Brunner:2013xna,Bhardwaj:2017xup,Komargodski:2020mxz,Gaiotto:2020iye,Huang:2021zvu} and for non-invertible 1-form symmetries in 2+1d (see  \cite{Kaidi:2021gbs,Buican:2021axn,Yu:2021zmu,Benini:2022hzx} for recent discussions). 
\item 
When there is an obstruction to the generalized gauging, the non-invertible symmetries have generalized 't Hooft anomalies.  
These anomalies lead to non-trivial dynamical constraints on the renormalization group flows \cite{Chang:2018iay,Thorngren:2019iar,Komargodski:2020mxz,Thorngren:2021yso,Choi:2021kmx},  generalizing the conventional 't Hooft anomaly matching argument. 
\item In  quantum gravity, there are two pieces of lore: the absence of global symmetries and the completeness of spectrum of the gauge group \cite{Misner:1957mt,Polchinski:2003bq,Banks:2010zn}.  As pointed out in \cite{Harlow:2018tng}, the two statements are not equivalent in the context of QFT. It was then realized that, under general assumptions, the absence of invertible and non-invertible symmetries is equivalent to the completeness of spectrum \cite{Rudelius:2020orz,Heidenreich:2021xpr,McNamara:2021cuo}.
\end{itemize}

While non-invertible symmetries are  ubiquitous in 1+1d QFT, it is not yet clear how common they are in higher spacetime dimensions.  
Motivated by this question, we provide a positive answer to this question for  (discrete) non-invertible symmetries.
We will argue that, loosely speaking,  \textit{non-invertible symmetries are at least as common as the (non-anomalous) higher-form symmetries.}  
More precisely, given any  higher-form global symmetry, one can engineer another (generally non-invertible) topological operator in the same QFT via  higher gauging, which we will define momentarily. 
This implies that non-invertible symmetries are generally an inevitable consequence of the higher-form symmetries.

\subsection{Higher gauging and condensation defects}

\textit{Higher gauging} is defined as gauging a discrete $q$-form symmetry on a codimension-$p$ manifold $M^{(D-p)}$ in spacetime. 
More precisely, higher gauging is implemented by summing over insertions of the codimension-($q+1$) topological defects on the codimension-$p$ manifold (and thus $p$ cannot be greater than $q+1$). 
Higher gauging does not change the bulk system, but engineers a new topological defect out of the old ones. 
Motivated by the discussion in \cite{Kong:2013aya,Kong:2014qka,Else:2017yqj,Gaiotto:2019xmp,Kong:2020cie}, the resulting topological defect will be called a \textit{condensation defect}.\footnote{In Section \ref{sec:example}, we will discuss examples of these defects where they are realized as  the condensation of a Higgs field that only live on a higher codimensional manifold. These examples include  the $U(1)$ Maxwell theory, the $U(1)$ Chern-Simons theory, and the $\mathbb{Z}_2$ gauge theory.  This Higgs presentation of the condensation defect justifies the terminology at least in the above examples. \label{fn:condensation}}   
Interestingly, even when the input $q$-form symmetry is invertible, the output condensation defect is generally non-invertible.

Similar to the ordinary gauging, there can be obstructions to gauging a $q$-form symmetry on a codimension-$p$ manifold. 
A $q$-form symmetry is \textit{$p$-gaugeable} if it can be gauged on a codimension-$p$ manifold, otherwise it is \textit{$p$-anomalous}.
 
In this paper, we focus on gauging 1-form global symmetries on a 2-dimensional surface in general 2+1d bosonic QFT.\footnote{
The observables of a  bosonic/non-spin QFT   do not require a choice of the spin structure.  In contrast, a fermionic/spin QFT can only be defined on spin manifolds and its observables depend on the choice of the spin structure. }  
Let $\eta$ be the topological line that generates   a $\mathbb{Z}_N$ 1-form symmetry. 
The condensation surface defect  $S(\Sigma)$ is defined as summing over insertions of the 1-form symmetry lines over a 2-dimensional manifold $\Sigma$ in spacetime:
\ie \label{condensationoperator}
S(\Sigma)  \equiv {1\over \sqrt{|H_1(\Sigma, \mathbb{Z}_N)|}} \sum_{\gamma\in H_1(\Sigma,\mathbb{Z}_N)} \eta(\gamma) \,.
\fe

Given a 1-gaugeable 1-form symmetry, we determine the universal fusion rule of the resulting condensation defects. 
The fusion rule is generally non-invertible, and depends on the braiding phase (i.e., the 0-anomaly) of the input 1-form symmetry.  
The simplest example of a 1-gaugeable 1-form symmetry is a  $\mathbb{Z}_2$ symmetry generated either by a boson or a fermion line. 
Interestingly,   the $\mathbb{Z}_2$ fermion 1-form symmetry is 1-gaugeable (i.e., can be gauged on a surface), but is not 0-gaugeable (i.e., cannot be gauged in the whole spacetime).\footnote{In contrast, a $\mathbb{Z}_2$ symmetry generated by a semion or an anti-semion is not even 1-gaugeable, and the would-be condensation defect is inconsistent. }
The fusion rule of the condensation defect $S$ is (see Section \ref{sec:z2}):
\ie\label{z2fusionintro}
&S(\Sigma)\times S(\Sigma) = \left(\text{1+1d}~\mathbb{Z}_2~\text{gauge theory}\right)\, S(\Sigma)\,,~~~~&\text{if $\eta$ is a boson}\,,\\
&S(\Sigma)\times S(\Sigma)=1\,,~~~&\text{if $\eta$ is a fermion}\,.
\fe

The condensation surface from a  boson line is our first encounter of a non-invertible defect from higher gauging. 
It obeys the non-invertible fusion rule of the Cheshire string \cite{Else:2017yqj,Johnson-Freyd:2020twl}. 
Interestingly, the fusion ``coefficient" in this non-invertible fusion rule is not a number, but rather it is the 1+1d $\mathbb{Z}_2$ gauge theory. 
More generally in a $D$-dimensional QFT, we expect the fusion of the $d$-dimensional topological defects to form an algebra with coefficients being $d$-dimensional topological quantum field theories (TQFTs), with the latter possibly enriched with global symmetries.\footnote{Mathematically, $d$-dimensional topological defects form a module over the fusion ring of $d$-dimensional TQFTs. This is because given a $d$-dimensional topological defect we can stack a decoupled $d$-dimensional TQFT to it and get another topological defect.}

In contrast, when the $\mathbb{Z}_2$ 1-form symmetry line $\eta$ is  a fermion, the resulting condensation surface is invertible and generates a $\mathbb{Z}_2$ 0-form symmetry. 
In particular, we show that the charge conjugation symmetry of  Chern-Simons theories and the  $\mathbb{Z}_2$ electromagnetic  symmetry of the 2+1d $\mathbb{Z}_2$ gauge theory  are both condensation defects of this kind.

We further compute the fusion rules between the lines, the condensation surfaces, and lines living on the surfaces for   $\mathbb{Z}_N$ and  $\mathbb{Z}_{N_1}\times \mathbb{Z}_{N_2}$ one-form symmetries. 
Some special cases of our fusion rules for the condensation surfaces were computed in the context of fusion 2-category~\cite{douglas2018fusion}   in \cite{2022arXiv220310331D} (see also \cite{2021arXiv210315150D}).\footnote{We thank T.\ D.\ D\'ecoppet for pointing out these reference to us.}

Higher gauging is related to many other ideas in the literature of topological phases of matter and TQFT. 
Similar process was discussed in \cite{Kong:2013aya,Kong:2014qka,Kong:2020cie} as a proliferation of topological excitations, where the resulting condensation defects are called ``descendants" of the underlying $q$-form symmetry defect. 
As we will discuss in Section \ref{sec:z2gauge}, the Cheshire string \cite{Else:2017yqj,Johnson-Freyd:2020twl} is also related to the higher gauging of a higher-form symmetry.  
(See also \cite{Hsin:2019fhf} for  discussions.) 
Finally, in the context of Reshetikhin-Turaev TQFT, condensation surfaces and their gauging/orbifold have been discussed in  \cite{Carqueville:2017ono,Carqueville:2018sld,Mulevicius:2020bat,Koppen:2021kry,Carqueville:2021dbv,Carqueville:2021edn}.

We emphasize that our results on higher gauging are applicable to general 2+1d non-topological QFT as well. In fact, we will give several examples of non-invertible condensation defects in non-topological  QFTs, such as the free $U(1)$ Maxwell theory and  QED.

We leave the systematic analysis of condensation defects in higher spacetime dimensions for future investigations. 
Some of the results in 3+1d will be reported in \cite{Choi:2022zal}.   
In the 3+1d toric code, the condensation defects have been discussed in \cite{Kong:2020wmn}.
In 3+1d, non-invertible duality defect $\cal D$ \cite{Koide:2021zxj,Choi:2021kmx,Kaidi:2021xfk} that were recently discovered can be thought of as a ``square root" of the condensation defect $S$ of a $\mathbb{Z}_N$ 1-form symmetry, in the sense that:
\ie
{\cal D}\times \overline{\cal D} = \overline {\cal D} \times {\cal D} = S\,.
\fe
(Here $\overline{\cal D}$ is the orientation-reversal of $\cal D$.)
For a general 3+1d QFT with a $\mathbb{Z}_N$ 1-form symmetry,  $\cal D$ is an interface between a QFT and its gauged version. When the QFT is invariant under gauging a $\mathbb{Z}_N$ 1-form symmetry, ${\cal D}$ becomes a defect in the same QFT.

\subsection{What qualifies as a global symmetry?}

What qualifies as a 0-form global symmetry? 
More precisely, by a 0-form global symmetry, we mean a global symmetry generated by a codimension-1 topological operator $U(M^{(D-1)})$ \cite{Gaiotto:2014kfa}.\footnote{Here and throughout we define a $q$-form symmetry to be  a symmetry  generated by a codimenion-$(q+1)$ topological operator, rather than by the requirement that it acts on   $q$-dimensional obejcts.  For an ordinary $q$-form symmetry, the two definitions coincide since the action of the symmetry on the charged objects is given by the canonical linking   \cite{Gaiotto:2014kfa}.  Condensation defects, however,  do not act   by  canonical linkings, and we need to clarify which definition we are using. We thank Daniel Harlow for discussions on related points. } 
There is a variety of different definitions of global symmetries, which are useful and relevant in their own ways depending on the context. 
Most conservatively, a 0-form global symmetry has to act faithfully on local operators. For example, this is the perspective adopted in \cite{Harlow:2018tng}.  In conformal field theory (CFT) or TQFT, this is equivalent to demanding the symmetry operator $U(S^{(D-1)})$ to be nontrivial when acting on the Hilbert space ${\cal H}(S^{D-1})$ on a sphere. 

 The above perspective, however, excludes a large class of interesting global symmetries. 
 For instance, consider the charge conjugation symmetry $A\to -A$   in the 2+1d $U(1)_{2N}$ Chern-Simons theory:
 \ie
{\cal L} =  {2iN\over 4\pi}AdA \,.
\fe
The theory has no (gauge-invariant) local operator because $F=0$ by the equation of motion. 
The charge conjugation symmetry does not  act on local operators, but it acts nontrivially on the   Wilson lines, i.e., $\exp(i n \oint A)\to \exp(-in \oint A)$. 
This means that while the charge conjugation symmetry operator $U(S^2)$ is trivial on a sphere, it is nontrivial on more general spatial manifolds, such as the torus $U(T^2)$. 
More generally, any 0-form global symmetry in a TQFT without local operators is of this kind:\footnote{In this paper, as it is  common in the condensed matter physics literature, we only restrict ourselves to robust TQFTs with no nontrivial local operators. This excludes BF theories of a 2-form gauge field and a compact scalar, which describes symmetry-breaking phases.} it acts nontrivially on the extended defects, but trivially on the local operators. 
Nonetheless, they are still interesting and have  been discussed extensively in condensed matter physics, under the name of Symmetry Enriched Topological (SET) Order \cite{PhysRevB.87.104406,PhysRevB.87.155115,2010arXiv1012.4470Y,PhysRevB.87.165107,PhysRevB.88.235103,2015PhRvX...5d1013C,Barkeshli:2014cna},\footnote{In fact, in the context of 2+1d SET, it is common that the global symmetries act nontrivially in the microscopic model, but   act trivially both on the line defects and the local operators in the continuum TQFT limit.} in mathematics under the name of $G$-crossed modular tensor categories \cite{2009arXiv0909.3140E,Barkeshli:2014cna}, and in high energy physics (see, for instance,  \cite{Fuchs:2014ema,Gaiotto:2014kfa,Aharony:2016jvv,Cordova:2017vab,Benini:2018reh,Delmastro:2019vnj}).

\begin{figure}[t]
    \centering
     \begin{tikzpicture}	
    	\draw[->-=0.6 rotate 0] (0.3,-2) -- (0.3,2);
    	\node[circle,inner sep=2pt,draw, fill, color = white] at (0.3,-0.2) {};
    	\draw[color = violet] (0.15,0.5) .. controls +(-0.8,-0.1) and +(-0.8,0) .. (0.3,-0.2) .. controls +(0.8,0) and +(0.8,-0.1) .. (0.44,0.5);  	
    	\draw[->-=0.6 rotate 0, color = violet] (0.29,-0.2) -- (0.31,-0.2);
    	\node at (0.5,-1.8) {$L$};
    	\node at (0.7,-0.4) {$a$};
    \end{tikzpicture}
    \hskip 4cm
     \begin{tikzpicture}[scale = 1]
    	\node[purple] at (-3,1.5) {$S$};
    	\node[cylinder, 
    		draw = violet, 
    		text = purple,
    		cylinder uses custom fill, 
    		cylinder body fill = magenta!10, 
    		cylinder end fill = magenta!40,
    		minimum width = 1.2cm,
    		minimum height = 3.1cm,
    		shape border rotate = 90] (c) at (-2,0) {};    		
		\node[circle,inner sep=0pt,draw] at (-2.2,1) (a) {};
		\node[circle,inner sep=0pt,draw] at (-1.5,0.7) (b) {};
		\node[circle,inner sep=0pt,draw] at (-1.8,0.3) (c) {};
		\node[circle,inner sep=0pt,draw] at (-2.5,0.5) (d) {};
		\node[circle,inner sep=0pt,draw] at (-1.7,-0.4) (e) {};
		\node[circle,inner sep=0pt,draw] at (-2.3,-0.5) (f) {};
		\node[circle,inner sep=0pt,draw] at (-2.1,-1) (g) {};
		\draw[color = violet] (a) to (b);
		\draw[color = violet] (b) to (c);
		\draw[color = violet] (c) to (d);
		\draw[color = violet] (c) to (e);
		\draw[color = violet] (a) to (d);
		\draw[color = violet] (d) to (-2.6,0);
		\draw[color = violet] (e) to (f);
		\draw[color = violet] (e) to (-1.4,-0.5);
		\draw[color = violet] (f) to (-2.6,-0.4);
		\draw[color = violet] (f) to (g);
		\draw[color = violet] (g) to (-2.6,-1);
		\draw[color = violet] (g) to (-1.5,-1.35);
		\draw[color = violet] (a) to (-2.2,1.4);
		\draw[color = violet] (b) to (-1.4,0.7);
		\draw[dashed, color = violet] (-2.5,-1.3) .. controls (-2,-1.1) .. (-1.4,-1.3);
    	\node at (-1.7,-1.7) {$L$};
    	\draw[->-=0.5 rotate 0, dashed] (-2,-1.4) -- (-2,1.5);
    	\draw[] (-2,1.3) -- (-2,2);
    	\draw[] (-2,-2) -- (-2,-1.4);
    	\draw[->-=1 rotate 0, color = red, thick, dashed] (-1.4,1.48) -- (-1.9,1.48);
    \end{tikzpicture}
    \caption{The left figure shows the action of a 1-form symmetry line $a$ on a line $L$, while the right figure shows the action $S \cdot L$ of a condensation surface $S$ (e.g., the charge conjugation symmetry) on a line $L$. The condensation defect is defined by summing over insertions of topological lines. The 1-form symmetry acts on lines by a phase without changing the type of lines. In contrast, the condensation surface defect changes the  types of lines.}
\label{fig: surface vs line action}
\end{figure}
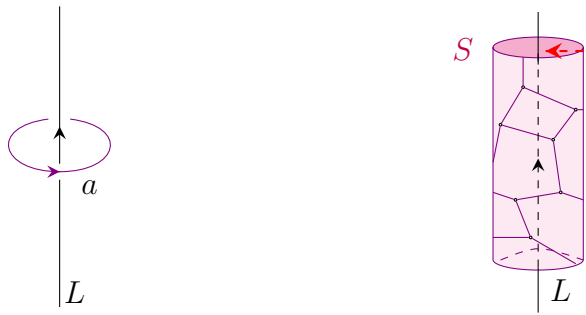

Even though the charge conjugation symmetry in Chern-Simons theory only acts on the lines, it is not a 1-form symmetry, at least not in the sense of \cite{Gaiotto:2014kfa}.  
Let us compare the two. 
The charge conjugation symmetry permutes different types of  Wilson line, while the 1-form symmetry acts on the line by a phase without changing its type.\footnote{Here we assume $L$ is a simple line, i.e., it cannot be decomposed into non-negative linear combinations of other lines. Also, throughout the paper, we assume that the only bulk topological local operator is the identity operator. }  
See Figure \ref{fig: surface vs line action}. 
So, how should we think about the charge conjugation symmetry given that it only acts on lines but  is not a 1-form symmetry?

Our work elucidates the nature of some of these 0-form symmetries that act trivially on all local operators:   
the condensation defects from higher gauging are exactly symmetries of this kind. 
In 2+1d, they are surfaces  made out of lines,  and therefore they are ``porous" to the local operators and act trivially on the latter. 
Instead, they  act nontrivially on the line defects and generally change the type of the latter. 
In fact, as will be discussed in Section \ref{sec:symTQFT}, we will argue that all 0-form symmetries in a 2+1d TQFT, including the charge conjugation symmetry, are condensation surface defects realized  by the higher gauging of 1-form symmetries.\footnote{In the current paper, we only determine the action of the surface defects (which include the 0-form symmetries) on the topological lines (i.e., the low-energy limit of the anyons). We leave  the freedom in choosing the symmetry fractionalization class, which is an essential piece of data in SET and in the $G$-crossed modular tensor category,  for future studies. }

\subsection{Outline}

For first time readers, they are encouraged to start with Section \ref{sec:pgaugeable} for the definition of higher gauging and Section \ref{sec:z2} for the simplest possible example. These two sections are largely written in a self-contained fashion, demonstrating the main point of this paper. 

This paper is organized as follows. 
In Section \ref{sec:pgaugeable}, we define the higher gauging of a higher-form symmetry and introduce the notion of higher anomaly in general spacetime dimensions.

In Section \ref{sec:1gaugeable}, we discuss the simplest possible case of higher gauging: gauging a 1-form symmetries on a surface in 2+1d. We include a brief review of $1$-form symmetries and we give the criterion for a 1-form symmetry to be 1-gaugeable. 
We give a general discussion on condensation surface defects from the higher gauging of 1-form symmetries. 
 We then move on to discuss the action of these condensation defects on line operators. We also introduce the notion of higher quantum symmetry generated by lines living on on these defects. 

In Section \ref{sec:z2}, we study in detail the case of a QFT with a 1-gaugeable $\mathbbm{Z}_2$ 1-form symmetry. We give explicit expressions for the condensation surface  and determine its  fusion rule. 
We then compute the action of the surfaces on the lines. In addition, we  study the higher quantum symmetry on the surface.  
At the end of this section, we present the full fusion algebra between the bulk line, the surface, and the line living only on the surface. 

Section \ref{sec:znn}  generalizes the discussion in the previous section for the case of a QFT with a 1-gaugeable $\mathbbm{Z}_N$ or $\mathbbm{Z}_{N_1} \times \mathbbm{Z}_{N_2}$ 1-form symmetry. Although the discussion is a direct generalization of the $\mathbbm{Z}_2$ case, it is more involved with many new features and interesting subtleties which are not present in the $\mathbbm{Z}_2$ case. One new feature is the higher gauging of $\mathbbm{Z}_{N_1} \times \mathbbm{Z}_{N_2}$  with discrete torsion on a surface, which leads to a family of  condensation defects  labeled by $H^2(B(\mathbbm{Z}_{N_1} \times \mathbbm{Z}_{N_2}), U(1))$.

In Section \ref{sec:example}, we study several examples of invertible and non-invertible condensation defects in 2+1d topological and non-topological QFTs. 
The TQFT examples are $U(1)$ Chern-Simons gauge theories,  the 2+1d $\mathbb{Z}_p$ gauge theory for prime $p$, and the $\mathcal{G}_k \times \mathcal{G}_{-k}$ Chern-Simons theory. 
In particular, the charge conjugation symmetry in $U(1)$ Chern-Simons theories and the $\mathbb{Z}_2$ electromagnetic symmetry of 2+1d $\mathbb{Z}_2$ gauge theory are both realized as the condensation defects of 1-form symmetries. 
The non-topological QFT examples are the free Maxwell theory and QED with a Dirac fermion of even charge.

Finally, in Section \ref{sec:symTQFT}, we comment on the higher gauging  of non-invertible 1-form symmetries and we make contact with the mathematical literature. We then present an argument why all $0$-form symmetries in any 2+1d TQFT can be realized from higher gauging.

\section{$p$-gaugeable $q$-form symmetries}\label{sec:pgaugeable}

It is common to gauge a generalized global symmetry in the whole spacetime to map one quantum system to its gauged version.
One can also gauge the symmetry only in a codimension-0 region of the spacetime (e.g., half of the spacetime), producing an interface or a boundary condition.\footnote{One needs to choose an appropriate boundary condition, such as the topological Dirichlet boundary condition, for the discrete gauge fields  in defining such an interface.}
(See, for example, \cite{Kaidi:2021gbs,Choi:2021kmx}   for recent  discussions.) 
In this paper,  we consider gauging a discrete 
$q$-form symmetry on a codimension-$p$ manifold $M^{(D-p)}$ in $D$ spacetime dimensions, which we refer to as the \textit{higher gauging}, or \textit{higher condensation}.\footnote{In the condensed matter literature \cite{Bais:2008ni, 2014NuPhB.886..436K,Burnell:2017otf} in 2+1d, gauging a 1-form symmetry in a codimension-0 region is commonly referred to  as anyon condensation. For this reason, it is natural to define the gauging of a higher-form symmetry on a higher codimensional manifold as higher condensation. See also footnote \ref{fn:condensation} for further justification of this terminology.}
More specifically, this is referred to as the \textit{$p$-gauging} of a $q$-form global symmetry.

While higher gauging does not change the bulk of the quantum system, it defines a topological defect $S$ supported on $M^{(D-p)}$, which we will call the \textit{condensation defect}. 
The condensation defect  is generally non-invertible. 
In other words, the higher gauging of a $q$-form symmetry gives an invertible or non-invertible $(p-1)$-form symmetry.

A  $q$-form global symmetry is called \textit{$p$-gaugeable},  or \textit{$p$-condensable}, if it  can be gauged on a codimension-$p$ manifold in spacetime.  
Otherwise, it is \textit{$p$-anomalous}. 
In particular, a  0-gaugeable symmetry is free of  't Hooft anomalies in the usual sense and can be gauged in the whole spacetime. 
The 0-anomaly of  a $q$-form symmetry is the ordinary 't Hooft anomaly, i.e., obstruction to gauging the $q$-form symmetry in the whole spacetime.  We will use 0-anomalies and 't Hooft anomalies interchangeably. 
The obstruction to gauging a $q$-form symmetry on a higher codimensional manifold will be referred to as the \textit{higher anomaly}.

Importantly, a  0-anomalous higher-form symmetry can be $p$-gaugeable with $p>0$: while it cannot be gauged  in the whole spacetime,  it can  be gauged on a higher codimensional manifold in spacetime. More generally, a $p$-gaugeable symmetry is $p'$-gaugeable for all $p'\ge p$.  
Conversely, a $p$-anomalous symmetry is $p'$-anomalous for all $p'\le p$.

Higher gauging of a $q$-form symmetry is defined by summing over insertions of the codimension-$(q+1)$ topological defects over a codimension-$p$ manifold in spacetime. 
For this procedure to make sense, we need
\ie\label{inequality}
p\le q+1\,.
\fe
In particular, when $p=q+1$, a $q$-form symmetry is always $(q+1)$-gaugeable in a trivial way. 
The resulting condensation defect is  a superposition of the $q$-form symmetry defects on the same manifold. 
It is not simple in the sense that it can be written in terms of other defects of the same dimensionality. 
We will henceforth focus on cases $p<q+1$, for which the condensation defects  are simple in the  sense above.

Just like the  ordinary gauging, there are generally different ways to gauge a symmetry on a higher codimensional manifold $M^{(D-p)}$. 
These choices  differ  by  $(D-p)$-dimensional invertible field theories on $M^{(D-p)}$ involving the higher-form gauge fields \cite{Gaiotto:2014kfa}. 
In the condensed matter physics terminology, this is the choice of a Symmetry Protected Topological (SPT) Phase on $M^{(D-p)}$, while in the high energy physics terminology, this is a choice of the discrete torsion. 
Different choices of the discrete torsion on $M^{(D-p)}$ give rise to different condensation defects.

Finally, the condensation defects themselves can have their own 't Hooft anomalies. 
Curiously, even when the underlying $q$-form symmetry is anomalous, the resulting condensation defects can still be  free of 't Hooft anomalies (or more precisely, free of 0-anomalies).  
Therefore, the gauging of the resulting condensation defects gives an indirect gauging of the original anomalous $q$-form symmetry. 
We will encounter many such examples below. 

\section{Gauging 1-form symmetries on surfaces in 2+1d}\label{sec:1gaugeable}

In this paper, we focus on  gauging 1-form symmetries on a surface  in 2+1d. 
 The resulting condensation defects are invertible and non-invertible topological surfaces in spacetime.   
For simplicity, we will only consider 2+1d bosonic/non-spin QFTs which do not require a choice of the spin structure of the spacetime manifold.\footnote{The only exception is in Section \ref{sec:QED}, where we discuss higher gauging in QED with a Dirac fermion.}  
 We also assume that the surface is orientable. 

In this section, we will discuss general aspects of condensation surfaces,   their fusion rules, and the action of surfaces on the lines. 
We will also discuss lines that only live on the surface, which will be called higher quantum symmetry lines. 
For readers who are interested in the simplest example of condensation defect, they are welcome to skip the review and general discussion  in the current section and move directly to   Section \ref{sec:z2}.
  
\subsection{Review of 1-form symmetries}\label{sec:general}

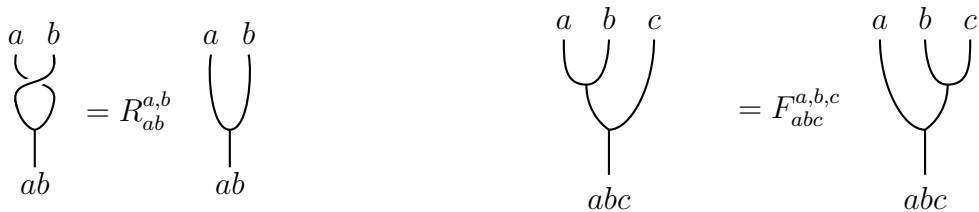
\begin{figure}[h]
\centering
\begin{minipage}{.5\textwidth}
  \centering
  \begin{tikzpicture}[scale = 0.5]
	\begin{knot}[
		clip width=5,
		flip crossing=1,
		]
		\strand[thick] (0,0) .. controls +(-0.2,-1) and +(0.2,0.8) .. (1,-1.3) .. controls 		                     +(-0.1,-0.5) and +(0.1,0) .. (0.5,-2);
		\strand[thick] (1,0) .. controls +(0.2,-1) and +(-0.2,0.8) ..  (0,-1.3).. controls  		  					 +(+0.1,-0.5) and   +(-0.1,0) .. (0.5,-2);
		\draw[thick] (0.5,-2) to  (0.5,-3);
	\end{knot}
	\node[anchor=south] at (0,0) {$a$};
	\node[anchor=south] at (1,0) {$b$};
	\node[anchor=south] at (0.5,-4) {$ab$};
	\node at (3,-1.5) {$= R^{a,b}_{ab}$};
\end{tikzpicture}
\begin{tikzpicture}[scale = 0.5]
	\begin{knot}[
		clip width=5,
		flip crossing=1,
		]
		\strand[thick] (0,0) .. controls +(-0.1,-0.5) and +(-0.5,0) .. (0.5,-2);
		\strand[thick] (1,0) .. controls +(+0.1,-0.5) and   +(0.5,0) .. (0.5,-2);
		\draw[thick] (0.5,-2) to  (0.5,-3);
	\end{knot}
	\node[anchor=south] at (0,0) {$a$};
	\node[anchor=south] at (1,0) {$b$};
	\node[anchor=south] at (0.5,-4) {$ab$};
\end{tikzpicture}
\end{minipage}~
\begin{minipage}{.5\textwidth}
  \centering
    \begin{tikzpicture}[scale = 0.6]
	\begin{knot}[
		clip width=5,
		flip crossing=1,
		]
		\strand[thick] (0,0) .. controls +(0,-1) and +(0.5,0) .. (-1,-2);
		\strand[thick] (-2,0) .. controls +(0,-0.5) and +(-0.5,0) .. (-1.5,-1) .. controls 		                     						+(0,-.7) and +(0,0) .. (-1,-2);
		\strand[thick] (-1,0) .. controls +(-0,0) and +(0.5,0) .. (-1.5,-1);
		\strand[thick] (-1,-2) .. controls +(0,0) and +(0,0) .. (-1,-3);
	\end{knot}
	\node[anchor=south] at (0,0) {$c$};
	\node[anchor=south] at (-1,0) {$b$};
	\node[anchor=south] at (-2,0) {$a$};
	\node[anchor=south] at (-1,-4) {$abc$};
	\node at (3,-1.5) {$= F^{a,b,c}_{abc}$};
  \end{tikzpicture}
  \begin{tikzpicture}[scale = 0.6]
	\begin{knot}[
		clip width=5,
		flip crossing=1,
		]
		\strand[thick] (0,0) .. controls +(0,-1) and +(-0.5,0) .. (1,-2);
		\strand[thick] (2,0) .. controls +(0,-0.5) and +(0.5,0) .. (1.5,-1) .. controls 		                     						+(0,-.7) and +(0,0) .. (1,-2);
		\strand[thick] (1,0) .. controls +(0,0) and +(-0.5,0) .. (1.5,-1);
		\strand[thick] (1,-2) .. controls +(0,0) and +(0,0) .. (1,-3);
	\end{knot}
	\node[anchor=south] at (0,0) {$a$};
	\node[anchor=south] at (1,0) {$b$};
	\node[anchor=south] at (2,0) {$c$};
	\node[anchor=south] at (1,-4) {$abc$};
  \end{tikzpicture}
\end{minipage}
\caption{The $R$- and $F$-symbols captures the braiding and crossing relations between topological lines in 2+1d. Here we only discuss the $R$- and $F$-symbols for invertible lines (i.e. abelian anyons), where $ab$ denotes the fusion of $a$ with $b$.} \label{fig:FRSymbol}
\end{figure}

Let us first review  discrete  1-form global symmetries in general 2+1d unitary QFT \cite{Gaiotto:2014kfa}.  
A 1-form symmetry  $G$, which is an abelian finite group, in 2+1d is generated by a topological line operator/defect. 
We will denote these invertible topological lines   by $a,b,c,\cdots$. 
Topological lines in 2+1d can have nontrivial \textit{braiding} as shown on the left of Figure \ref{fig:FRSymbol}.  
They can also have nontrivial \textit{crossing} relations with lines splitting and rejoining as shown on the right of Figures \ref{fig:FRSymbol}.

In a general unitary QFT with a discrete 1-form global symmetry, the associated topological lines are characterized by a unitary braided fusion category \cite{Moore:1988qv,Kitaev:2005hzj,drinfeld2010braided,Barkeshli:2014cna}, with their braiding and crossing given by the $R$- and $F$-symbols, respectively.\footnote{The set of topological lines in a non-topological QFT only  forms a braided tensor category, but not necessarily a modular tensor category. Generally, there are other non-topological lines that have nontrivial correlation functions with the topological lines. For example, 2+1d QED with a charge 2 scalar has a non-anomalous $\mathbb{Z}_2$ 1-form global symmetry, which forms a braided fusion category that is not modular. The line  charged under this $\mathbb{Z}_2$ 1-form symmetry is a non-topological Wilson line. } 
The $R$- and $F$-symbols have to obey certain consistency conditions, known as the pentagon and hexagon identities \cite{Moore:1988qv}. 
Furthermore, the individual values of the $R$- and $F$-symbols  can be changed by redefining the phases of the junctions between the lines, which is sometimes referred to as a ``gauge transformation".  
Two sets of $R$- and $F$-symbols that differ by this redefinition of the junction phases are  equivalent. See, for example,  \cite{Kitaev:2005hzj,Barkeshli:2014cna} for more details.

If the braiding is trivial, then the crossing is necessarily trivial from the above consistency conditions. 
However, there can be lines with trivial crossing and yet nontrivial braiding, such as a fermion line as we will discuss below. 

\begin{figure}[h]
\centering
\begin{tikzpicture}
\draw[thick] (0,0) .. controls +(0,1.8) and +(0,0.3) .. (0.5,1) .. controls +(0,-0.5) and +(0,0) .. (0.15,0.9);
\draw[thick] (0.05,1) .. controls +(-0.1,0) and +(0,0) .. (0,2);
\node[black, anchor = north] at (0,0) {$a$};  
\draw[->-=0.5 rotate 0, color = black, ultra thin] (0,0.2) -- (0,0.25);
\end{tikzpicture}
\begin{tikzpicture}
\node[black] at (-1,1) {$= \theta(a)$};  
\draw[->-=0.5 rotate 0, color = black, thick] (0,0) -- (0,2);
\node[black, anchor = north] at (0,0) {$a$};
\node[black] at (1,1) {$=$};    
\end{tikzpicture}
\begin{tikzpicture}
\draw[thick] (0,0) .. controls +(-0,-1.8) and +(0,-0.3) .. (-0.5,-1) .. controls +(0,0.5) and +(0,0) .. (-0.15,-0.9);
\draw[thick] (-0.05,-1) .. controls +(0.1,0) and +(0,0) .. (0,-2);
\node[black, anchor = north] at (0,-2) {$a$};  
\draw[->-=0.5 rotate 0, color = black, ultra thin] (0,-1.8) -- (0,-1.7);
\end{tikzpicture}
\caption{Topological spin $\theta(a)$ of the line $a$.}
\label{fig:spin}
\end{figure}
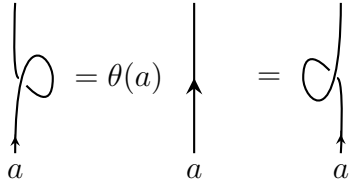

An important property of a topological line $a$ that can be derived from the $R$-symbols is the \textit{topological spin}  $\theta(a)$, which is a root of unity \cite{Vafa:1988ag}, defined in Figure \ref{fig:spin}.  
For invertible lines, the topological spin $\theta(a)$ is related to the $R$-symbols as
\ie
\theta(a) =\theta(a^{-1}) =( R^{a ,a^{-1}}_{1})^{-1}\,.
\fe
where $a^{-1}$ is the inverse of $a$.\footnote{For more general non-invertible lines, the topological spin is defined as $\theta(a) = (R^{a, \bar a}_1)^{-1}$ where $\bar a$ is the orientation-reversal of $a$.}  
Note that the twisting in Figure \ref{fig:spin} is nontrivial because the topological line  should be viewed as a ribbon of infinitesimal width.
For Wilson lines in Chern-Simons gauge theory, the ribbon structure can be understood from the point-splitting regularization that is needed to define the Wilson line \cite{Witten:1988hf}. 

The braiding phase $B(a,b)$   of two invertible topological lines $a$ and $b$, is defined as
\begin{equation}
	\raisebox{-3.4em}{\begin{tikzpicture}
	\begin{knot}[
		clip width=5,
		]
		\strand[thick, black] (0,0) .. controls +(0,1.5) and +(0,-1.5) .. (1,2);
		\strand[thick, black] (1,0) .. controls +(0,1.5) and +(0,-1.5) ..  (0,2);
	\end{knot}
	\node[black, anchor = north] at (0,0) {$a$};
	\node[black, anchor = north] at (1,0) {$b$};
	\draw[->-=0.5 rotate 0, color = black, ultra thin] (0,0.2) -- (0,0.25);
	\draw[->-=0.5 rotate 0, color = black, ultra thin] (1,0.2) -- (1,0.25);
	\draw[->-=0.5 rotate 10, color = black, ultra thin] (0,1.70) -- (0,1.75);
	\draw[->-=0.5 rotate -10, color = black, ultra thin] (1,1.70) -- (1,1.75);
\end{tikzpicture}}  = B(a,b) 
\raisebox{-3.4em}{\begin{tikzpicture}
		\begin{knot}[
			clip width=5,
			flip crossing=1,
			]
			\strand[thick, black] (0,0) .. controls +(0,1.5) and +(0,-1.5) .. (1,2);
			\strand[thick, black] (1,0) .. controls +(0,1.5) and +(0,-1.5) ..  (0,2);
		\end{knot}
		\node[black, anchor = north] at (0,0) {$a$};
		\node[black, anchor = north] at (1,0) {$b$};
		\draw[->-=0.5 rotate 0, color = black, ultra thin] (0,0.2) -- (0,0.25);
		\draw[->-=0.5 rotate 0, color = black, ultra thin] (1,0.2) -- (1,0.25);
		\draw[->-=0.5 rotate 10, color = black, ultra thin] (0,1.70) -- (0,1.75);
		\draw[->-=0.5 rotate -10, color = black, ultra thin] (1,1.70) -- (1,1.75);
\end{tikzpicture}}~.
\end{equation}
It can be expressed in term of the topological spins as follows\footnote{Because of this relation, $B(a,b)$ is sometimes referred to as the double-braiding phase (or monodromy phase), and $R$ as the braiding. We will loosely refer to both of them as braiding.} 
\begin{equation}\label{Bab}
 B(a,b)= R^{a,b}_{ab} R^{b,a}_{ab}= {	\theta(ab)\over  \theta(a) \theta(b) }~.
\end{equation}
See, for example, \cite{Barkeshli:2014cna}, for a derivation of this formula. 

\subsection{Higher gauging of 1-form symmetries}\label{sec:generalhighergauging}

  \begin{figure}[h]
    \centering
    \begin{tikzpicture}[scale = 0.7]
		\draw[color = violet, thick] (0,0) to (5,2);
		\draw[color = violet, thick] (0,5) to (5,7);
		\draw[color = violet, thick] (0,-0.01) to (0,5.01);
		\draw[color = violet, thick] (5,2) to (5,7);
		\node[circle,inner sep=0pt,draw] at (1.2,1.5) (a) {};
		\node[circle,inner sep=0pt,draw] at (1.5,2.5) (b) {};
		\node[circle,inner sep=0pt,draw] at (0.8,3) (c) {};
		\node[circle,inner sep=0pt,draw] at (2,3.5) (d) {};
		\node[circle,inner sep=0pt,draw] at (2.8,4.3) (e) {};
		\node[circle,inner sep=0pt,draw] at (3.5,5.3) (f) {};
		\node[circle,inner sep=0pt,draw] at (4,4) (g) {};
		\node[circle,inner sep=0pt,draw] at (3.5,2) (h) {};
		\node[circle,inner sep=0pt,draw] at (2.2,1.8) (j) {};
		\node[circle,inner sep=0pt,draw] at (3,2.5) (i) {};
		\node[circle,inner sep=0pt,draw] at (3.2,5.1) (k) {};
		\draw[color = violet] (a) to (b);
		\draw[color = violet] (b) to (c);
		\draw[color = violet] (c) to (d);
		\draw[color = violet] (d) to (e);
		\draw[color = violet] (e) to (k);
		\draw[color = violet] (k) to (f);
		\draw[color = violet] (e) to (i);
		\draw[color = violet] (i) to (j);
		\draw[color = violet] (i) to (h);
		\draw[color = violet] (h) to (g);
		\draw[color = violet] (g) to (f);
		\draw[color = violet] (j) to (b);
		\draw[color = violet] (a) to (1.2,0.5);
		\draw[color = violet] (a) to (0,1.6);
		\draw[color = violet] (c) to (0,3);
		\draw[color = violet] (d) to (1,5.4);
		\draw[color = violet] (j) to (2.6,1.05);
		\draw[color = violet] (h) to (3.3,1.3);
		\draw[color = violet] (g) to (5,3);
		\draw[color = violet] (f) to (5,6.8);
		\draw[color = violet] (k) to (3.2,6.3);
    \end{tikzpicture}
    \caption{
    Gauging a $q$-form symmetry on a  codimension-$p$ manifold is equivalent to inserting  a network of symmetry defects along the dual graph of a triangulation of the codimension-$p$ manifold.}
    \label{fig:mesh}
\end{figure}
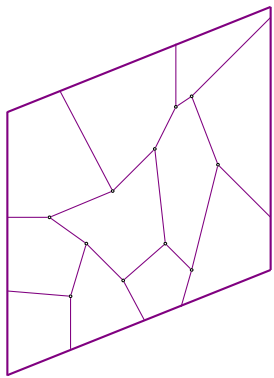

Gauging a 1-form symmetry $G$ in the whole spacetime corresponds to summing over insertions of  these lines along
a trivalent graph dual to the triangulation of the spacetime manifold \cite{Gaiotto:2014kfa}.   
For the gauging to be unambiguous, all observables have to be independent of the choice of the triangulation, which is equivalent to demanding  the \textit{braiding} of these lines to be trivial.  
This implies that the topological spins of these lines are all trivial, i.e., they are boson lines. 
Hence, a 1-form global symmetry is free of 't Hooft anomalies if the braiding of the associated topological lines is trivial \cite{Gaiotto:2014kfa,Gomis:2017ixy,Hsin:2018vcg}. 
See \cite{Hsin:2018vcg} (which was based on  \cite{Moore:1988ss,Moore:1989yh}) for more discussions on the gauging of 1-form symmetries.

Next, we discuss gauging a 1-form symmetry on a codimension-1 surface, i.e.\ higher gauging. 
This is done by summing over insertions  of the 1-form symmetry lines along the dual graph of a triangulation of an oriented 2-dimensional surface $\Sigma$ (see Figure  \ref{fig:mesh}).
In fact, it is as if we are gauging a 0-form symmetry in 1+1d \cite{Frohlich:2009gb,Brunner:2014lua,Bhardwaj:2017xup,Lin:2019kpn}. 
For this higher gauging to be consistent, we need the answer to be independent of the triangulation of the surface. 
This requires the \textit{crossing} of these lines to be trivial, which is characterized by the $F$-symbols (see Figure \ref{fig:FRSymbol}).

A 1-form symmetry is non-anomalous in the usual sense (i.e., 0-gaugeable) if the braiding of the symmetry lines is trivial. 
This in particular implies that the crossing is also trivial.  
A 1-form symmetry is 1-gaugeable if the crossing of the symmetry lines is trivial, but they are allowed to have nontrivial braiding.

In any 2+1d  QFT with a 1-gaugeable 1-form global symmetry $G$, we can  construct the condensation surface defects  $S(\Sigma)$ by gauging $G$ on an oriented 2-dimensional manifold $\Sigma$.\footnote{We will sometimes suppress the dependence of $S(\Sigma)$ on the 2-dimensional manifold $\Sigma$  and write it simply as $S$.}  
As already mentioned above, this higher gauging is not unique, and different choices are related by discrete torsion. 
Since the procedure of gauging a 1-form symmetry on a surface in 2+1d is identical to gauging a 0-form symmetry in 1+1d, the discrete torsion is classified by $H^2(BG,U(1))$. 
We obtain a family of condensation surface defects labeled by $H^2(BG,U(1))$.\footnote{Note that generally there is no canonical way to gauge a 1-form symmetry on a surface, similar to the situation of gauging in 1+1d.  In other words, the choice of the discrete torsion forms a torsor, rather than a group with a natural identity.  For example, see the discussion around equation \eqref{torsor}.} 
We will encounter such examples in Section \ref{sec:ZN1ZN2}.

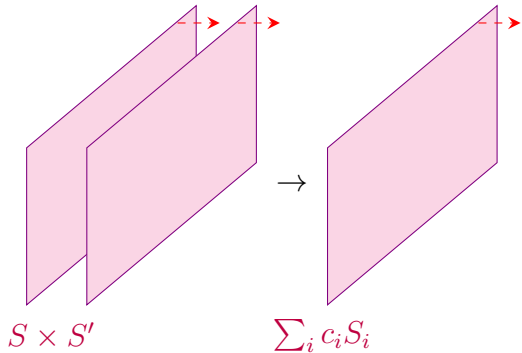
\begin{figure}[h]
	\centering
	\begin{tikzpicture}[scale = 0.8]
		\node[purple] at (4,-2.5) {$S\times S'$};
		\node[purple] at (8.5,-2.5) {$\sum_i c_i S_i$};
		\node[	trapezium, 
		draw = violet, 
		trapezium left angle=50, 
		trapezium right angle=130, 
		rotate = 40, 
		trapezium stretches=false,
		minimum height=1.6cm, 
		minimum width=0.8cm, 
		fill = magenta!20,
		]
		at (5,0.5) {};
		\draw[->-=1 rotate 0, color = red, dashed] (6.1,2.7) -- (6.8,2.7); 
		\node[	trapezium, 
		draw = violet, 
		trapezium left angle=50, 
		trapezium right angle=130, 
		rotate = 40, 
		trapezium stretches=false,
		minimum height=1.6cm, 
		minimum width=0.8cm, 
		fill = magenta!20,
		]
		at (6,0.5) {};
		\draw[->-=1 rotate 0, color = red, dashed] (7.1,2.7) -- (7.8,2.7);  
		\node at (8,0) {$\rightarrow$};    
		\node[	trapezium, 
		draw = violet, 
		trapezium left angle=50, 
		trapezium right angle=130, 
		rotate = 40, 
		trapezium stretches=false,
		minimum height=1.6cm, 
		minimum width=0.8cm, 
		fill = magenta!20,
		]
		at (10,0.5) {};
		\draw[->-=1 rotate 0, color = red, dashed] (11.1,2.7) -- (11.8,2.7);  
	\end{tikzpicture}
	\caption{The fusion of two parallel surface defects $S(\Sigma)$ and $S'(\Sigma)$ on an oriented manifold $\Sigma$. The red dashed arrows denote the normal vector of $\Sigma$. The fusion ``coefficients" $c_i$ are generally 1+1d TQFTs, rather than numbers.}\label{fig:fusion}
\end{figure}

Given two condensation surfaces $S$ and $S'$, we can define their fusion by bringing them parallelly to each other on an oriented manifold $\Sigma$:
\ie
S(\Sigma)\times S'(\Sigma) = \sum_i c_i(\Sigma) S_i(\Sigma)\,.
\fe
See Figure \ref{fig:fusion}.
Note that we define the fusion such that the normal vectors are pointing in the same direction.
As we shall see explicitly in examples in Section \ref{sec:example}, the fusion gives a sum of other surfaces, where the fusion ``coefficients" $c_i(\Sigma)$ are generally partition functions of 1+1d TQFTs which depend on the surface $\Sigma$.

Finally, given a condensation defect $S$, its orientation-reversal $\overline{S}$ is defined in the same way as for a general defect as follows. 
The insertion of $S$ on an oriented manifold $\Sigma$ is equivalent to the insertion of $\overline{S}$ on $\overline{\Sigma}$, where $\overline{\Sigma}$ is the orientation-reversal of the manifold $\Sigma$:
\ie\label{orientationrev}
\overline{S}(\overline{\Sigma} ) = S(\Sigma)\,. 
\fe

\subsection{Action on the lines}

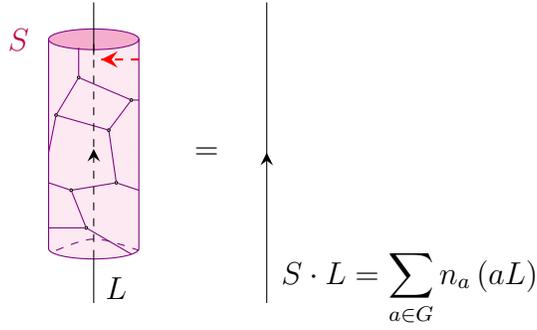
\begin{figure}[h]
	\centering
	\begin{tikzpicture}[scale = 1]
		\node[purple] at (-3,1.5) {$S$};
		\node[cylinder, 
		draw = violet, 
		text = purple,
		cylinder uses custom fill, 
		cylinder body fill = magenta!10, 
		cylinder end fill = magenta!40,
		minimum width = 1.2cm,
		minimum height = 3.06cm,
		shape border rotate = 90] (c) at (-2,0) {};    		
		\node[circle,inner sep=0pt,draw] at (-2.2,1) (a) {};
		\node[circle,inner sep=0pt,draw] at (-1.5,0.7) (b) {};
		\node[circle,inner sep=0pt,draw] at (-1.8,0.3) (c) {};
		\node[circle,inner sep=0pt,draw] at (-2.5,0.5) (d) {};
		\node[circle,inner sep=0pt,draw] at (-1.7,-0.4) (e) {};
		\node[circle,inner sep=0pt,draw] at (-2.3,-0.5) (f) {};
		\node[circle,inner sep=0pt,draw] at (-2.1,-1) (g) {};
		\draw[color = violet] (a) to (b);
		\draw[color = violet] (b) to (c);
		\draw[color = violet] (c) to (d);
		\draw[color = violet] (c) to (e);
		\draw[color = violet] (a) to (d);
		\draw[color = violet] (d) to (-2.6,0);
		\draw[color = violet] (e) to (f);
		\draw[color = violet] (e) to (-1.4,-0.5);
		\draw[color = violet] (f) to (-2.6,-0.4);
		\draw[color = violet] (f) to (g);
		\draw[color = violet] (g) to (-2.6,-1);
		\draw[color = violet] (g) to (-1.5,-1.35);
		\draw[color = violet] (a) to (-2.2,1.4);
		\draw[color = violet] (b) to (-1.4,0.7);
		\draw[dashed, color = violet] (-2.5,-1.3) .. controls (-2,-1.1) .. (-1.4,-1.3);
		\node at (-1.7,-1.8) {$L$};
		\draw[->-=0.5 rotate 0, dashed] (-2,-1.4) -- (-2,1.5);
		\draw[] (-2,1.3) -- (-2,2);
		\draw[] (-2,-2) -- (-2,-1.4);
		\node[] at (-0.5,0) {$=$};
		\draw[->-=0.5 rotate 0] (0.3,-2) -- (0.3,2);
		\node at (2.2,-1.8) {$S \cdot L = \displaystyle \sum_{a\in G}n_a\,(a L)$};
		\draw[->-=1 rotate 0, color = red, thick, dashed] (-1.4,1.24) -- (-1.9,1.24);
	\end{tikzpicture}
	\caption{A topological condensation surface $S$ acts on a (generally non-topological) line $L$ by enclosing a tubular neighborhood of the latter. The action maps the original line $L$ to a sum of other lines $\sum_{a\in G} n_a (aL)$, where $n_a\in \mathbb{Z}_{\ge0}$ and $aL$ is the fusion of the 1-form topological symmetry line $a$ with the (non-topological) line $L$. We denote this action as $S \cdot L$. The red arrow in the figure denotes the normal vector of the surface $S$.  }
	\label{fig:cylinder S}
\end{figure}

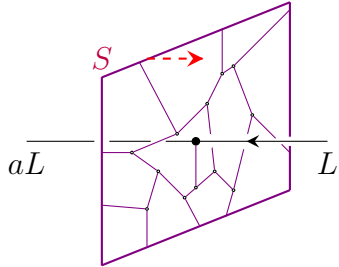
\begin{figure}[h]
	\centering
	\begin{tikzpicture}[scale = 0.5]
		\draw[color = violet, thick] (0,0) to (5,2);
		\draw[color = violet, thick] (0,5) to (5,7);
		\draw[color = violet, thick] (0,-0.01) to (0,5.01);
		\draw[color = violet, thick] (5,2) to (5,7);
		\node[circle,inner sep=0pt,draw] at (1.2,1.5) (a) {};
		\node[circle,inner sep=0pt,draw] at (1.5,2.5) (b) {};
		\node[circle,inner sep=0pt,draw] at (0.8,3) (c) {};
		\node[circle,inner sep=0pt,draw] at (2,3.5) (d) {};
		\node[circle,inner sep=0pt,draw] at (2.8,4.3) (e) {};
		\node[circle,inner sep=0pt,draw] at (3.5,5.3) (f) {};
		\node[circle,inner sep=0pt,draw] at (4,4) (g) {};
		\node[circle,inner sep=0pt,draw] at (3.5,2) (h) {};
		\node[circle,inner sep=0pt,draw] at (2.2,1.8) (j) {};
		\node[circle,inner sep=0pt,draw] at (3,2.5) (i) {};
		\node[circle,inner sep=0pt,draw] at (3.2,5.1) (k) {};
		\draw[color = violet] (a) to (b);
		\draw[color = violet] (b) to (c);
		\draw[color = violet] (c) to (d);
		\draw[color = violet] (d) to (e);
		\draw[color = violet] (e) to (k);
		\draw[color = violet] (k) to (f);
		\draw[color = violet] (e) to (i);
		\draw[color = violet] (i) to (j);
		\draw[color = violet] (i) to (h);
		\draw[color = violet] (h) to (g);
		\draw[color = violet] (g) to (f);
		\draw[color = violet] (j) to (b);
		\draw[color = violet] (a) to (1.2,0.5);
		\draw[color = violet] (a) to (0,1.6);
		\draw[color = violet] (c) to (0,3);
		\draw[color = violet] (d) to (1,5.4);
		\draw[color = violet] (j) to (2.6,1.05);
		\draw[color = violet] (h) to (3.3,1.3);
		\draw[color = violet] (g) to (5,3);
		\draw[color = violet] (f) to (5,6.8);
		\draw[color = violet] (k) to (3.2,6.3);
		\node[circle,inner sep=2pt,draw, fill, color = white] at (3.8,3.3) {};
		\node[circle,inner sep=1.1pt,draw, fill, color = white] at (5,3.3) {};
		\node[circle,inner sep=2pt,draw, fill, color = white] at (2.8,3.3) {};
		\node[circle,inner sep=1pt,draw, fill, color = white] at (4.7,3.3) {};
		\draw[->-=0.5 rotate 0, color = black] (6,3.3) -- (1.7,3.3);
		\draw[color = black] (1.3,3.3) -- (0.2,3.3);
		\draw[color = black] (-0.2,3.3) -- (-2,3.3);
		\node[anchor = north] at (6,3.3) {$L$};
		\node[anchor = north] at (-2,3.3) {$aL$};
		\node[purple] at (0,5.5) {$S$};
		\draw[->-=1 rotate 0, color = red, thick, dashed] (1.2,5.5) -- (2.7,5.5);
		\node[circle,inner sep=1pt,draw, fill, color = black] at (2.5,3.3) (a1) {};
		\node[circle,inner sep=0pt,draw, fill, color = white] at (2.5,2.07) {};
		\node[circle,inner sep=0pt,draw] at (2.5,2.07) (a2) {};
		\draw[color = violet] (a1) to (a2);
	\end{tikzpicture}
	\caption{A junction between the line $L$ and $aL$ on the condensation surface $S$. Such a junction exists if $n_a\neq0$ in the action $S \cdot L=\sum_{a\in G}n_a (aL)$. This configuration can be obtained by topologically deforming that in Figure \ref{fig:cylinder S}.}
	\label{fig:junction}
\end{figure}

We now discuss the action of the condensation surface $S$ on the operators.  
Since the condensation surface  is made out of topological lines, it is ``porous" to local operators and therefore act trivially on them. 
Rather, it can act nontrivially on the  lines $L$ by enclosing a tubular neighborhood of the latter as in Figure \ref{fig:cylinder S}.  
These line defects $L$ include not only the topological 1-form symmetry lines (which were used to construct the condensation surface), but also other non-topological lines in a general QFT.\footnote{Throughout the paper, $L$ always denotes a general, not necessarily topological line, while $a$ or $\eta^a$ denote   topological lines.}
Note that the condensation surface $S$ is oriented and in the figure we assume the orientation is pointing inwards. We denote this action of the surface $S$ on the line $L$ as $S \cdot L$. This action generally maps the line  $L$ to a finite sum of other lines:
 \ie \label{action.on.lines}
 S \cdot L = \sum_{a\in G} n_a (aL)\,,
 \fe
 where $n_a\in \mathbb{Z}_{\ge0}$ and $aL$ is the fusion of the topological 1-form symmetry line $a$ with $L$. 
This action  gives a junction between $L$ and $aL$ on the surface for all $a$ such that $n_a\neq0$. 
This is shown in Figure  \ref{fig:junction}, which can be deformed to Figure \ref{fig:cylinder S} topologically. 

Note that the action $\cdot$ of a surface on a line and the fusion product $\times$ of the surfaces obey the following relation:
\ie
(S \times S') \cdot L = S \cdot (S' \cdot L)\,.
\fe

 If instead the orientation of $S$ in Figure \ref{fig:cylinder S} is pointing outwards, then we denote the action as $\overline {S}\cdot L$, where $\overline{S}$ is the orientation-reversal of $S$ \eqref{orientationrev}.

We can recast the above Euclidean discussion into the action of the condensation surface operator on the Hilbert space. 
Let the space be a closed 2-manifold $\Sigma$ and ${\cal H}(\Sigma)$ be the associated Hilbert space. 
For simplicity, we assume the underlying QFT is a conformal field theory, so there is a map between local operators and states on $S^2$. 
The fact that the condensation surface acts trivially on the local operators implies that it is a trivial operator on ${\cal H}(S^2)$. 
However, it can still act nontrivially on other Hilbert spaces. In particular, the action of the condensation surface on the  Hilbert space ${\cal H}(T^2)$   captures its action on line defects.

Let us compare the action of a 1-form symmetry line  on a charged line with that of the condensation surface (see Figure \ref{fig: surface vs line action}). In the former case, the 1-form symmetry action does not change the type of the charged line, but only gives an overall phase \cite{Gaiotto:2014kfa}. 
In the latter case, the action generally maps one  line to a different type of line, or a sum of multiple types of lines with non-negative integer coefficients.
There are examples of condensation surfaces that do not act on any lines, and therefore is considered trivial. 
We will discuss such examples in Section \ref{sec:morita}.

\subsection{Higher quantum symmetry on condensation surfaces} \label{sec:quantum.symmetry}

It is well-known that after gauging an ordinary (non-anomalous) $\mathbb{Z}_N$ symmetry in a 1+1d QFT, one obtains a quantum  $\mathbb{Z}_N$ symmetry in the gauged theory \cite{Vafa:1989ih}, which acts on the twisted sector operators.\footnote{Quantum symmetry is also known as the dual or the orbifold symmetry.}  
 More generally, by gauging an abelian symmetry $G$ we obtain a quantum $\widehat{G}=\mathrm{Hom}(G,U(1))$ symmetry in the gauged theory.\footnote{See \cite{Bhardwaj:2017xup} for a review and generalizations to the cases of gauging non-abelian discrete symmetries and non-invertible lines.}

Quantum symmetry can be  generalized to higher gauging of 1-form symmetry lines on a 2-dimensional defect in 2+1d QFT. 
They are generated by topological lines that only live on the condensation surface defects. 
We will call them \textit{higher quantum symmetry lines}. 
Let us explain this point further below. 

As discussed above, the condensation defect is defined by summing over insertions of a mesh of $G$ 1-form symmetry lines on a 2-dimensional surface $\Sigma$ (see Figure \ref{fig:mesh}). 
We define $A \in H^1(\Sigma,G)$ to be the Poincare dual of these 1-form symmetry lines on  $\Sigma$. 
We emphasize that $A$ is not the Poincare dual with respect to the whole 2+1d spacetime, but with respect to the 2-dimensional surface $\Sigma$.

Let $Z[M;\Sigma,A]$  be the partition function of the theory on the 2+1d spacetime manifold $M$ coupled to the gauge field $A$. 
Since we  sum over insertions of the lines on a surface, $A$ is a dynamical 1-form gauge field on the surface $\Sigma$. 
The higher quantum  symmetry $\widehat{G}$ is defined  by coupling the partition function of the condensation defect to a background gauge field $\hat A \in H^1(\Sigma,\widehat{G})$ as follows: 
\begin{equation} \label{defect.coupled.to.gauge.fields}
	Z_\mathrm{defect}[M;\Sigma,\hat{A}] = \frac{1}{\sqrt{|H^1(\Sigma,G)|}} \sum_{A \in H^1(\Sigma,G)} e^{2\pi i \langle \hat{A}, A \rangle} \, Z[M;\Sigma,A]~.
\end{equation}
Here $\langle \cdot, \cdot \rangle: H^1(\Sigma,\widehat{G}) \times H^1(\Sigma,G) \to \mathbb{R}/\mathbb{Z} $ is the intersection pairing on cohomology.
The coupling of the defect to $\hat{A} \in H^1(\Sigma,\widehat{G})$ is equivalent to inserting $\widehat{G}$ topological lines on the Poincare dual of $\hat{A}$ in $\Sigma$. 
We conclude that the condensation defect possesses a higher quantum  symmetry $\widehat{G}$ generated by topological lines living only on the defect.

Gauging an ordinary quantum symmetry  takes us back to the original theory. 
Similarly, gauging the higher quantum symmetry $\widehat{G}$ on the defect returns the trivial defect (leaving the bulk theory intact):
\begin{equation}
	\frac{1}{\sqrt{|H^1(\Sigma,G)|}} \sum_{\hat A \in H^1(\Sigma,\widehat{G})} Z_\mathrm{defect}[M;\Sigma,\hat{A}] = Z[M;\Sigma,0]~. 
\end{equation}
We will give  explicit expressions for the quantum symmetry lines in later sections.

\bigskip\centerline{\it Fusion of lines with the condensation defect}\bigskip

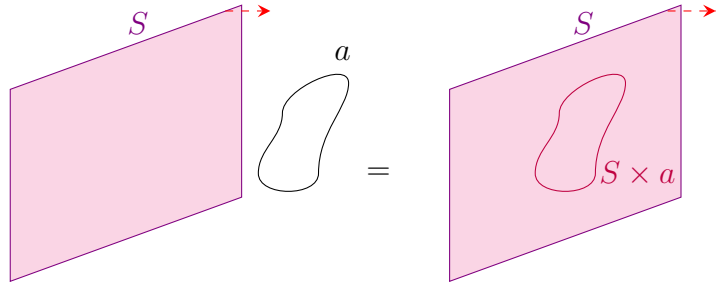
\begin{figure}
	\centering
	\begin{tikzpicture}[scale = 0.8]
		\node[	trapezium, 
		draw = violet, 
		minimum width=3.2cm,
		trapezium left angle=70, 
		trapezium right angle=110, 
		rotate = 20, 
		trapezium stretches=false,
		minimum height=2.4cm, 
		fill = magenta!20,
		]
		at (0.8,0.5) {};
		
		\draw[] (3,0) .. controls +(0,-0.4) and +(0,-0.4) .. (4,0).. controls +(0,0.7) and +(0,-0.4) .. (4.5,1.5).. controls +(0,0.4) and +(0,0.4) .. (3.4,1).. controls +(0,-0.4) and +(0,0.4) .. (3,0);
		\node[] at (5,0) {$=$};
		\node[] at (4.4,2) {$a$};
		\node[color = violet] at (1,2.5) {$S$};
		\draw[->-=1 rotate 0, color = red, dashed] (2.45,2.7) -- (3.2,2.7);
	\end{tikzpicture}
	\hskip 0.5cm
	\begin{tikzpicture}[scale = 0.8]
		\node[] at (0,0) {};
		\node[	trapezium, 
		draw = violet, 
		minimum width=3.2cm,
		trapezium left angle=70, 
		trapezium right angle=110, 
		rotate = 20, 
		trapezium stretches=false,
		minimum height=2.4cm, 
		fill = magenta!20,
		]
		at (0.5,0.5) {};
		
		\draw[color = purple] (0,0) .. controls +(0,-0.4) and +(0,-0.4) .. (1,0).. controls +(0,0.7) and +(0,-0.4) .. (1.5,1.5).. controls +(0,0.4) and +(0,0.4) .. (0.4,1).. controls +(0,-0.4) and +(0,0.4) .. (0,0);
		\node[color = violet] at (0.8,2.5) {$S$};
		\node[color = purple] at (1.7,0) {$S \times a $};
		\draw[->-=1 rotate 0, color = red, dashed] (2.2,2.7) -- (3,2.7);
	\end{tikzpicture}
	\caption{Fusing a line $a$ from the right with a surface $S$ creates a line defect on the latter. We denote this line defect living on $S$ by $S\times a$. Analogously, fusing a line $a$ from the left we create a line defect $a \times S$ on $S$. Generally the two ways of fusing $a$ with $S$ lead to different lines living on $S$.}
	\label{fig:Fusion of line to surface}
\end{figure}

Generally, the fusion of a bulk topological line with a surface defect gives a topological line living only on the latter as shown in Figure \ref{fig:Fusion of line to surface}.
Consider a 1-form symmetry line $a$ which is close and parallel to the condensation defect $S$. We denote the fusion of $a$ with $S$ from the left and the right, respectively, as 
\begin{equation}
	a \times S  \qquad \text{and} \qquad S \times a ~. \label{q.symm}
\end{equation}
These lines become the higher quantum symmetry lines on $S$.\footnote{More precisely, \eqref{q.symm} specifies a junctions between $a$ and the higher quantum symmetry lines on $S$.} Therefore, the fusion defines maps (or more precisely a homomorphisms) from the bulk 1-form symmetry to the higher quantum symmetry on the condensation defect.\footnote{These maps are called $\alpha$-induction in the mathematical literature, and are denoted by $\alpha^+(a) = a \times S$ and $\alpha^-(a) =S \times a$. They first appeared in subfactor theory~\cite{Longo:1994xe,Bockenhauer:1998ca}, and then in category theory~\cite{muger2003subfactors,2010arXiv1009.2117D}. Finally, for their interpretation in terms of topological defects see \cite{Fuchs:2002cm,Kapustin:2010if,Komargodski:2020mxz}.}  
However, the homomorphism is generally not surjective: there are higher quantum symmetry lines that cannot be obtained by fusing a bulk line with the surface. 
We will see such examples in the case of a $\mathbb{Z}_2$ boson line in Section \ref{sec:z2}.
 
\begin{figure}
	\centering
	\begin{tikzpicture}[scale = 0.8]
		\node[	trapezium, 
		draw = violet, 
		minimum width=3.2cm,
		trapezium left angle=70, 
		trapezium right angle=110, 
		rotate = 20, 
		trapezium stretches=false,
		minimum height=2.4cm, 
		fill = magenta!20,
		]
		at (0.8,0.5) {};

		\draw[->-=0.7 rotate 0] (4,.5) -- (1,0.5);
		\draw[dashed] (-1.2,.5) to (1,0.5);
		\draw[->-=0.5 rotate 0] (-1.2,.5) -- (-2.5,0.5);
		\node[] at (4,1) {$a$};
		\node[] at (-2,1) {$b$};
		\node[color = violet] at (1,2.5) {$S$};
		\draw[->-=1 rotate 0, color = red, dashed] (2.45,2.7) -- (3.2,2.7);
		\node[circle,inner sep=1pt,draw, fill, color = black] at (0.9,0.5) {};
		
		\draw[] (-0.63,2.4) circle (0.6cm and 0.6cm);
		\fill[left color=gray!50!black,right color=gray!50!black,middle color=gray!50,shading=axis,opacity=0.25] (-0.1,2.7) -- (0.9,0.5) -- (-1.1,2) arc (225:360:0.63cm and 0.69cm);
		
		\draw[->-=0.7 rotate 0 , thick] (-0.1,2.4) -- (-0.6,2.4);
		\draw[->-=0.7 rotate 0, thick] (-0.6,2.4) -- (-1.2,2.4);
		\draw[color = violet, thick] (-0.6,2.4) -- (-0.6,2.9);
		\node[circle,inner sep=1pt,draw, fill, color = black] at (-0.6,2.4) {};
		\node[] at (-0.3,2.1) {$ \scriptstyle a $};
		\node[] at (-0.9,2.1) {$ \scriptstyle b $};
		\node[] at (5,0.5) {$=$};
	\end{tikzpicture}
	\begin{tikzpicture}[scale = 0.8]
		\node[	trapezium, 
		draw = violet, 
		minimum width=3.2cm,
		trapezium left angle=70, 
		trapezium right angle=110, 
		rotate = 20, 
		trapezium stretches=false,
		minimum height=2.4cm, 
		fill = magenta!20,
		]
		at (0.8,0.5) {};

		\draw[->-=0.7 rotate 0] (4,1.5) -- (1,1.5);
		\draw[dashed] (-1.2,-0.5) to (1,-0.5);
		\draw[->-=0.5 rotate 0] (-1.2,-0.5) -- (-2.5,-0.5);
		\draw[->-=0.6 rotate 0, color = purple, thick] (1,1.5) -- (1,0.5);
		\draw[->-=0.6 rotate 0, color = purple, thick] (1,0.5) -- (1,-0.5);
		\node[] at (4,2) {$a$};
		\node[] at (-2,0) {$b$};
		\node[color = purple] at (1.8,1) {$S \times a$};
		\node[color = purple] at (0,0) {$b \times S $};
		\node[color = violet] at (1,2.5) {$S$};
		\draw[->-=1 rotate 0, color = red, dashed] (2.45,2.7) -- (3.2,2.7);
		\node[circle,inner sep=1pt,draw, fill, color = black] at (1,1.5) {};
		\node[circle,inner sep=1pt,draw, fill, color = purple] at (1,0.5) {};
		\node[circle,inner sep=1pt,draw, fill, color = black] at (1,-0.5) {};
	\end{tikzpicture}
	\caption{A junction between topological lines $b$ and $a$ meeting at the surface defect $S$ is equivalent to a junction between the lines $b \times S$ and $S \times a$ living on $S$. The lines $b \times S$ and $S \times a$ denote the fusion of the bulk lines with surface as in Figure \ref{fig:Fusion of line to surface}.}
	\label{fig:junction on S}
\end{figure}
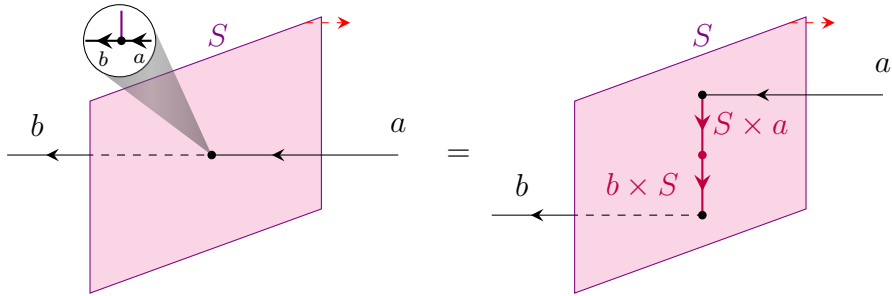

As we now explain, the action of a surface defect $S$ on lines is determined by the fusion of lines with $S$. As argued before, a junction between the bulk lines $b$ and $S \cdot a$ is equivalent to a junction between $b$ (going to the left of $S$) and $a$ (coming from the right of $S$) that meet on $S$ as in Figure \ref{fig:junction on S}.
Furthermore, the latter can be obtained by first fusing the lines $b$ and $a$ with surface and then morphing them into each other by a junction on the surface as shown in Figure \ref{fig:junction on S}. Therefore, there is a junction between the bulk lines $b$ and $S \cdot a$ if any only if there is a junction between $b \times S$ and $S \times a$. More specifically we find\footnote{This is because $b\times S$ and $S \times a$ are simple lines (i.e., they cannot be decomposed into other lines living on the surface), therefore they have a junction between them if and only if they are identical. For general non-invertible topological lines $\ell,\ell'$ we have $S \cdot \ell = \sum_{\ell'} \mathrm{dim Hom}(\ell' \times S, S \times \ell) \, \ell'$.}
\begin{equation} \label{action.and.fusion}
	S \cdot a = \sum_{b \in G} \delta_{b \times S,S \times a} \, b ~.
\end{equation}

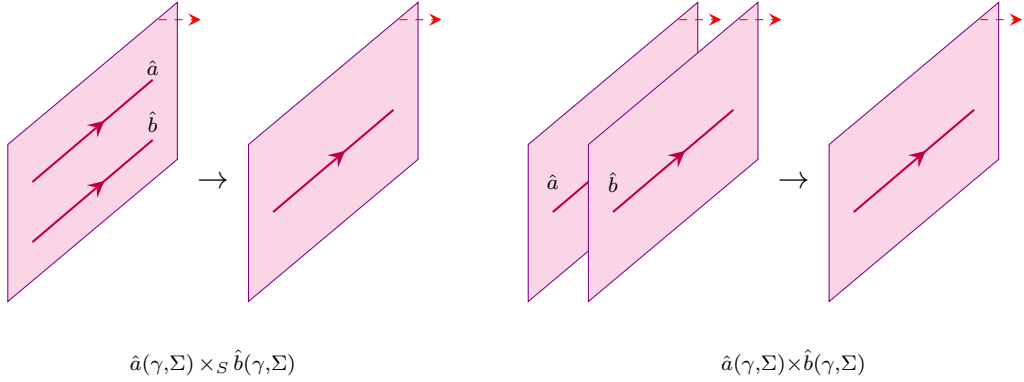
\begin{figure}
\centering
\begin{tikzpicture}[scale = 0.8]
\node[	trapezium, 
		draw = violet, 
		trapezium left angle=50, 
		trapezium right angle=130, 
		rotate = 40, 
		trapezium stretches=false,
		minimum height=1.6cm, 
		minimum width=0.8cm, 
		fill = magenta!20,
		]
    at (0,0.5) {}; 
\node at (2,0) {$\rightarrow$}; 
\draw[->-=1 rotate 0, color = red, dashed] (1.1,2.7) -- (1.8,2.7);    
\draw[->-=0.6 rotate 0, color = purple, thick] (-1,0) -- (1,1.7);
\node[] at (1,1.9) {$\scriptstyle \hat{a}$}; 
\draw[->-=0.6 rotate 0, color = purple, thick] (-1,-1) -- (1,0.7);
\node[] at (1,1) {$\scriptstyle \hat{b}$}; 
\node[] at (2,-3) {$\scriptstyle \hat{a}(\gamma, \Sigma) \, \times_S \, \hat{b}(\gamma, \Sigma)$};  
\node at (2,-2.5) {$~$}; 

\node[	trapezium, 
		draw = violet, 
		trapezium left angle=50, 
		trapezium right angle=130, 
		rotate = 40, 
		trapezium stretches=false,
		minimum height=1.6cm, 
		minimum width=0.8cm, 
		fill = magenta!20,
		]
    at (4,0.5) {}; 
\draw[->-=1 rotate 0, color = red, dashed] (5.1,2.7) -- (5.8,2.7); 
\draw[->-=0.6 rotate 0, color = purple, thick] (3,-0.5) -- (5,1.2);
\node[] at (4,-2.5) {$~$}; 
\end{tikzpicture}
\hskip 1cm
\begin{tikzpicture}[scale = 0.8]
\node[	trapezium, 
		draw = violet, 
		trapezium left angle=50, 
		trapezium right angle=130, 
		rotate = 40, 
		trapezium stretches=false,
		minimum height=1.6cm, 
		minimum width=0.8cm, 
		fill = magenta!20,
		]
    at (5,0.5) {};
\draw[->-=1 rotate 0, color = red, dashed] (6.1,2.7) -- (6.8,2.7); 
\draw[->-=0.6 rotate 0, color = purple, thick] (4,-0.5) -- (6,1.2);    
\node[	trapezium, 
		draw = violet, 
		trapezium left angle=50, 
		trapezium right angle=130, 
		rotate = 40, 
		trapezium stretches=false,
		minimum height=1.6cm, 
		minimum width=0.8cm, 
		fill = magenta!20,
		]
    at (6,0.5) {};
\draw[->-=1 rotate 0, color = red, dashed] (7.1,2.7) -- (7.8,2.7);  
\draw[->-=0.6 rotate 0, color = purple, thick] (5,-0.5) -- (7,1.2); 
\node[] at (4,0) {$\scriptstyle \hat{a}$}; 
\node[] at (5,0) {$\scriptstyle \hat{b}$}; 
\node at (8,0) {$\rightarrow$};    
\node[] at (8,-3) {$\scriptstyle \hat{a}(\gamma, \Sigma) \times \hat{b}(\gamma, \Sigma)$};  
\node at (8,-2.5) {$~$}; 
\node[	trapezium, 
		draw = violet, 
		trapezium left angle=50, 
		trapezium right angle=130, 
		rotate = 40, 
		trapezium stretches=false,
		minimum height=1.6cm, 
		minimum width=0.8cm, 
		fill = magenta!20,
		]
    at (10,0.5) {};
\draw[->-=1 rotate 0, color = red, dashed] (11.1,2.7) -- (11.8,2.7);  
\draw[->-=0.6 rotate 0, color = purple, thick] (9,-0.5) -- (11,1.2); 
\node[] at (10,-2.5) {$~$}; 
\end{tikzpicture}
\caption{The figure on the left depicts the fusion of two lines living on the same surface $\Sigma$, which we denote by $\hat{a}(\gamma, \Sigma) \times_S \hat{b}(\gamma, \Sigma) $. 
The figure on the right depicts the fusion of the two surfaces with lines, which we denote by $\hat{a}(\gamma, \Sigma) \times \hat{b}(\gamma, \Sigma)$.}
\label{fig: fusion of lines on S}
\end{figure}

Before we end this section let us introduce some extra notation.
 We denote a general higher quantum symmetry line on $S(\Sigma)$ by 
\begin{equation}
\hat{a}(\gamma, \Sigma)\,.
\end{equation}
We denote the fusion of two such lines $\hat{a}(\gamma, \Sigma)$  and $\hat{b}(\gamma, \Sigma)$ on the surface $S(\Sigma)$  as  (left figure in \ref{fig: fusion of lines on S}):
\begin{equation}
\hat{a}(\gamma, \Sigma) \times_S \hat{b}(\gamma, \Sigma)\,.
\end{equation}
On the other hand,  the fusion  of  two parallel  surfaces  with lines is denoted as (right figure in \ref{fig: fusion of lines on S})
\begin{equation}
\hat{a}(\gamma, \Sigma) \times \hat{b}(\gamma, \Sigma)\,.
\end{equation}
More generally, one can bring together two such higher quantum symmetry lines on different 1-cycles $\gamma_1,\gamma_2 \in H_1(\Sigma,\mathbb{Z})$. Since the fusion product is typically defined for parallel defects, we will always assume the two lines are on the same 1-cycle in the fusion products $\times_S$ and $\times$.

\section{Higher gauging of $\mathbb{Z}_2$ 1-form symmetries} \label{sec:z2}

We start with the simplest possible higher gauging:  gauging a $\mathbb{Z}_2$ 1-form symmetry on a 2-dimensional surface in a 2+1d QFT. 
As we will see below, there are two possible outcomes for the condensation surface, one is invertible while the other is non-invertible. 

\subsection{1-gaugeable $\mathbb{Z}_2$ 1-form symmetries}

 Let $\eta$ be the topological line generating the $\mathbb{Z}_2$ 1-form symmetry. It obeys the $\mathbb{Z}_2$ fusion rule $\eta^2 = 1$.  
Let the topological spin of the $\mathbb{Z}_2$ line be $\theta(\eta)$, defined as in Figure \ref{fig:spin}.  For a $\mathbb{Z}_2$ line, the topological spin $\theta(\eta)$ can take four possible values in $\{\pm 1,\pm i\}$ \cite{Hsin:2018vcg}.\footnote{This can also be seen using \eqref{Bab} as follows. Since $1=B(\eta, \eta^2)= B(\eta,\eta)^2 = 1/\theta(\eta)^4$, we find that $\theta(\eta)$ must be  a fourth root of unity.} 
The four values correspond to the possible anomaly of the 1-form symmetry classified by $H^4(B^2\mathbb{Z}_2,U(1))=\mathbb{Z}_4$. 
The 1-form symmetry is anomaly free (0-gaugeable) if and only if $\theta(\eta)=1$.\footnote{Gauging a fermion line in the whole spacetime results in a spin QFT. Since we only consider non-spin QFT in most of this paper, we do not allow ourselves to gauge a fermion line in the whole spacetime.  Instead, we will see momentarily that we can gauge the fermion line on a 2-dimensional surface, which gives a topological surface defect in a non-spin QFT.}
 See Table \ref{z2.condensability}.

For the higher gauging of the 1-form symmetry, the braiding data are no longer relevant on the 2-dimensional surface. 
The higher gauging is implemented by   summing over insertions  of line defects along the dual graph of a triangulation of the 2d surface,  as if we are gauging a 0-form symmetry in 1+1d. 
For the higher gauging to be independent of the triangulation, i.e., for the 1-form symmetry to be 1-gaugeable, we demand the crossing of the lines, which is captured by the $F$-symbols, to be trivial 
(see \cite{Freed:1987qk,Bhardwaj:2017xup,Chang:2018iay,Lin:2019kpn,Lin:2021udi} for discussions on the obstruction to gauging in 1+1d).

\begin{figure}[h]
  \centering
  \begin{tikzpicture}[scale = 0.8] 
  \draw[ color = black,  thick] (0,0) .. controls +(0.7,0.7) and +(0.7,-0.7) .. (0,2);
  \draw[color = black,  thick] (2,0) .. controls +(-0.7,0.7) and +(-0.7,-0.7) .. (2,2);
  \node[] at (3,1) {$= F$};
  \node[anchor = north] at (2,0) {$\eta$};
  \node[anchor = north] at (0,0) {$\eta$};
  \end{tikzpicture}
    \begin{tikzpicture}[scale = 0.8] 
  \draw[color = black,  thick] (0,0) .. controls +(0.7,0.7) and +(-0.7,+0.7) .. (2,0);
   \draw[color = black,  thick] (0,2) .. controls +(0.7,-0.7) and +(-0.7,-0.7) .. (2,2);
  \node[anchor = north west] at (2,0) {$\eta$};
  \node[anchor = north west] at (2,2) {$\eta$};
  \end{tikzpicture}
  \caption{The crossing phase $F$ for a $\mathbb{Z}_2$ topological line $\eta$. }\label{fig:crossingz2}
\end{figure}

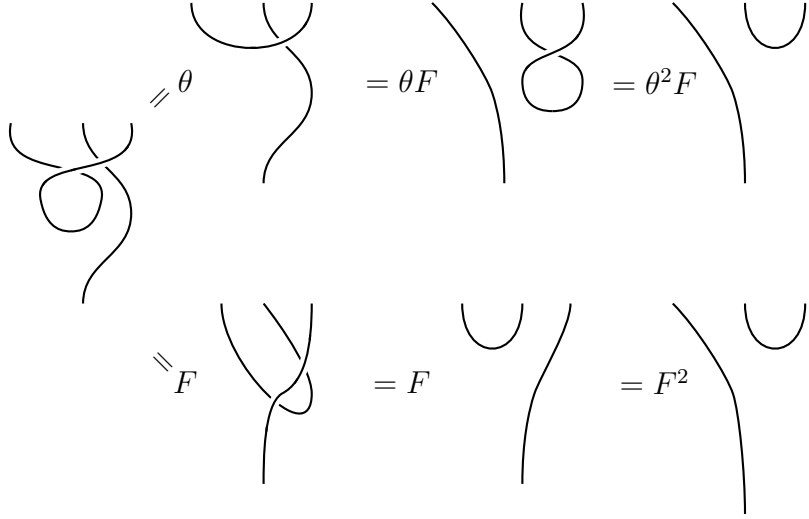
\begin{figure}[h]
  \centering
  \begin{tikzpicture}[scale = 0.8] 
	\begin{knot}[
		clip width=5,
		flip crossing=1,
		]
		\strand[thick] (-1.5,0) .. controls +(-0.2,-1) and +(0.2,0.8) .. (0,-1.3) .. controls 		                     +(-0.1,-0.5) and +(0.1,0) .. (-0.5,-1.8);
		\strand[thick] (0.5,0) .. controls +(0.2,-1) and +(-0.2,0.8) ..  (-1,-1.3).. controls  		  					 +(+0.1,-0.5) and   +(-0.1,0) .. (-0.5,-1.8);
		\strand[thick] (-0.3,0) .. controls +(0,-0.7) and +(0,0.7) .. (0.5,-1.5) .. controls 		                    		+(0,-0.7) and +(0,0.7) .. (-0.3,-3);
	\end{knot}
	
	\node[] at (1,0.4) {\rotatebox{45}{$=$}};
	\node[] at (1.4,0.7) {$\theta$};
	
	\begin{knot}[
		clip width=5,
		flip crossing=0,
		]
		\strand[thick] (1.5,2) .. controls +(0,-1) and +(0,-1) .. (3.5,2);
		\strand[thick] (2.7,2) .. controls +(0,-0.7) and +(0,0.7) .. (3.5,0.5) .. controls 		                    		+(0,-0.7) and +(0,0.7) .. (2.7,-1);
	\end{knot}

	\node[] at (5,0.7) {$=\theta F $};
	
	\begin{knot}[
		clip width=5,
		flip crossing=1,
		]
		\strand[thick] (5.5,2) .. controls +(0.4,-0.4) and +(-0.1,0.3) .. (6.5,0.5) .. controls 		                    		+(0.1,-.3) and +(0,0.7) .. (6.7,-1);
		\strand[thick] (7,2) .. controls +(-0.2,-1) and +(0,0.6) .. (8,0.7) .. controls 		                     +(0,-0.5) and +(0.1,0) .. (7.5,0.2);
		\strand[thick] (8,2) .. controls +(0.2,-1) and +(0,0.6) ..  (7,0.7).. controls  		  					 +(+0,-0.5) and   +(-0.1,0) .. (7.5,0.2);
	\end{knot}
	
	\node[] at (9.2,0.7) {$=\theta^2 F $};
	
	\begin{knot}[
		clip width=5,
		flip crossing=1,
		]
		\strand[thick] (9.5,2) .. controls +(0.4,-0.4) and +(-0.1,0.3) .. (10.5,0.5) .. 											controls +(0.1,-.3) and +(0,0.7) .. (10.7,-1);
		\strand[thick] (10.7,2) .. controls +(0,-1) and +(0,-1) .. (11.7,2);
	\end{knot}
	
	\node[] at (1,-4) {\rotatebox{-45}{$=$}};
	\node[] at (1.4,-4.3) {$F $};
	
	\begin{knot}[
		clip width=5,
		]
		\strand[thick] (3.5,-3) .. controls +(0,-0.7) and +(0.4,0.2) .. (3,-4.5) .. controls										+(-0.3,-0.2) and +(0,0.4) .. (2.7,-6);
		\strand[thick] (2.0,-3) .. controls +(0,-1) and +(0,-1).. (3.5,-4.5) .. controls 											+(0,+0.5) and +(0,0).. (2.7,-3);
	\end{knot}

	\node[] at (5,-4.3) {$= F $};
	
	\begin{knot}[
		clip width=5,
		flip crossing=1,
		]
		\strand[thick] (7.8,-3) .. controls +(0,-0.4) and +(0.1,0.3) .. (7.2,-4.5) .. 									controls +(-0.1,-.3) and +(0,0.7) .. (7,-6);
		\strand[thick] (6,-3) .. controls +(0,-1) and +(0,-1) .. (7,-3);
	\end{knot}

	\node[] at (9.2,-4.3) {$ =F^2 $};
	
	\begin{knot}[
		clip width=5,
		flip crossing=1,
		]
		\strand[thick] (9.5,-3) .. controls +(0.4,-0.4) and +(-0.1,0.3) .. (10.5,-4.5) .. 											controls +(0.1,-.3) and +(0,0.7) .. (10.7,-6.5);
		\strand[thick] (10.7,-3) .. controls +(0,-1) and +(0,-1) .. (11.7,-3);
	\end{knot}
  \end{tikzpicture}
  \caption{The consistency condition (hexagon identity) between the $F$-symbol and the topological spin $\theta$ of a $\mathbb{Z}_2$ line implies that $F= \theta^2$.}
  \label{fig:spin-F}
\end{figure}

For a $\mathbb{Z}_2$ topological line, the only nontrivial crossing phase is shown in Figure \ref{fig:crossingz2}.
It is determined from the topological spin $\theta(\eta)$ following the consistency condition (i.e., the hexagon identity)  
 in Figure \ref{fig:spin-F}. We obtain:
\ie
F =\theta(\eta)^2\,.
\fe

\begin{table}[h!]
\begin{align*}
\left.\begin{array}{|c|c|c|c|c|}
\hline
~~\text{Anyon} ~~& ~~\text{Boson}~~ & ~~~~\text{Semion}~~ ~~&~~~~ \text{Fermion}~~~~ &~~ \text{Anti-semion} ~~\\
\hline 
\text{Spin} &~~ \theta(\eta)=1 ~~& \theta(\eta)=i & \theta(\eta)=-1 & \theta(\eta)=-i \\
\hline 
F\text{-symbol} & F=1 &F=-1 & F=1 &F=-1 \\
\hline  
~~0\text{-gaugeable}~~& ~~ \checkmark~~& ~~\times~~ &~~  \times~~ &~~\times~~ \\
\hline
~~1\text{-gaugeable}~~& ~~~\checkmark ~~~& ~~\times~~ &~~\checkmark  ~~ &~~\times~~ \\
\hline 
\end{array}\right.
\end{align*}
\caption{Gaugeability of a general $\mathbb{Z}_2$ 1-form symmetry line in 2+1d QFT. Here $\eta$ denotes the nontrivial $\mathbb{Z}_2$ line. A boson line is anomaly-free in the usual sense (0-gaugeable), and is in particular also 1-gaugeable. A fermion line is 0-anomalous, but it is 1-gaugeable. The semion and the anti-semion lines are not only 0-anomalous, but also 1-anomalous.  Here $F$ stands for the value of the only nontrivial component $F^{\eta\eta\eta}_\eta$ of the $F$-symbol. \label{z2.condensability}}
\end{table}

We summarize the possible values of  the $F$-symbol in Table \ref{z2.condensability}. By demanding the trivial crossing $F=1$, we see that the $\mathbb{Z}_2$ 1-form symmetry is 1-gaugeable if and only if the topological spin of $\eta$ satisfies
\begin{equation}\label{z21gaugeable}
	\text{1-gaugeable:}~~~~\theta(\eta) = \pm 1 \,.
\end{equation}
That is, a $\mathbb{Z}_2$ 1-form symmetry is 1-gaugeable iff it is generated by a boson or a fermion line.

The 0-anomaly of a $\mathbb{Z}_2$ 1-form symmetry in 2+1d is classified by $H^4(B^2\mathbb{Z}_2 , U(1))=\mathbb{Z}_4$.  
On the other hand, the 1-anomaly of a $\mathbb{Z}_2$ 1-form symmetry is classified by the ordinary anomaly of gauging a 0-form symmetry in 1+1d, which is given by $H^3(B\mathbb{Z}_2,U(1)) =\mathbb{Z}_2$. 
By forgetting the braiding, this defines a map from the 0-anomaly to the 1-anomaly of the 1-form symmetry:
\ie
H^4(B^2\mathbb{Z}_2,U(1)) \to H^3(B\mathbb{Z}_2,U(1)) \,,
\fe
which maps the generator of the former to the generator of the latter. 
This map  was studied in \cite{Kaidi:2021xfk}.

When the  condition \eqref{z21gaugeable} is satisfied we can gauge the $\mathbb{Z}_2$ 1-form symmetry on a surface to construct a topological condensation surface defect $S$. 
The condensation defect on a  surface $\Sigma$ is given by the sum over insertions of the 1-form symmetry  lines:
\begin{equation}\label{Z2condsurface}
	S(\Sigma) = \frac{1}{\sqrt{|H_1(\Sigma,\mathbb{Z}_2)|}} \sum_{\gamma \in H_1(\Sigma,\mathbb{Z}_2)} \eta(\gamma)  \, .
\end{equation}
Given any topological surface defect, we can always redefine it by a topological Euler counterterm, which rescales the normalization of the defect by $\lambda^{\chi(\Sigma)}$ for any real number $\lambda$. 
Here we have made a choice for this Euler counterterm to fix the normalization for later convenience.

\subsection{Fusion rules of condensation surfaces}\label{sec:z2fusion}

From this point on, we assume the $\mathbb{Z}_2$ 1-form symmetry to be 1-gaugeable and hence it is either a boson or a fermion, i.e., $\theta(\eta)=\pm1$. 

We now  determine the fusion rule of the condensation surface  $S$. For simplicity, we  take the surface to be a 2-torus at a fixed time  with generating 1-cycles ${\bf A},{\bf B} \in H_1(T^2,\mathbb{Z})$ and $\langle {\bf A},{\bf B}\rangle=1$, and leave the more general fusion rule for Section \ref{sec:ZN}. 
The fusion $S\times S$  is given by
\begin{equation}
	S(T^2) \times S (T^2) = \left[ \frac{1}{2} \big( 1+\eta({\bf A})+\eta({\bf B})+\eta({\bf A}+{\bf B}) \big) \right]^2 \, .
\end{equation}
 Since $\eta$ is a $\mathbb{Z}_2$ line, we have $\eta({\bf A})^2 = \eta({\bf B})^2 =1$.  
 The only nontrivial step is to express $\eta({\bf A}+{\bf B})$  in terms of $\eta({\bf A}) $ and $\eta({\bf B})$.
Using the braiding relation, we obtain

\begin{equation}
\centering
\raisebox{-4.5em}{\begin{tikzpicture}
\draw[thick] (0,0) to (0,2);
\draw[thick] (0,2) to (2,2);
\draw[thick] (2,2) to (2,0);
\draw[thick] (2,0) to (0,0);
\draw[violet] (1,0) .. controls +(0,0.5) and +(-0.5,0) .. (2,1);
\draw[violet] (0,1) .. controls +(0.5,0) and +(0,-0.5) .. (1,2);
\node[] at (1,-0.3) {\rotatebox{90}{$\,=$}};
\node[anchor = south] at (1,-1) {$\eta({\bf A+B})$};
\end{tikzpicture}} ~ =\theta(\eta) ~
\raisebox{-4.5em}{\begin{tikzpicture}
\draw[thick] (0,0) to (0,2);
\draw[thick] (0,2) to (2,2);
\draw[thick] (2,2) to (2,0);
\draw[thick] (2,0) to (0,0);
\draw[violet] (1,0) to (1,2);
\node[circle,inner sep=2pt,draw, fill, color = white] at (1,1) {};
\draw[violet] (0,1) to (2,1);
\node[] at (1,-0.3) {\rotatebox{90}{$\,=$}};
\node[anchor = south] at (1,-1) {$\eta({\bf A})\,\eta({\bf B})$};
\node[anchor = south] at (0.4,1) {$\bf A$};
\node[anchor = south] at (1.3,0) {$\bf B$};
\end{tikzpicture}}
\end{equation}
Therefore,
\ie \label{z2.braiding}
\eta({\bf A+B}) = \theta(\eta) \, \eta({\bf A})\eta({\bf B}) = \theta(\eta) \, \eta({\bf B})\eta({\bf A})\,.
\fe
More generally, we have
\begin{equation}\label{z2quantumtorus}
	\eta\left( \gamma \right) \eta\left( \gamma' \right) = \theta(\eta)^{ \langle \gamma,\gamma' \rangle} \, \eta\left( \gamma + \gamma' \right) \,,
\end{equation}
where $\langle \gamma, \gamma' \rangle $ is the intersection number between the 1-cycles $\gamma,\gamma' \in H_1(\Sigma,\mathbb{Z}_2)$.

Using the above relation, we find the fusion rule of the condensation surface
\begin{equation}
	S(T^2)  \times S(T^2) = 1+\eta({\bf A})+\eta({\bf B})+\eta({\bf A+B}) = 2 S(T^2) \, ,~~~\theta(\eta)=1\,,
\end{equation}
 when the $\mathbb{Z}_2$ line $\eta$ is a boson.  
When $\eta$ is a fermion, the fusion rule is
\begin{equation}
	S(T^2)\times S(T^2) = 1 \, ,~~~\theta(\eta)=-1\,.
\end{equation}
Interestingly,  the condensation surface is non-invertible if the underlying $\mathbb{Z}_2$ 1-form symmetry line is a boson, and is an invertible ordinary $\mathbb{Z}_2$ 0-form symmetry if the 1-form symmetry line is a fermion. 
The invertible $\mathbb{Z}_2$ surface in the case of  a fermion line was mentioned in \cite{Aharony:2016jvv}.\footnote{We thank Po-Shen Hsin for pointing this reference to us.} 

More precisely, the fusion ``coefficient" 2  in the non-invertible fusion rule should be interpreted as the 1+1d $\mathbb{Z}_2$ gauge theory, rather than a number. 
In Section \ref{sec:ZN}, we will compute the fusion rule on a general 2-dimensional surface $\Sigma$ and find
\begin{equation}
	S (\Sigma) \times S (\Sigma) = 	\sqrt{|H_1(\Sigma,\mathbb{Z}_2)|} \,S (\Sigma) \,,~~~\theta(\eta) = 1 ~. \label{Z2 S fusion}
\end{equation}
Indeed, the partition function of the 1+1d $\mathbb{Z}_2$ gauge theory is $\sqrt{|H_1(\Sigma,\mathbb{Z}_2)|}$ (up to the usual Euler counterterm ambiguity). Therefore, for $\theta(\eta)=1$, we write the general fusion rule as
\ie\label{higherprojection}
&S\times S= (\mathcal{Z}_2)\, S \,,~~~\theta(\eta) = 1 \,,
\fe
where we have defined
 \ie
 (\mathcal{Z}_N)\equiv  \text{1+1d $\mathbb{Z}_N$ gauge theory}  \,.
 \fe
We will verify this fusion rule explicitly using the explicit worldsheet Lagrangian for the surface in various examples in Section \ref{sec:example}.
Note that this is precisely the non-invertible fusion rule of the Cheshire string \cite{Else:2017yqj,Johnson-Freyd:2020twl}.

This non-invertible fusion rule \eqref{higherprojection} is analogous to the ordinary projection algebra.
Consider a QFT  with  an ordinary $\mathbb{Z}_2$ symmetry operator $U$  with $U^2=1$. 
We can define a new conserved operator $P_+\equiv 1+U$, which is not simple because it is the sum of two other operators of the same dimensionality.\footnote{As discussed below \eqref{inequality}, $P_+$ can be thought of as the condensation of a 0-form symmetry $U$ on a codimension-1 manifold. } 
The non-simple operator $P_+$ is twice the projection operator, and it obeys a non-invertible fusion rule, ${P_+}\times P_+= 2P_+$. 
In contrast, the condensation surface $S$ here is a simple surface defect because it cannot be written as the sum of other surface defects. 
The non-invertible fusion rule \eqref{higherprojection} is similar to the projection algebra, but with the fusion  ``coefficient"
 replaced by the 1+1d $\mathbb{Z}_2$ gauge theory $(\mathcal{Z}_2)$.

We conclude that the fusion rule of the condensation surface is
\begin{equation}\label{genusgfusionZ2}
	S \times S = \begin{cases}
			(\mathcal{Z}_2)\,S &\,,~ \theta(\eta) = 1 \\
		~~~~~~1 &\,,~ \theta(\eta) = -1 \\
	\end{cases} ~.
\end{equation}
In Section \ref{sec:example} we will give several examples realizing these condensation defects. 
For instance, both kinds of condensation defects are realized in the 2+1d $\mathbb{Z}_2$ gauge theory (see Section \ref{sec:z2gauge}). 
The invertible condensation defect is also realized as the charge conjugation symmetry in $U(1)_4$ Chern-Simons theory (see Section \ref{sec:U1CS}). 
The free $U(1)$ gauge theory on the other hand realizes the non-invertible condensation defect (see Section \ref{sec:Maxwell}).

Finally, we make an interesting comment on the 't Hooft anomalies of the 1-form symmetry and the condensation defect. 
In the case of a fermion, the underlying 1-form $\mathbb{Z}_2$ symmetry is 0-anomalous and cannot be gauged in the whole spacetime in a bosonic QFT.  
However, the resulting condensation surface $S$ generates a $\mathbb{Z}_2$ 0-form symmetry, which is always free of 't Hooft anomaly.\footnote{This is because the cohomology group classifying the anomaly of a $\mathbb{Z}_N$ 0-form symmetry  is given by $H^4(B\mathbb{Z}_N,U(1))=1$, which is trivial.}
We therefore arrive at a curious observation:  an anomalous 1-form symmetry can give rise to a non-anomalous 0-form symmetry via higher gauging.

\subsection{Action on the lines}

Since the condensation surface \eqref{Z2condsurface} is a mesh of $\mathbb{Z}_2$ 1-form symmetry lines, it acts trivially on all local operators.  
Instead, it admits a nontrivial action on the line defects. 
In a general QFT, these line defects include the topological $\mathbb{Z}_2$ 1-form symmetry lines, as well as other non-topological lines. 
Here we compute the action of the condensation surface on line defects.

\begin{figure}[t]
    \centering
    \includegraphics[scale=0.25]{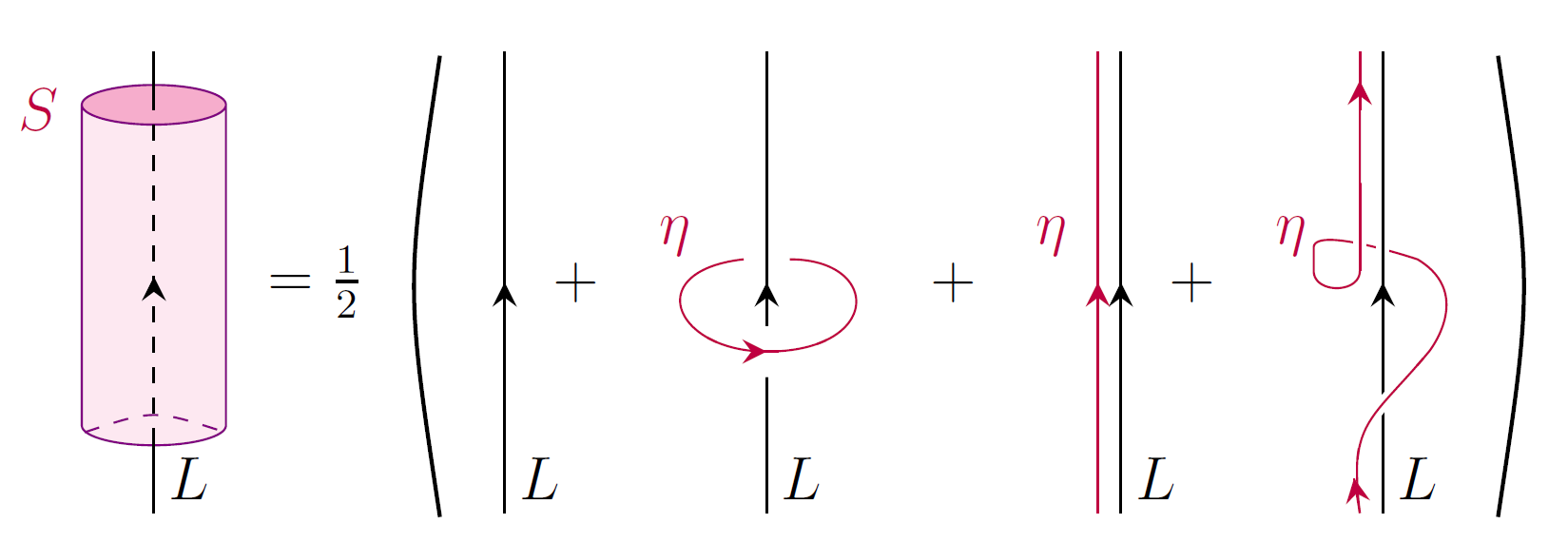}
\caption{The action of the condensation defect  generated by a $\mathbb{Z}_2$ topological line $\eta$ on a line defect $L$. Note that here $L$ does not intersect  with the surface $S$. \label{fig:z2action}}
\end{figure}

This action is given by wrapping the surface $S$ on the tubular neighborhood of the line $L$ and then shrinking it on the line to get a new line defect, as shown in Figure \ref{fig:z2action}.  
We obtain
\begin{equation}
	S \cdot L = \left(\frac{1+B(\eta,L)}{2}\right) L + \left(\frac{1+\theta(\eta)B(\eta,L)}{2}\right) \eta L~,
\end{equation}
where $B(\eta,L)$ is the braiding phase between $\eta$ and $L$, i.e.,\ the $\mathbb{Z}_2$ 1-form symmetry charge of $L$. 
We emphasize that $L$ can be any line defect and need not  be topological. 
Here $\eta L$ is the fusion of the $\mathbb{Z}_2$ 1-form symmetry line $\eta$ and the line $L$.  
Note that $S\cdot L$ is generally not a simple line, but a sum of different lines.

More explicitly, the surface from condensing a boson $\theta(\eta)=1$ acts on a general  line $L$ as
\ie
\theta(\eta)=1\,,~~~S\cdot L = \begin{cases}
L+\eta L\,,~~~~\text{if}~~B(\eta,L)=1\,,\\
~~~0\,,~~~~~~~~~\text{if}~~B(\eta,L)=-1\,.
\end{cases}
\fe
The surface from condensing a fermion $\theta(\eta)=-1$ acts on a general line $L$ as
\ie \label{action.of.fermioncondensation}
\theta(\eta)=-1\,,~~~~S\cdot L = \begin{cases}
~\,~L\,,~~~~\text{if}~~B(\eta,L)=1\,,\\
~\eta L\,,~~~~\text{if}~~B(\eta,L)=-1\,.
\end{cases}
\fe

\subsection{Higher quantum symmetry lines}

In addition to the bulk lines, there are also topological lines that only live on the condensation surface $S$. 
They are the higher quantum symmetry lines arising from higher gauging. 
Let $\hat{\eta}$  be the topological line on $S$ that generates the $\widehat{\mathbb{Z}}_2$ higher quantum symmetry. Following the discussion around equation \eqref{defect.coupled.to.gauge.fields}, the higher quantum symmetry line $\hat{\eta}$ on $S(\Sigma)$ is given by
\begin{equation}
	\hat{\eta}( \gamma, \Sigma ) = \frac{1}{\sqrt{|H_1(\Sigma,\mathbb{Z}_2)|}} \sum_{\gamma' \in H_1(\Sigma,\mathbb{Z}_2)} (-1)^{\langle {\gamma}, \gamma '\rangle} \, \eta(\gamma')  \, ,
\label{Z2 eta hat}
\end{equation}
where ${\gamma} \in H_1(\Sigma,\widehat{\mathbb{Z}}_2)$ is the Poincare dual of $\hat{A} \in H^1(\Sigma,\widehat{\mathbb{Z}}_2)$ in \eqref{defect.coupled.to.gauge.fields}.  
Unlike $S(\Sigma)$, the higher quantum symmetry $\hat\eta(\gamma,\Sigma)$ depends not only on the choice of a 2-manifold $\Sigma$, but also on a 1-cycle $\gamma$ on $\Sigma$.

The higher quantum symmetry line $\hat{\eta}$ is a $\mathbb{Z}_2$ line because two such lines on the same surface pair annihilate with each other, leaving the surface with no extra line. We denote this fusion by
\begin{equation}
\hat{\eta}^2(\gamma,\Sigma) \equiv\hat{\eta}(\gamma,\Sigma) \times_S \, \hat{\eta}(\gamma,\Sigma) = S(\Sigma) ~.
\end{equation}
Here the subscript in $\times_S$ denotes the fusion of two lines living on the surface $S(\Sigma)$, as in the left of Figure \ref{fig: fusion of lines on S}.

On the other hand, there is a closely related but different fusion process of  two surfaces, each with a line living on it (see the right of Figure \ref{fig: fusion of lines on S}). We denote this fusion product by  $\hat{\eta}(\gamma,\Sigma) \times \hat{\eta}(\gamma,\Sigma)$.  
It is given by 
\ie
	\hat{\eta}(\gamma, \Sigma) \times \hat{\eta}(\gamma, \Sigma) 
	&={1\over |H_1(\Sigma,\mathbb{Z}_2)|}
	\sum_{\gamma' ,\gamma''\in H_1(\Sigma,\mathbb{Z}_2)}
	(-1)^{\langle \gamma ,\gamma'+\gamma''\rangle  }\theta(\eta)^{\langle\gamma' , \gamma''\rangle}
	\eta(\gamma'+\gamma'')\\
	&=\begin{cases}
	(\mathcal{Z}_2)~ \hat{\eta}(\gamma,\Sigma)\,,~~~~~&\theta(\eta)=1\\
	1\,,~~~~~~~~~~~~~~~~~~~~~~~~~~~~~~~~~~~~~~~~~~~&\theta(\eta)=-1
	\end{cases}
	\,,
\fe
where we have used \eqref{z2quantumtorus} and $\sum_{\gamma'} (-1)^{\langle\gamma,\gamma'\rangle} = |H_1(\Sigma,\mathbb{Z}_2)| \delta_{\gamma,0}$.

We now compute the fusion of the bulk line $\eta$ with $S$ in terms of the higher quantum symmetry lines on the surface. 
Using the relation \eqref{z2quantumtorus}, we have
\ie
\eta(\gamma)\times S(\Sigma) = S(\Sigma) \times \eta(\gamma) &
={1\over \sqrt{|H_1(\Sigma,\mathbb{Z}_2)|}} \sum_{\gamma'\in H_1(\Sigma,\mathbb{Z}_2)}
\theta(\eta)^{\langle \gamma,\gamma'\rangle} \eta(\gamma+\gamma')\\
&=\begin{cases}
		S (\Sigma)~, & \theta(\eta) = 1 \\
		\hat{\eta}(\gamma,\Sigma) ~, & \theta(\eta) = -1
	\end{cases}~.
\fe

We can also compute the fusion of the bulk line $\eta$ and the quantum symmetry line $\hat{\eta}$ on the surface:
\ie
\eta(\gamma)\times \hat{\eta}(\gamma,\Sigma) =
 \hat{\eta}(\gamma,\Sigma) \times \eta(\gamma)
 =\begin{cases}
 		\hat{\eta}(\gamma,\Sigma) ~, & \theta(\eta) = 1 \\
		S (\Sigma)~, & \theta(\eta) = -1
 \end{cases}~.
\fe

Finally, let us compute the fusion of the condensation defect $S$ with the defect carrying a higher quantum symmetry line $\hat{\eta}(\gamma,\Sigma)$:
\ie\label{Shateta}
S(\Sigma) \times \hat{\eta}(\gamma,\Sigma)  = \hat{\eta}(\gamma,\Sigma)\times S(\Sigma)&=
{1\over |H_1(\Sigma,\mathbb{Z}_2)|}
\sum_{\gamma',\gamma''\in H_1(\Sigma,\mathbb{Z}_2)}
(-1)^{\langle\gamma,\gamma''\rangle}
\theta(\eta)^{\langle\gamma',\gamma''\rangle} \eta(\gamma'+\gamma'')\\
&=\begin{cases}
\sqrt{|H_1(\Sigma,\mathbb{Z}_2)|}~ \delta_{\gamma,0} \,S(\Sigma)\,,~~~&\theta(\eta)=1\,,\\
\eta(\gamma)\,,~~~~~&\theta(\eta)=-1\,.
\end{cases}
\fe
The factor $\sqrt{|H_1(\Sigma,\mathbb{Z}_2)|} \, \delta_{\gamma,0} $ should be interpreted as the partition function of the 1+1d $\mathbb{Z}_2$ gauge theory on $\Sigma$ with a $\mathbb{Z}_2$ Wilson line inserted along the cycle $\gamma\in H_1(\Sigma,\mathbb{Z}_2)$.

\subsection{Summary of the fusion rules}

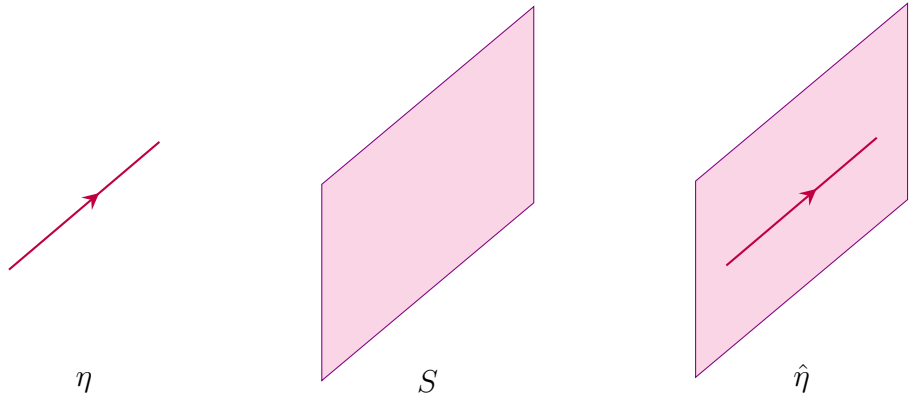
\begin{figure}
\centering 
\begin{tikzpicture}[scale = 1]
\draw[->-=0.6 rotate 0, color = purple, thick] (4,-0.5) -- (6,1.2);
\node[] at (5,-2) {$\eta$}; 
\end{tikzpicture}
\hskip 2cm
\begin{tikzpicture}[scale = 1]
\node[	trapezium, 
		draw = violet, 
		trapezium left angle=50, 
		trapezium right angle=130, 
		rotate = 40, 
		trapezium stretches=false,
		minimum height=2cm, 
		minimum width=1cm, 
		fill = magenta!20,
		]
    at (5,0.5) {}; 
\node[] at (5,-2) {$S$}; 
\end{tikzpicture}
\hskip 2cm
\begin{tikzpicture}[scale = 1]
\node[	trapezium, 
		draw = violet, 
		trapezium left angle=50, 
		trapezium right angle=130, 
		rotate = 40, 
		trapezium stretches=false,
		minimum height=2cm, 
		minimum width=1cm, 
		fill = magenta!20,
		]
    at (5,0.5) {}; 
\draw[->-=0.6 rotate 0, color = purple, thick] (4,-0.5) -- (6,1.2);
\node[] at (5,-2) {$\hat{\eta}$}; 
\end{tikzpicture}
  \caption{Topological defects arising from  a 1-gaugeable $\mathbbm{Z}_2$ 1-form symmetry. Here $\eta$ is the bulk 1-form topological symmetry line, $S$ is the condensation surface, and $\hat\eta$ is a line living on the surface $S$, which we call the higher quantum symmetry line. Their fusion rules are presented in \eqref{bosonsummary} and \eqref{z2fermion} for the cases of a boson and a fermion line, respectively. }
 \label{eta hat}
\end{figure}

 To conclude the discussion of the higher gauging of a $\mathbb{Z}_2$ 1-form symmetry, below we summarize the fusion between the bulk $\mathbb{Z}_2$ line $\eta(\gamma)$, the condensation surface $S(\Sigma)$, and the higher quantum symmetry line $\hat{\eta}(\gamma,\Sigma)$ (which lives only on the surface).   (These defects are summarized in  Figure \ref{eta hat}.) 
 If the $\mathbb{Z}_2$ 1-form symmetry line is a boson $\theta(\eta)=1$, then
 \ie\label{bosonsummary}
 &S \times S = (\mathcal{Z}_2)\, S \,,\\
  &\eta \times \eta =1\,,~~~\\
&  \hat{\eta}\times\hat{\eta} = (\mathcal{Z}_2) \,\hat{\eta}\,,~~~\\
 &\eta  \times S  =S \times \eta  =S\,,\\
&  \eta\times \hat{\eta} = \hat{\eta}\times \eta= \hat{\eta}\,,\\
&  S  \times \hat{\eta}   = \hat{\eta} \times S  =0\,,\\
  &\hat{\eta} \times_S \hat{\eta} =S\,.
 \fe
Recall that $(\mathcal{Z}_2)$ stands for the 1+1d $\mathbb{Z}_2$ gauge theory. 
The fusion product $\times_S$   denotes the fusion of two lines living on the same surface $S(\Sigma)$ (see   Figure \ref{fig: fusion of lines on S}).  
 In the fusion $S  \times \hat{\eta}   = \hat{\eta} \times S  =0$, we assume that the 1-cycle $\gamma$ of $\hat{\eta}(\gamma,\Sigma)$ is nontrivial, so that $\hat{\eta} \neq S$. 
 The zero here can be interpreted as the partition function of the 1+1d $\mathbb{Z}_2$ gauge theory with a $\mathbb{Z}_2$ Wilson line $\hat{\eta}$ inserted along a nontrivial cycle (see \eqref{Shateta}).

 If the $\mathbb{Z}_2$ 1-form symmetry line is a fermion $\theta(\eta)=-1$, then 
\ie\label{z2fermion}
&S \times S = 1\,,\\
  &\eta \times \eta =1\, ,~~~\\
&   \hat{\eta}\times\hat{\eta} = 1,~~~\\
 &\eta  \times S  =S \times \eta =\hat{\eta} \,,\\
&   \eta\times \hat{\eta} = \hat{\eta}\times \eta= S\,,\\
   &  S  \times \hat{\eta}   = \hat{\eta} \times S  =\eta\,,\\
   &  \hat{\eta} \times_S \hat{\eta} =S\,.
\fe
Formally, $S,\eta, \hat{\eta}$ generate a $\mathbb{Z}_2\times \mathbb{Z}_2$ algebra under the fusion product $\times$, but the two factors of $\mathbb{Z}_2$ have different form degrees: $S$ generates a $\mathbb{Z}_2$ 0-form symmetry while $\eta$ generates a $\mathbb{Z}_2$ 1-form symmetry. 
The fusion rules \eqref{bosonsummary} and \eqref{z2fermion} of these lines, surfaces, and lines living only on the surfaces are some of the most basic data for a full-fledged fusion 2-category.

\section{Higher gauging of $\mathbb{Z}_N$ and $\mathbb{Z}_{N_1}\times \mathbb{Z}_{N_2}$ 1-form symmetries}\label{sec:znn}

In this section we generalize the condensation surfaces and higher quantum symmetry lines for a $\mathbb{Z}_2$ 1-form symmetry in the previous section to $\mathbb{Z}_N$ and $\mathbb{Z}_{N_1}\times \mathbb{Z}_{N_2}$ 1-form symmetries. 
There are several new elements, including the choice of the discrete torsion in the higher gauging of  $\mathbb{Z}_{N_1}\times \mathbb{Z}_{N_2}$ 1-form symmetries.

\subsection{Higher gauging of  $\mathbb{Z}_N$ 1-form symmetries}\label{sec:ZN}

Consider a $\mathbb{Z}_N$ 1-form symmetry generated by a topological line $\eta$ with $\eta^{N}=1$. 
In a bosonic QFT, the possible values of the topological spin are  \cite{Hsin:2018vcg}
\ie \label{anomaly.spin}
\theta(\eta) = e^{\frac{2\pi i k}{N}}\,,~~~~k\sim k+N\,,~~~~
k\in 
\begin{cases}
 \mathbb{Z}~,~~~~~~~~\text{odd}~N\,,\\
 \mathbb{Z}/2~,~~~\text{even}~N\,,
\end{cases}\,.
\fe 
Here $k$ labels the 't Hooft anomaly of the $\mathbb{Z}_N$ 1-form symmetry generated by $\eta$. 
Fixing a particular representative, the $F$-symbols and $R$-symbols are~\cite{Moore:1988qv,Barkeshli:2014cna}
\begin{equation}
	F^{a,b,c}_{a+b+c} = e^{2\pi i k  a \frac{b+c-[b+c]_N}{N}}\,, \qquad R^{a,b}_{a+b} = e^{2\pi i k \frac{ab}{N}}\, . \label{FandR-symbols}
\end{equation}
Here we use $a\in \{0,1,\dots,N-1\}$ to label the line $\eta^a$.\footnote{Compared to the general discussion in Section \ref{sec:general}, the labels $a,b,\cdots$ for the invertible topological lines here and below are  additive rather than multiplicative. We hope this change of convention will not cause too much confusions. } 
The integer $[x]_N \in \{0,1,\dots,N-1\}$ is defined such that $[x]_N = x \pmod{N}$ for any integer $x$. 
(Note that $x-[x]_N$ is always a multiple of $N$.) 
When $k$ is an integer the $F$-symbols are all trivial. 
In contrast, when $k$ is a half-integer, they cannot be made trivial by choosing a different representative for the $F$- and $R$-symbols~\cite{Barkeshli:2014cna}.

Therefore, the $\mathbb{Z}_N$ 1-form symmetry is 0-gaugeable (i.e., anomaly-free in the usual sense) if the spin is trivial, i.e., $k=0$.  
It is 1-gaugeable (i.e., can be gauged on a surface) if the crossing is trivial, i.e.,   $k\in \mathbb{Z}$. 
In other words, a $\mathbb{Z}_N$ 1-form symmetry is 1-gaugeable iff the spin of its generator $\eta$ is an $N$-th root of unity:\footnote{This is related to the spin selection rule in 1+1d CFT.   
In  \cite{Chang:2018iay,Lin:2019kpn,Lin:2021udi}, it is shown that an ordinary $\mathbb{Z}_N$ global symmetry in 1+1d is anomaly-free if and only if the operator living at the end of the line (e.g., the disorder operator) has Lorentz spin $s=h-\bar h\in \mathbb{Z}/N$.}
\ie\label{ZN1cond}
\text{1-gaugeable}:~~\theta(\eta)^N=1\,.
\fe
This condition was discussed in \cite{Moore:1988qv}. 
We henceforth assume the $\mathbb{Z}_N$ 1-form symmetry is 1-gaugeable and discuss the corresponding condensation surface defect.

Gauging the $\mathbb{Z}_N$ 1-form symmetry on a surface $\Sigma$ gives the condensation defect $S_N$:
\ie
S_N (\Sigma) = {1\over \sqrt{|H_1 (\Sigma ,\mathbb{Z}_N)|}}
\sum_{\gamma\in H_1(\Sigma, \mathbb{Z}_N)} \eta(\gamma) \, .
\label{zn.condensation}
\fe
Note that the defect $S_N$ is  equal to its orientation-reversal $\overline{S_N}$  since the sum over the cycles does not depend on the choice of the orientation of the surface $\Sigma$.  (See \eqref{orientationrev} for the definition of the orientation reversal of a defect.)

Mathematically, the 0-anomaly (i.e., the ordinary obstruction to gauging in the whole spacetime) of a $\mathbb{Z}_N$ 1-form symmetry in 2+1d is classified by 
\ie
H^4(B^2\mathbb{Z}_N ,U(1))=
\begin{cases}
\mathbb{Z}_N\,,&\text{odd~}N\,,\\
\mathbb{Z}_{2N}\,,&\text{even~}N\,.
\end{cases}
\fe
The 1-anomaly (i.e., the obstruction to gauging in a codimension-1 manifold) of a $\mathbb{Z}_N$ 1-form symmetry is in the image of the map:
\ie
H^4(B^2\mathbb{Z}_N ,U(1)) \to H^3( B\mathbb{Z}_N,U(1))=\mathbb{Z}_N\,.
\fe 
For odd $N$, this is a trivial map where the images are always the identity (i.e., all 1-form symmetries with odd $N$ are 1-gaugeable). For even $N$, it maps the generator of $\mathbb{Z}_{2N}$ to the generator of the $\mathbb{Z}_2$ subgroup of $\mathbb{Z}_N$.

\bigskip\centerline{\it Fusion rule of the surfaces}\bigskip

Since the $F$-symbols are trivial and  $R^{a,b}_{a+b} = e^{2\pi i k \frac{ab}{N}}$, we have
\begin{equation}
	\raisebox{-2.2em}{\begin{tikzpicture}
			\draw [thick, decoration = {markings, mark=at position .7 with {\arrow[scale=1.5]{stealth}}}, postaction=decorate] (0,-.5) node[below] {\small $a+b$} to (0,0);
			\draw [thick] (0,0) arc [radius=.3, start angle=240, end angle=120];
			\draw [thick, decoration = {markings, mark=at position .75 with {\arrow[scale=1.5]{stealth}}}, postaction=decorate] (0,0.52) to +(30:.5) node[right] {\small $b$};
			\draw [thick] (0,0) arc [radius=.3, start angle=-60, end angle=42];
			\draw [thick, decoration = {markings, mark=at position .8 with {\arrow[scale=1.5]{stealth}}}, postaction=decorate] (0,0.52) ++(150:.1) to +(150:.4) node[left] {\small $a$};
	\end{tikzpicture}}
	\!\!\! = e^{2\pi i k \frac{ab}{N}} \!\!\!\!\!
	\raisebox{-2.2em}{\begin{tikzpicture}
			\draw [thick, decoration = {markings, mark=at position .5 with {\arrow[scale=1.5]{stealth}}}, postaction=decorate] (0,-.8) node[below] {\small $a+b$} to (0,0);
			\draw [thick, decoration = {markings, mark=at position .7 with {\arrow[scale=1.5]{stealth}}}, postaction=decorate] (0,0)  to (-.6,.5) node[left] {\small $a$};
			\draw [thick, decoration = {markings, mark=at position .7 with {\arrow[scale=1.5]{stealth}}}, postaction=decorate] (0,0)--(.6,.5) node[right] {\small $b$};
	\end{tikzpicture}}
	\quad\;\Rightarrow\quad\;
	\raisebox{-1.8em}{\begin{tikzpicture}
			\draw [thick, decoration = {markings, mark=at position .9 with {\arrow[scale=1.5]{stealth}}}, postaction=decorate] (-.6,-.6) node[left] {\small $b$} to (.6,.6) node[right] {\small $b$};
			\draw [thick] (.6,-.6) node[right] {\small $a$} to (.1,-.1);
			\draw [thick, decoration = {markings, mark=at position .75 with {\arrow[scale=1.5]{stealth}}}, postaction=decorate] (-.1,.1) to (-.6,.6) node[left] {\small $a$};
	\end{tikzpicture}}
	=  e^{2\pi i k \frac{ab}{N}} \!\!\!
	\raisebox{-2.4em}{\begin{tikzpicture}
			\draw [thick, decoration = {markings, mark=at position .7 with {\arrow[scale=1.5]{stealth}}}, postaction=decorate] (-.5, -.5) node[left] {\small $b$} to (0,0);
			\draw [thick, decoration = {markings, mark=at position .7 with {\arrow[scale=1.5]{stealth}}}, postaction=decorate] (.5, -.5) node[right] {\small $a$} to (0,0);
			\draw [thick, decoration = {markings, mark=at position .7 with {\arrow[scale=1.5]{stealth}}}, postaction=decorate] (0,0) to node[midway, right] {\footnotesize $a+b$} (0,.7);
			\draw [thick, decoration = {markings, mark=at position .7 with {\arrow[scale=1.5]{stealth}}}, postaction=decorate] (0, .7) to (-.5,1.2) node[left] {\small $a$};
			\draw [thick, decoration = {markings, mark=at position .7 with {\arrow[scale=1.5]{stealth}}}, postaction=decorate] (0, .7) to (.5,1.2) node[right] {\small $b$};
	\end{tikzpicture}}~.
\end{equation}
Using these relations, we find that the 1-form symmetry lines obey a non-commutative algebra, similar to \eqref{z2quantumtorus}, on the surface $\Sigma$:
\begin{equation}\label{quantumtorus}
	\eta\left( \gamma \right) \eta\left( \gamma' \right) = e^{2\pi i \frac{k}{N} \langle \gamma,\gamma' \rangle} \, \eta\left( \gamma + \gamma' \right) \,,
\end{equation}
where $\langle \gamma, \gamma' \rangle \in \mathbb{Z}_N$ is the intersection number between the 1-cycles $\gamma,\gamma' \in H_1(\Sigma,\mathbb{Z}_N)$.

We are now ready to compute the fusion of $S_N$ with itself on a general 2-dimensional surface $\Sigma$. This fusion is given by
\ie
S_N(\Sigma) \times S_N(\Sigma) &= {1\over |H_1(\Sigma,\mathbb{Z}_N)|}\sum_{\gamma,\gamma'\in H_1(\Sigma ,\mathbb{Z}_N)}
e^{2\pi i {k\over N} \langle\gamma,\gamma'\rangle}
\eta(\gamma+\gamma')\, ,\\
&={1\over |H_1(\Sigma,\mathbb{Z}_N)|}
\sum_{\gamma\in H_1(\Sigma,\mathbb{Z}_N)}
\left(\sum_{\gamma'\in H_1(\Sigma ,\mathbb{Z}_N)} e^{2\pi i {k\over N} \langle\gamma,\gamma'\rangle}\right)
\eta(\gamma)\,,
\fe
where in the first line we have used \eqref{quantumtorus}, and in the second line we have redefined $\gamma\to \gamma-\gamma'$. 
The sum over $\gamma'$ then gives a Kronecker delta, multiplied by a normalization factor $|H_1(\Sigma,\mathbb{Z}_N)|$,  that restricts the sum of $\gamma$ to those satisfying $k\gamma =0\in H_1(\Sigma,\mathbb{Z}_N)$.  
We can thus rewrite the fusion as
\ie
S_N(\Sigma) \times S_N(\Sigma) & = \sum_{\gamma\in H_1(\Sigma, \mathbb{Z}_{\text{gcd}(k,N)})} \eta(\gamma)^{N\over \text{gcd}(k,N)}\, ,\\
& =  \sqrt{|H_1(\Sigma ,\mathbb{Z}_{\text{gcd}(k,N)})|}\, \, S_{\text{gcd}(k,N)}(\Sigma) \, ,
\label{SN fusion}\fe
where for any positive divisor $n$ of $N$, the surface defect $S_n$ is defined as
\ie
S_n = {1\over \sqrt{|H_1(\Sigma ,\mathbb{Z}_n)|}} \sum_{\gamma\in H_1(\Sigma, \mathbb{Z}_n)} \eta(\gamma)^{N\over n}\, .
\fe
It is the  surface defect from condensing the $\mathbb{Z}_n$ subgroup of the $\mathbb{Z}_N$ 1-form symmetry on $\Sigma$.  
In particular, $S_1$ is the trivial surface defect.

The fusion coefficient $\sqrt{H_1(\Sigma ,\mathbb{Z}_{\text{gcd}(k,N)})}$ is the partition function of the 1+1d $\mathbb{Z}_{\text{gcd}(k,N)}$ gauge theory on $\Sigma$ (up to the usual Euler counterterm). Interestingly, when $\text{gcd}(k,N)\neq 1$, this fusion coefficient is a non-invertible TQFT. Using $S_N=\overline{S_N}$, we conclude that  the condensation surface $S_N$ is non-invertible when $\text{gcd}(k,N)\neq 1$. (Recall that for an invertible defect $U^{-1}=\overline U$ and  $U\times \overline U=\overline U\times U=1$.)
We therefore conclude that
\begin{equation}
	S_N \times S_N = 
(\mathcal{Z}_{\mathrm{gcd}(k,N)})~ S_{\mathrm{gcd}(k,N)} \,.
\end{equation}
Recall that $k=0,1,\cdots, N-1$ is an integer that labels the  0-anomaly of the $\mathbb{Z}_N$ 1-form symmetry. 
Here $(\mathcal{Z}_{\mathrm{gcd}(k,N)})$ stands for the 1+1d $\mathbb{Z}_{\mathrm{gcd}(k,N)}$ gauge theory. 
We again see that the fusion ``coefficient" of the condensation surface is a 1+1d TQFT, rather than a number.

More generally,  we compute  the fusion of condensation surface defects $S_n$ and $S_{n'}$ where $n$ and $n'$ are positive divisors of $N$ in Appendix \ref{app:ZN}:
\begin{equation}\label{generalSSZN}
	S_n \times S_{n'} = 
	(\mathcal{Z}_{\mathrm{gcd}(n,n',k\ell)})~S_{\frac{\mathrm{gcd}(n,n',k\ell)nn'}{\mathrm{gcd}(n,n')^2}} \,,
\end{equation}
where $\ell\equiv {N\over\text{lcm}(n,n')}$.  
In the $k=1$ case, this reduces to the fusion rule discussed in \cite{Fuchs:2002cm,Kapustin:2010if}. 
 
\bigskip\centerline{\it Action on the lines}\bigskip

Now let us discuss the action of the condensation surface $S_N$ on the line defects (which are not necessarily topological) as  on the left-hand side of Figure \ref{fig:z2action}. 
The condensation surface $S_N$ is supported on a torus that encloses the line $L$. 
Let $\bf A, B$ be a basis for 1-cycles of the torus, such that $\bf A$ links nontrivially with $L$ and $\bf B$ is parallel to $L$. 
Using \eqref{quantumtorus}, the condensation surface on the torus can be written as
\begin{align}
	S_N(T^2) =
	{1\over N}\sum_{a,b=0}^{N-1} \eta(a \,{\bf A} +b\,{\bf B})
	&= \frac{1}{N} \sum_{a , b=0}^{N-1} e^{2\pi i k \frac{ab}{N}} \, \eta(b \, {\bf B}) \eta( a \, {\bf A})
	\notag\\
	& = \frac{1}{N} \sum_{a, b=0}^{N-1} e^{2\pi i k \frac{ab}{N}} ~ \begin{gathered}
		\begin{tikzpicture}[scale = 0.9]
			\draw (0,0) -- (2,0) -- (2,2) -- (0,2) -- (0,0);
			\draw [decoration = {markings, mark=at position 0.6 with {\arrow[scale=1.3]{stealth}}},  postaction=decorate] (0,1) -- (0.5,1) node[above] {$a$} -- (0.9,1);
			\draw (1.1,1) -- (2,1);
			\draw [decoration = {markings, mark=at position 0.3 with {\arrow[scale=1.3]{stealth}}}, postaction=decorate] (1,0) -- (1,0.5) node[right] {$b$} -- (1,2);
		\end{tikzpicture}
	\end{gathered} ~.
\end{align}
When acting on the line $L$, the line $\eta(a \,{\bf A})$  yields a braiding phase, while $\eta(b\,{\bf B})$ changes $L$ to $\eta^b L$. 
Using the above expression for $S_N$, we find
\begin{equation}
	S_N \cdot L = \sum_{b=0}^{N-1} \left( \frac{ \sum_{a=0}^{N-1} \left( e^{2\pi i k \frac{b}{N}} B(\eta,L) \right)^a }{N} \right) \eta^{b}  L ~.
\end{equation}
Here $B(\eta,L)$ is the $\mathbb{Z}_N$ 1-form symmetry charge of $L$.  
Writing $B(\eta,L)= \exp(2\pi i Q/N)$, we can simplify the action of $S_N$ on this charge $Q$ line $L$ as
\ie \label{zn.action.on.lines}
S_N \cdot L = \sum_{\substack{a=0 \\k a+ Q=0~\text{mod}~N}}^{N-1} \eta^a L\,.
\fe

The above formula can be easily generalized to $S_n$ with $n$ a positive divisor of $N$. 
Recall that $S_n$ is the condensation defect from gauging the $\mathbb{Z}_n$ subgroup of the $\mathbb{Z}_N$ 1-form symmetry. 
This is done by performing the substitution $\eta,N,k,Q \mapsto \eta^{N/n},n,(N/n)k,Q$ in the above formula:
\ie \label{Snaction}
S_n \cdot L = \sum_{\substack{a=0 \\(N/n) k a+ Q=0~\text{mod}~n}}^{n-1} \eta^{aN/n} L\,.
\fe

\bigskip\centerline{\it Higher quantum symmetry}\bigskip

We now discuss the higher quantum symmetry lines living only on the condensation surface $S_n$ and their fusion rules.  
The higher quantum symmetry on $S_n$ is $\widehat{\mathbb{Z}}_n \times \mathbb{Z}_N/\mathbb{Z}_n$.\footnote{In general the quantum symmetry might be a non-trivial extension of $\mathbb{Z}_N/\mathbb{Z}_n$ by $\widehat{\mathbb{Z}}_n$. As explained in \cite[Section 5.3.2]{Bhardwaj:2017xup} (see also \cite{Tachikawa:2017gyf}), the extension is given by the mixed anomaly between the $\mathbb{Z}_n$ subgroup and $\mathbb{Z}_N/\mathbb{Z}_n$ before gauging. However, this anomaly is trivial in this case since the $\mathbb{Z}_N$ 1-form symmetry is 1-gaugeable.} Let $\hat{\eta}_n$ and $\tilde{\eta}_n$ denote, respectively, the generators of the $\widehat{\mathbb{Z}}_n$ and $\mathbb{Z}_N/\mathbb{Z}_n$ higher quantum symmetries. 
The subscript $n$ implies that they are lines living only on the condensation surface $S_n$. 
They are given by
\begin{equation} \label{hat.eta.n}
	\hat{\eta}_n( \gamma, \Sigma ) = \frac{1}{\sqrt{|H_1(\Sigma,\mathbb{Z}_n)|}} \sum_{\gamma' \in H_1(\Sigma,\mathbb{Z}_n)} e^{\frac{2\pi i }{n}\langle \gamma, \gamma' \rangle} \, \eta^{\frac{N}{n}}(\gamma')  \, ,
\end{equation}
and
\begin{equation} \label{eta.n}
	\tilde{\eta}_n( \gamma, \Sigma ) = \frac{1}{\sqrt{|H_1(\Sigma,\mathbb{Z}_n)|}} \sum_{\gamma' \in H_1(\Sigma,\mathbb{Z}_n)} \eta\left( \frac{N}{n} \gamma' + \gamma \right)  \, ,
\end{equation}
for $\gamma \in H_1(\Sigma,\mathbb{Z})$. The 1-gaugeable $\mathbb{Z}_N$ 1-form symmetry line $\eta(\gamma)$  depends only on the 1-cycle  $\gamma\in H_1(\Sigma, \mathbb{Z})$   modulo $N$. On the other hand, it is straightforward to see from \eqref{hat.eta.n} and \eqref{eta.n} that $\hat\eta_n(\gamma,\Sigma)$ and $\tilde\eta_n(\gamma,\Sigma)$ depend on $\gamma$ modulo $n$ and $N/n$, respectively. One way to arrive at \eqref{hat.eta.n} and \eqref{eta.n} is to consider the fusion of the line $\eta$ with the condensation surface $S_n$ (see below). From that, one can extract the above combinations $\hat{\eta}_n$ and $\tilde{\eta}_n$  which generate the $\widehat{\mathbb{Z}}_n \times \mathbb{Z}_{N/n}$ higher quantum symmetry with respect to the fusion product $\times_S$  in the sense that  $\hat{\eta}_n^n=S_n$ and $ \tilde{\eta}_n^{N/n}= S_n$.  Here we have defined
	\ie
	&\hat{\eta}_n^a\equiv    \underbrace{
	\hat{\eta}_n  \times_{S_n} \cdots \times_{S_n} \hat{\eta}_n  }_{a\text{ times} }\,,\\
	& \tilde{\eta}_n^{b}\equiv  \underbrace{
	\tilde{\eta}_n  \times_{S_n} \cdots \times_{S_n} \tilde{\eta}_n  }_{b\text{ times} }
	\fe
	(The product $\times_{S_n}$ is defined in  Figure \ref{fig: fusion of lines on S}.)  
	
Note that $\hat{\eta}_1  = 1$ is the trivial defect, $ \tilde{\eta}_1 = \eta$ is the bulk $\mathbb{Z}_N$ 1-form symmetry line, and $\tilde{\eta}_N =S_N$ is the condensation surface without any line.

The most general  higher quantum symmetry line living only on the condensation defect $S_n$ is  a composite line of  $\hat{\eta}_n$ and $\tilde{\eta}_n$:
\ie \label{hat.eta.tilde.eta}
\hat{\eta}_n^a \tilde{\eta}_n^b(\gamma,\Sigma)
&=  \hat{\eta}_n^a (\gamma,\Sigma)\times_{S_n} \tilde{\eta}_n^b(\gamma,\Sigma)
 = \frac{1}{\sqrt{|H_1(\Sigma,\mathbb{Z}_n)|}} \sum_{\gamma' \in H_1(\Sigma,\mathbb{Z}_n)} e^{\frac{2\pi i a}{n}\langle \gamma, \gamma' \rangle} \, \eta\left( \frac{N}{n} \gamma' + b \gamma \right)  \, .
\fe
Note that $\hat{\eta}_n^a\tilde{\eta}_n^b=\hat{\eta}_n^{a+n}\tilde{\eta}_n^b =\hat{\eta}_n^a\tilde{\eta}_n^{b+N/n}$.  
In this notation, every defect we have discussed so far is a special case of $\hat{\eta}_n^a \tilde{\eta}_n^b$. 
In particular, we have $S_n=  \hat{\eta}_n^{0} \tilde{\eta}_n^{0}$.

Importantly, from the explicit dependence on $\eta$ in \eqref{hat.eta.tilde.eta}, the higher quantum symmetry line $\hat\eta_n \tilde\eta_n(\gamma,\Sigma)$ generally depends on the 1-cycle $\gamma$ modulo $N$, despite the fact that it is a line of order $\text{lcm}(n,N/n)$ with respect to the fusion $\times_{S_n}$ on the surface $S_n$.  
The mod $N$ dependence of $\hat\eta_n \tilde\eta_n(\gamma,\Sigma)$ on $\gamma$ comes from a possible mixed anomaly between the two constituent lines, in the sense that they cannot be gauged on the surface.\footnote{A simple example is to choose $N=4$ and $n=2$. The order 2 higher quantum symmetry line $\hat\eta_2\tilde\eta_2$ is anomalous and  depends on $\gamma$ modulo 4, rather than 2.}  Indeed, generally  an anomalous $\mathbb{Z}_m$ symmetry defect depends not only on the cycles modulo $m$, but on a choice of the uplift \cite{Lin:2019kpn,Lin:2021udi}.

This anomaly can be explained from the fact that the extension and mixed-anomaly classes are exchanged under gauging subgroups~\cite{Tachikawa:2017gyf}. 
More specifically, when $\mathrm{gcd}(n,N/n)\neq 1$, the $\mathbb{Z}_N$ symmetry before gauging is a non-trivial extension of $\bZ_N/\bZ_n$ by $\bZ_n$. 
Therefore, after gauging, the nontrivial extension results in a mixed anomaly between the higher quantum $\bZ_N/\bZ_n$ and $\widehat{\bZ}_n$ symmetries.\footnote{The symmetry groups $\bZ_N/\bZ_n$ and $\widehat{\bZ}_n$ are individually anomaly free and there is only a mixed anomaly between them in the sense that the product group  $\bZ_N/\bZ_n \times \widehat{\bZ}_n$ is anomalous.}
The mixed anomaly implies an anomaly in the $\bZ_{\mathrm{lcm}(n,N/n)}$ subgroup generated by $\hat\eta_n \tilde\eta_n$ on $S_n$.

Finally, we compute the fusion between $\hat{\eta}_{n}^a \tilde{\eta}_{n}^b$ and $\hat{\eta}_{n'}^{a'} \tilde{\eta}_{n'}^{b'}$ for $n$ and $n'$ positive divisors of $N$ in Appendix \ref{app:ZN}, and find:
\begin{equation} \label{zn.general.fusion}
	\hat{\eta}_{n}^a \tilde{\eta}_{n}^b \times \hat{\eta}_{n'}^{a'} \tilde{\eta}_{n'}^{b'} = \left( \mathcal{Z}_{\mathrm{gcd}(n,n',k\ell)}, W^{c-c'} \right) \, \hat{\eta}_{x}^{\frac{pcn'+p'c'n}{\mathrm{gcd}(n,n')}} \tilde{\eta}_{x}^{{\ell(c-c') \over \mathrm{gcd}(n,n',k\ell)}\,\left(\frac{k\ell}{\mathrm{gcd}(n,n',k\ell)}\right)^{-1}_{N \over x\ell}\,+b+b'} \,,
\end{equation}
where
\ie
	\ell &= \frac{N}{\mathrm{lcm}(n,n')} \,, \qquad &x &= \frac{\mathrm{gcd}(n,n',k\ell)nn'}{\mathrm{gcd}(n,n')^2} \,,\\
	c&=a-kb'\,, \qquad &c' &= a'+kb \,.\\
\fe
The integers $p$ and $p'$ are chosen such that $p n' + p'n = \mathrm{gcd}(n,n')$, but the fusion does not depend on the particular choice of $p$ and $p'$. 
Here, $(y)_m^{-1}$ is the multiplicative inverse of $y$ modulo $m$. 
Since $\frac{k\ell}{\mathrm{gcd}(n,n',k\ell)}$ is coprime with respect to ${\mathrm{gcd}(n,n') \over \mathrm{gcd}(n,n',k\ell)} = {N \over x\ell}$, its multiplicative inverse modulo ${N \over x\ell}$ always exists.
Finally, $\left( \mathcal{Z}_{N}, W^{a} \right)$ denotes 1+1d $\mathbb{Z}_N$ gauge theory with an insertion of the Wilson line $W^a$.
 
The partition function of the fusion ``coefficient" in \eqref{zn.general.fusion} is given by\footnote{Here   $\gamma = 0 \mod{n}$ for a 1-cycle $\gamma$ means that there exists another 1-cycle $\gamma'\in H_1(\Sigma, \mathbb{Z})$ such that $\gamma = n \gamma'$.}
\begin{equation*}
	{\left( \mathcal{Z}_{\mathrm{gcd}(n,n',k\ell)}, W^{c-c'} \right) }(\gamma, \Sigma) = \begin{cases}
		\sqrt{|H_1(\Sigma,\mathbb{Z}_{\mathrm{gcd}(n,n',k\ell)})|} \,, & (c-c')\gamma = 0 \mod{\mathrm{gcd}(n,n',k\ell)} \\
		0\,, & \text{otherwise}
	\end{cases} \,,
\end{equation*}
see Appendix \ref{app:ZN} for more details. We see that ${\ell(c-c') \over \mathrm{gcd}(n,n',k\ell)} \, \gamma$ (which appears in the power  of $\tilde{\eta}_x$ in \eqref{zn.general.fusion}) is a properly normalized 1-cycle whenever the fusion ``coefficient" is nonzero. 
Therefore, the right hand side of \eqref{zn.general.fusion} is always well-defined.
In particular, when $k=0$, we have $N/x = \ell$ and the argument of $\tilde{\eta}_x$ only needs to be well-defined modulo $\ell$. Indeed, this is automatically satisfied since $\ell {(c-c') \gamma \over \mathrm{gcd}(n,n',k\ell)} = 0 \mod{\ell}$.

\bigskip\centerline{\it Summary of the fusion rules}\bigskip

Given any QFT with a $\mathbb{Z}_N$ 1-form symmetry with a 0-anomaly labeled by $k$, we have the following set of line and surface defects.  
The $\mathbb{Z}_N$ 1-form symmetry is generated by  a bulk line $\eta$, obeying $\eta^N=1$.  
The condensation surface $S_n$ arises from gauging the $\mathbb{Z}_n$ subgroup of the $\mathbb{Z}_N$ 1-form symmetry on a surface. 
More generally, there are lines $\hat\eta_n^a\tilde\eta_n^b$ that only live on the condensation surface $S_n$ (hence the subscript $n$ for them). 
They generate the $\widehat{\mathbb{Z}_n}\times \mathbb{Z}_N/\mathbb{Z}_n$ higher quantum symmetry on $S_n$ obeying  $\hat\eta_n^n =\tilde\eta_n^{N/n}=S_n$ under the fusion product $\times_{S_n}$.  
Both $\eta$ and $S_n$ are special cases of $\hat\eta_n^a \tilde\eta_n^b$, with $\eta=\tilde\eta_1$ and $S_n=\hat\eta_n^0\tilde\eta^0_n$.

The most general  fusion rule between these defects is given in \eqref{zn.general.fusion}. 
Some important special cases are:
\ie\label{zn.fusion.rule.summary}
&S_n \times S_{n'} = 
	(\mathcal{Z}_{\mathrm{gcd}(n,n',k\ell)})~S_{\frac{\mathrm{gcd}(n,n',k\ell)nn'}{\mathrm{gcd}(n,n')^2}} \,,\\
& \eta^a \times S_n = {\hat{\eta}_n}^{ka} \tilde{\eta}_n^a ~, \qquad S_n \times \eta^a = {\hat{\eta}_n}^{-ka}\tilde{\eta}_n^a \, .
\fe
Here $(\mathcal{Z}_n)$ stands for the 1+1d $\mathbb{Z}_n$ gauge theory. 
The fusion between $\eta$ and the higher quantum symmetry line  can be obtained as follows. We first fuse the bulk line $\eta$ with $S_n$ away from the quantum symmetry line and use \eqref{zn.fusion.rule.summary}, and then we fuse the two higher quantum symmetry lines on $S_n$ together using $\times_{S_n}$. 
In particular, we obtain
\begin{equation} \label{hateta.eta.from.hateta}
	\hat{\eta}_{n}^a \tilde{\eta}_{n}^b = \hat{\eta}_{n}^{a+kb} \times \eta^b = \eta^b \times \hat{\eta}_{n}^{a-kb} \,.
\end{equation}
The most general fusion between $\hat{\eta}_n^a$ and $\hat{\eta}_{n'}^{a'}$ is\footnote{When $a=a'$, this fusion rule simplifies to
\ie
\hat{\eta}_n^a \times \hat{\eta}_{n'}^{a} = \left( \mathcal{Z}_{\mathrm{gcd}(n,n',k\ell)} \right) \, \hat{\eta}_{\frac{\mathrm{gcd}(n,n',k\ell)nn'}{\mathrm{gcd}(n,n')^2}}^a \,.
\fe
Also when $k=0$ (i.e., when the $\mathbb{Z}_N$ symmetry is non-anomalous), the fusion simplifies:
\begin{equation}
	\hat{\eta}_{n}^a \times \hat{\eta}_{n'}^{a'} = \left( \mathcal{Z}_{\mathrm{gcd}(n,n')}, W^{a-a'} \right) \, \hat{\eta}_{\mathrm{lcm}(n,n')}^{\frac{pan'+p'a'n}{\mathrm{gcd}(n,n')}} \,.
\end{equation}
}
\begin{equation}\label{hatetahateta}
	\hat{\eta}_{n}^a  \times \hat{\eta}_{n'}^{a'} = \left( \mathcal{Z}_{\mathrm{gcd}(n,n',k\ell)}, W^{a-a'} \right) \, \hat{\eta}_{x}^{\frac{pan'+p'a'n}{\mathrm{gcd}(n,n')}} \tilde{\eta}_{x}^{{\ell(a-a') \over \mathrm{gcd}(n,n',k\ell)}\,\left(\frac{k\ell}{\mathrm{gcd}(n,n',k\ell)}\right)^{-1}_{N \over x\ell}} \,.
\end{equation}
See below \eqref{zn.general.fusion} for the definition of our notations. 
The other fusion rules can be obtained by combining the above equations.

An interesting special case of the general fusion rule \eqref{zn.general.fusion} is when $N$ is prime and  $k=1$. 
 Since the only positive divisors of a prime $N$ are 1 and $N$, we  have a trivial surface defect $S_1$ and a nontrivial condensation surface $S\equiv S_N$. 
The higher quantum symmetry line on $S$ is generated by $\hat\eta\equiv \hat\eta_N$, obeying $\hat\eta^N=S$. 
The fusion rule \eqref{hatetahateta} reduces to $\hat\eta^a \times \hat\eta^{a'} = \eta^{a-a'}$.
To summarize, for prime $N$ we have
\ie
&S\times S=1\,,\\
&\eta^N=1\,,\\
&S\times \eta = \eta^{-1} \times S\,,
\fe
and $\hat\eta = \eta \times S$. We see that the fusion rule is the dihedral group $D_{2N}$ of order $2N$.\footnote{In fact, whenever gcd$(N,k)=1$, $S_N$ and $\eta$ form a dihedral group $D_{2N}$ even when $N$ is not prime, but there are additional non-invertible surfaces $S_n$ with $n|N$.}  
This generalizes the fusion rule in \eqref{z2fermion}.

\subsection{Higher gauging of $\mathbb{Z}_{N_1}\times \mathbb{Z}_{N_2}$ 1-form symmetries and discrete torsions}\label{sec:ZN1ZN2}

Next, we move on to  1-gaugeable $\mathbb{Z}_{N_1} \times \mathbb{Z}_{N_2}$ 1-form symmetries in a general 2+1d QFT. 
The new feature here is that there are multiple ways to gauge the $\mathbb{Z}_{N_1} \times \mathbb{Z}_{N_2}$ 1-form symmetry on a surface in 2+1d. 
These different choices differ by adding a 1+1d invertible field theory of the background $\mathbb{Z}_{N_1} \times \mathbb{Z}_{N_2}$ gauge fields on the surface when gauging. 
The possible invertible field theories are classified by  the discrete torsion $H^2(B( \mathbb{Z}_{N_1} \times \mathbb{Z}_{N_2}),U(1))=\mathbb{Z}_{\mathrm{gcd}(N_1,N_2)}$.

Let the generators of $\mathbb{Z}_{N_1}\times \mathbb{Z}_{N_2}$ be $\eta_1$ and $\eta_2$ with $\eta_1^{ N_1} = \eta_2^{N_2} = 1$. The 't Hooft anomaly (i.e., the 0-anomaly) of $\mathbb{Z}_{N_1}\times \mathbb{Z}_{N_2}$ 1-form symmetry is parameterized by $k_1, k_2$, and $k_{12}$ 
corresponding to the spins of the $\eta_1$ and $\eta_2$ lines and the braiding between them
\begin{equation}
	\theta(\eta_1) = e^{2\pi i \frac{k_1}{N_1}}\,, \quad \theta(\eta_2) = e^{2\pi i \frac{k_2}{N_2}}\,, \quad B(\eta_1,\eta_2) = e^{2\pi i \frac{k_{12}}{\mathrm{gcd}(N_1,N_2)}}\,. \label{braiding}
\end{equation}
The possible values of $k_1,k_2$ are given by \eqref{anomaly.spin} (where $N$ is equal to $N_1$ and  $N_2$), while $k_{12} \in \mathbb{Z}_{\mathrm{gcd}(N_1,N_2)}$.\footnote{Since $1=B(\eta_1^{N_1},\eta_2)= B(\eta_1,\eta_2)^{N_1}$ and similarly $1=B(\eta_1,\eta_2)^{N_2}$, we see that $k_{12}\in \mathbb{Z}_{\text{gcd}(N_1,N_2)}$.}

For  the $\mathbb{Z}_{N_1}\times \mathbb{Z}_{N_2}$ 1-form symmetry to be 1-gaugeable,  we require  $k_1 \in \mathbb{Z}_{N_1}, k_2 \in \mathbb{Z}_{N_2}$, and $k_{12} \in \mathbb{Z}_{\mathrm{gcd}(N_1,N_2)}$. 
We will henceforth assume $k_1,k_2,k_{12}$ to be all integers in which case the $F$-symbols can be chosen to be trivial. 
Following a similar discussion in Section \ref{sec:ZN}, these 1-form symmetry lines obey the following non-commutative algebra on a surface $\Sigma$:
\ie \label{commutation.relations}
	\eta_1\left( \gamma_1 \right) \eta_1\left( \gamma_1'\right) &= e^{2\pi i \frac{k_1}{N_1} \langle \gamma_1,\gamma_1' \rangle} \, \eta_1\left( \gamma_1 + \gamma_1' \right) \,, \\
	\eta_2\left( \gamma_2 \right) \eta_2\left( \gamma_2' \right) &= e^{2\pi i \frac{k_2}{N_2} \langle \gamma_2,\gamma_2 '\rangle} \, \eta_2\left( \gamma_2 + \gamma_2' \right) \,, \\
	\eta_1\left( \gamma_1 \right) \eta_2\left( \gamma_2 \right) &= e^{2\pi i \frac{k_{12}}{\mathrm{gcd}(N_1,N_2)} \langle \gamma_1,\gamma_2 \rangle} \, \eta_2\left( \gamma_2 \right) \eta_1\left( \gamma_1 \right) \,,
\fe
for 
$\gamma_1,\gamma_1' \in H_1(\Sigma , \mathbb{Z}_{N_1})$ and  $\gamma_2,\gamma_2'\in H_1(\Sigma , \mathbb{Z}_{N_2})$.\footnote{
The intersection number $\langle \gamma_1 ,\gamma_2\rangle\in \mathbb{Z}_{\text{gcd}(N_1,N_2)}$ is defined by viewing $\gamma_1, \gamma_2$ as elements in $H_1(\Sigma, \mathbb{Z}_{\text{gcd}(N_1,N_2)})$.}

The condensation surfaces from gauging either $\mathbb{Z}_{N_1}$ or $\mathbb{Z}_{N_2}$ proceed in the same way as in Section \ref{sec:ZN}. 
We will denote them as $S_{\mathbb{Z}_{N_1}}$ and $S_{\mathbb{Z}_{N_2}}$. 
We can further gauge the whole $\mathbb{Z}_{N_1} \times \mathbb{Z}_{N_2}$ 1-form symmetry on a surface. 
For simplicity, we will start with the condensation defect on a torus, while the generalization to an arbitrary surface is straightforward. 
When all the $F$-symbols are trivial, the condensation defect on a torus takes the form:
\begin{equation}
	S_{\mathbb{Z}_{N_1} \times \mathbb{Z}_{N_2},f}(T^2) = \frac{1}{N_1 N_2} \sum_{a,b \in \mathbb{Z}_{N_1} \times \mathbb{Z}_{N_2}} 
	e^{\, {2\pi i \, f\over \mathrm{gcd}(N_1,N_2)}(a_1b_2-a_2b_1) } ~
	\begin{gathered}
		\begin{tikzpicture}
			\draw (0,0) -- (2,0) -- (2,2) -- (0,2) -- (0,0);
			\draw [decoration = {markings, mark=at position 0.15 with {\arrow[scale=1.3]{stealth}}, mark=at position 0.55 with {\arrow[scale=1.3]{stealth}}, mark=at position 0.9 with {\arrow[scale=1.3]{stealth}}},  postaction=decorate] (1,0) -- (1,0.2) node[left] { $b$} -- (1,2);
			\draw [decoration = {markings, mark=at position 0.55 with {\arrow[scale=1.3]{stealth}}}, postaction=decorate] (0,1) -- (0.5,1.2) node[above] {$a$} -- (1,1.4);
			\draw [decoration = {markings, mark=at position 0.55 with {\arrow[scale=1.3]{stealth}}}, postaction=decorate] (1,0.6) -- (2,1);
			\draw [fill=black] (1,1.4) circle (0.04);
			\draw [fill=black](1,0.6) circle (0.04);
		\end{tikzpicture}
	\end{gathered}~, \label{gauging.znxzn.torus}
\end{equation}
where $a= (a_1,a_2), b=(b_1,b_2) \in \mathbb{Z}_{N_1} \times \mathbb{Z}_{N_2}$ correspond to topological lines $\eta_1^{a_1} {\eta_2}^{a_2}$ and $\eta_1^{b_1} \eta_2^{b_2}$, respectively. 
The phase in the sum corresponds to a choice of the discrete torsion, labeled by $f \in \mathbb{Z}_{\mathrm{gcd}(N_1,N_2)}$ \cite{Vafa:1986wx,Vafa:1994rv}.

\bigskip\centerline{\it Discrete torsion in higher gauging}\bigskip

One new feature of the condensation defect $S_{\mathbb{Z}_{N_1}\times \mathbb{Z}_{N_2},f}$ is that it involves trivalent junctions between the 1-form symmetry lines. This is because in the case of 1-gauging $\mathbb{Z}_{N_1}\times \mathbb{Z}_{N_2}$, the condensation defect involves a sum over the insertions of 1-cycles inside $H_1(\Sigma , \mathbb{Z}_{N_1})$ and $H_1(\Sigma , \mathbb{Z}_{N_2})$ that would necessarily intersect each other. As mentioned in Section \ref{sec:general}, one can always redefine the phases at the trivalent junction, which is sometimes referred to as a ``gauge transformation" of the $R$- and $F$-symbols. 
Two sets of $R$- and $F$-symbols are equivalent if they differ by this ``gauge transformation."  
Below we will make a particular choice of the trivalent junctions below to define our condensation defect unambiguously.

In our discussion, it is preferred to work with trivial $F$-symbols. 
The most general redefinition of the junction phases, keeping the $F$-symbols trivial, is given by a 2-cocyle $\alpha \in Z^2(\mathbb{Z}_{N_1} \times \mathbb{Z}_{N_2},U(1))$ as 
\begin{equation}\label{phaseredef}
	\raisebox{-2.2em}{\begin{tikzpicture}
			\draw [thick, decoration = {markings, mark=at position .6 with {\arrow[scale=1.5]{stealth}}}, postaction=decorate] (0,0) to (0,.7) node[above] {\small $a+b$};
			\draw [thick, decoration = {markings, mark=at position .7 with {\arrow[scale=1.5]{stealth}}}, postaction=decorate] (-.6,-.5) node[left] {\small $a$} to (0,0);
			\draw [thick, decoration = {markings, mark=at position .7 with {\arrow[scale=1.5]{stealth}}}, postaction=decorate] (.6,-.5) node[right] {\small $b$} -- (0,0);
			\draw [fill=black](0,0) circle (-0.05);
	\end{tikzpicture}} \mapsto \, \alpha(a,b)
	\raisebox{-2.2em}{\begin{tikzpicture}
			\draw [thick, decoration = {markings, mark=at position .6 with {\arrow[scale=1.5]{stealth}}}, postaction=decorate] (0,0) to (0,.7) node[above] {\small $a+b$};
			\draw [thick, decoration = {markings, mark=at position .7 with {\arrow[scale=1.5]{stealth}}}, postaction=decorate] (-.6,-.5) node[left] {\small $a$} to (0,0);
			\draw [thick, decoration = {markings, mark=at position .7 with {\arrow[scale=1.5]{stealth}}}, postaction=decorate] (.6,-.5) node[right] {\small $b$} -- (0,0);
			\draw [fill=black](0,0) circle (-0.05);
	\end{tikzpicture}} , \qquad
	\raisebox{-2.2em}{\begin{tikzpicture}
			\draw [thick, decoration = {markings, mark=at position .6 with {\arrow[scale=1.5]{stealth}}}, postaction=decorate] (0,-.7) node[below] {\small $a+b$} to (0,0);
			\draw [thick, decoration = {markings, mark=at position .7 with {\arrow[scale=1.5]{stealth}}}, postaction=decorate] (0,0)  to (-.6,.5) node[left] {\small $a$};
			\draw [thick, decoration = {markings, mark=at position .7 with {\arrow[scale=1.5]{stealth}}}, postaction=decorate] (0,0)--(.6,.5) node[right] {\small $b$};
			\draw [fill=black](0,0) circle (0.05);
	\end{tikzpicture}} \mapsto \, \alpha(a,b)^{-1}
	\raisebox{-2.2em}{\begin{tikzpicture}
			\draw [thick, decoration = {markings, mark=at position .6 with {\arrow[scale=1.5]{stealth}}}, postaction=decorate] (0,-.7) node[below] {\small $a+b$} to (0,0);
			\draw [thick, decoration = {markings, mark=at position .7 with {\arrow[scale=1.5]{stealth}}}, postaction=decorate] (0,0)  to (-.6,.5) node[left] {\small $a$};
			\draw [thick, decoration = {markings, mark=at position .7 with {\arrow[scale=1.5]{stealth}}}, postaction=decorate] (0,0)--(.6,.5) node[right] {\small $b$};
			\draw [fill=black](0,0) circle (0.05);
	\end{tikzpicture}} ,
\end{equation} 
The closeness condition  $\alpha(a,b) \alpha(a+b,c) = \alpha(a, b+c)\alpha(b,c)$ guarantees that the $F$-symbols remain trivial after the phase redefinition. 

This phase redefinition \eqref{phaseredef} modifies the summand in  \eqref{gauging.znxzn.torus} by a phase $\alpha(a,b-a)/\alpha(b-a,a)$. Note that shifting $\alpha$ by an exact term, $\alpha(a,b) \mapsto \alpha(a,b) \frac{\beta(a) \beta(b)}{\beta(a+b)}$, leaves the above ratio invariant. 
Therefore we can view $\alpha$ as an element  in the cohomology group, i.e.,  $[\alpha] \in H^2(B(\mathbb{Z}_{N_1} \times \mathbb{Z}_{N_2}),U(1)) \cong \mathbb{Z}_{\mathrm{gcd}(N_1,N_2)}$. 
Let
\begin{equation} \label{torsor}
	\alpha(a,b) = e^{2\pi i f'\frac{a_1b_2}{\mathrm{gcd}(N_1,N_2)}}\,, ~~~~f'\in \mathbb{Z}_{\text{gcd}(N_1,N_2)}\,,
\end{equation}
then the phase redefinition \eqref{phaseredef} changes the discrete torsion phase in \eqref{gauging.znxzn.torus} from $f$ to $f+f'$.  

To recap the discussion above, the discrete torsion phase, labeled by $f$, is  a torsor over $H^2(B(\mathbb{Z}_{N_1} \times \mathbb{Z}_{N_2}),U(1))$ with no natural zero if we do not make a choice of our convention for the phases at the trivalent junctions. 
In other words, a phase redefinition \eqref{phaseredef} (which is sometimes called a ``gauge transformation" of the $R$- and $F$-symbols) shifts the value of $f$.

\bigskip\centerline{\it Fusion rule of the surfaces}\bigskip

To compute the fusion rule, we make a particular choice of the junction phases  such that the $R$-symbols take the following form\footnote{Note that this is compatible with the braiding phases in \eqref{braiding}, since $R^{a,b}_{a+b}R^{b,a}_{a+b} = B(a,b) \equiv B(\eta_1^{a_1}\eta_2^{a_2}, \eta_1^{b_1}\eta_2^{b_2}) $.}
\begin{equation}\label{RZN1ZN2}
	R^{a,b}_{a+b} = e^{2\pi i \left(\frac{a_1 b_1 k_1}{N_1} + \frac{a_2 b_2 k_2}{N_2} + \frac{a_1b_2 k_{12}}{\mathrm{gcd}(N_1,N_2)}\right) } ~.
\end{equation}
With the above choice, we can simplify the condensation defect to a form that will be convenient for computing the fusion rule:
\begin{equation}
	S_{\mathbb{Z}_{N_1} \times \mathbb{Z}_{N_2}, f}(T^2) = \frac{1}{N_1 N_2} \sum_{\substack{\gamma_1 \in H_1(T^2 , \mathbb{Z}_{N_1}) \\ \gamma_2 \in H_1(T^2 , \mathbb{Z}_{N_2})}} e^{2\pi i \frac{f \langle \gamma_1,\gamma_2 \rangle}{\mathrm{gcd}(N_1,N_2)}} \, \eta_1(\gamma_1) \, \eta_2(\gamma_2)~. \label{znxzn.condensation}
\end{equation}
To see that this coincides with \eqref{gauging.znxzn.torus}, we first parametrize the two cycles as $\gamma_i = a_i {\bf A}+b_i{\bf B}$ with $a_i,b_i \in \mathbb{Z}_{N_i}$, with $\bf A,B$ being the generating 1-cycles of the torus.  
Next, we  rewrite $\eta_1(\gamma_1)\eta_2(\gamma_2)$ in \eqref{znxzn.condensation} as
\begin{align}\label{proof}
& \eta_1(a_1 {\bf A}+ b_1 {\bf B}) \, \eta_2(a_2 {\bf A}+ b_2 {\bf B}) \notag \\
=&e^{-2\pi i \left(\frac{a_1 b_1 k_1}{N_1} + \frac{a_2 b_2 k_2}{N_2} + \frac{b_1a_2 k_{12}}{\mathrm{gcd}(N_1,N_2)}\right) } \, \eta_1(a_1 {\bf A}) \eta_2(a_2 {\bf A}) \, \eta_1(b_1 {\bf B}) \eta_2(b_2 {\bf B}) \notag \\
= &e^{-2\pi i \left(\frac{a_1 b_1 k_1}{N_1} + \frac{a_2 b_2 k_2}{N_2} + \frac{b_1a_2 k_{12}}{\mathrm{gcd}(N_1,N_2)}\right) } \, ~
	\begin{gathered}
		\begin{tikzpicture}[scale = 0.9]
			\draw (0,0) -- (2,0) -- (2,2) -- (0,2) -- (0,0);
			\draw [decoration = {markings, mark=at position 0.6 with {\arrow[scale=1.3]{stealth}}},  postaction=decorate] (1,0) -- (1,0.5) node[right] {$b$} -- (1,0.9);
			\draw (1,1.1) -- (1,2);
			\draw [decoration = {markings, mark=at position 0.3 with {\arrow[scale=1.3]{stealth}}}, postaction=decorate] (0,1) -- (0.5,1) node[above] {$a$} -- (2,1);
		\end{tikzpicture}
	\end{gathered} 
~	=~
	\begin{gathered}
		\begin{tikzpicture}[scale = 0.9]
			\draw (0,0) -- (2,0) -- (2,2) -- (0,2) -- (0,0);
			\draw [decoration = {markings, mark=at position 0.2 with {\arrow[scale=1.3]{stealth}}, mark=at position 0.55 with {\arrow[scale=1.3]{stealth}}, mark=at position 0.9 with {\arrow[scale=1.3]{stealth}}},  postaction=decorate] (1,0) -- (1,0.25) node[right] { $b$} -- (1,2);
			\draw [decoration = {markings, mark=at position 0.55 with {\arrow[scale=1.3]{stealth}}}, postaction=decorate] (0,1) -- (0.5,1.2) node[above] {$a$} -- (1,1.4);
			\draw [decoration = {markings, mark=at position 0.55 with {\arrow[scale=1.3]{stealth}}}, postaction=decorate] (1,0.6) -- (2,1);
			\draw [fill=black] (1,1.4) circle (0.04);
			\draw [fill=black](1,0.6) circle (0.04);
		\end{tikzpicture}
	\end{gathered} ~ ,
\end{align}
where we have used \eqref{RZN1ZN2} for $R^{b,a}_{a+b}$ in the last equality.

More generally, for $n_1$ and $n_2$ positive divisors of $N_1$ and $N_2$ ,we can gauge the $\mathbb{Z}_{n_1} \times \mathbb{Z}_{n_2}$ 1-form symmetry subgroup with a choice of discrete torsion $f \in \mathbb{Z}_{\mathrm{gcd}(n_1,n_2)}$ on a general surface $\Sigma$. It is straightforward to show that the condensation defect is
\begin{equation} \label{zn.zn.general.genus}
	S_{\mathbb{Z}_{n_1} \times \mathbb{Z}_{n_2}, f}(\Sigma) = {1\over \sqrt{|H_1 (\Sigma ,\mathbb{Z}_{n_1} \times \mathbb{Z}_{n_2})|}} \sum_{\substack{\gamma_1 \in H_1(\Sigma , \mathbb{Z}_{n_1}) \\ \gamma_2 \in H_1(\Sigma , \mathbb{Z}_{n_2})}} e^{2\pi i \frac{f \langle \gamma_1,\gamma_2 \rangle}{\mathrm{gcd}(n_1,n_2)}} \, \eta_1^{{N_1}/{n_1}}(\gamma_1) \, \eta_2^{N_2/n_2}(\gamma_2)~.
\end{equation}
Its orientation-reversal is:
\begin{equation}
	\overline{ S_{\mathbb{Z}_{n_1} \times \mathbb{Z}_{n_2}, f} }= S_{\mathbb{Z}_{n_1} \times \mathbb{Z}_{n_2}, f_0-f}~,
\end{equation}
where $f_0 = -\frac{N_1N_2 \mathrm{gcd}(n_1,n_2)}{n_1n_2 \mathrm{gcd}(N_1,N_2)} k_{12}$.  This is because orientation-reversal changes the order of multiplication of $\eta_1$ with $\eta_2$, and by using \eqref{commutation.relations} this gives $f_0$. Moreover, $\langle \gamma_1,\gamma_2 \rangle$ changes sign under orientation-reversal which changes the sign of $f$ in the formula. 

The following fusion rule follows immediately from this presentation of the condensation surfaces:
\begin{equation}
	S_{\mathbb{Z}_{n_1}} \times S_{\mathbb{Z}_{n_2}} = S_{\mathbb{Z}_{n_1} \times \mathbb{Z}_{n_2}, 0}~, \qquad S_{\mathbb{Z}_{n_2}} \times S_{\mathbb{Z}_{n_1}} = S_{\mathbb{Z}_{n_1} \times \mathbb{Z}_{n_2}, f_0 }~.
\end{equation}
Here $S_{\mathbb{Z}_{N_i}}$ is the condensation defect constructed by gauging $\mathbb{Z}_{N_i}$ on a surface. 

The more general fusion rule of these surfaces and their higher quantum symmetry lines  is quite involved and depends on the number theoretic properties of $n_1$ and $n_2$. In Sections \ref{sec:z2gauge} and \ref{sec:zpgauge}, we will work out the complete fusion rule in the case of 2+1d $\mathbb{Z}_2$ and $\mathbb{Z}_p$ gauge theory with prime $p$.

\bigskip\centerline{\it Action on the lines}\bigskip

We begin by a more convenient expression for $S_{\mathbb{Z}_{N_1} \times \mathbb{Z}_{N_2}, f}$ on the torus:
\begin{equation}
	S_{\mathbb{Z}_{N_1} \times \mathbb{Z}_{N_2},f}(T^2) =\frac{1}{N_1 N_2} \sum_{a,b \in \mathbb{Z}_{N_1}\times\mathbb{Z}_{N_2}} e^{2\pi i \left(k_1 \frac{a_1 b_1}{N_1} + k_2 \frac{a_2 b_2}{N_2} + \frac{(f+k_{12}) a_1b_2 - f a_2b_1 }{\mathrm{gcd}(N_1,N_2)}\right) }  ~
	\begin{gathered}
		\begin{tikzpicture}[scale = 0.9]
			\draw (0,0) -- (2,0) -- (2,2) -- (0,2) -- (0,0);
			\draw [decoration = {markings, mark=at position 0.6 with {\arrow[scale=1.3]{stealth}}},  postaction=decorate] (0,1) -- (0.5,1) node[above] {$a$} -- (0.9,1);
			\draw (1.1,1) -- (2,1);
			\draw [decoration = {markings, mark=at position 0.3 with {\arrow[scale=1.3]{stealth}}}, postaction=decorate] (1,0) -- (1,0.5) node[right] {$b$} -- (1,2);
		\end{tikzpicture}
	\end{gathered} ~,
\end{equation}
which differs from \eqref{gauging.znxzn.torus} by a braiding move (see \eqref{proof}). 
Recall that $a=(a_1,a_2), b=(b_1,b_2)\in \mathbb{Z}_{N_1}\times \mathbb{Z}_{N_2}$ label the topological lines of the $\mathbb{Z}_{N_1}\times \mathbb{Z}_{N_2}$ 1-form symmetry. 
Using this expression we find the action of $S_{\mathbb{Z}_{N_1} \times \mathbb{Z}_{N_2}, f}$ on a line defect $L$ to be
\begin{equation}
	\sum_{b\in \mathbb{Z}_{N_1}\times \mathbb{Z}_{N_2}} \left( \frac{ \sum_{a\in \mathbb{Z}_{N_1}\times \mathbb{Z}_{N_2}} e^{2\pi i \left(k_1 \frac{a_1 b_1}{N_1} + k_2 \frac{a_2 b_2}{N_2} + \frac{(f+k_{12}) a_1b_2 - f a_2b_1 }{\mathrm{gcd}(N_1,N_2)}\right) } B(\eta_1^{a_1}\eta_2^{a_2},L) }{N_1N_2} \right) \eta_1^{b_1} \eta_2^{b_2} L ~.
\end{equation}
Here $B(\eta_1^{a_1}\eta_2^{a_2},L) = \exp(2\pi i a_1 Q_1/N_1)\exp(2\pi i a_2 Q_2/N_2)$, where $(Q_1,Q_2)$ are the $\mathbb{Z}_{N_1} \times \mathbb{Z}_{N_2}$ 1-form symmetry charges of $L$. The above action can be simplified to
\begin{equation} \label{action.of.znxzn.condensation}
	S_{\mathbb{Z}_{N_1} \times \mathbb{Z}_{N_2},f} \cdot L = \sum^{N_i-1}_{\substack{b_i =0\\ Q_1 + k_1 b_1 + \frac{(f+k_{12})b_2 N_1}{\mathrm{gcd}(N_1,N_2)} = 0 \mod{N_1} \\ Q_2 + k_2 b_2 - f\frac{b_1 N_2}{\mathrm{gcd}(N_1,N_2)} = 0 \mod{N_2}}} \eta_1^{b_1}\eta_2^{b_2}L ~.
\end{equation}

\section{Examples of   condensation defects}\label{sec:example}

Let us now see how the construction of condensation surfaces that we described in previous sections works in some familiar models. As we have already pointed out, this construction applies to both TQFTs as well as non-topological QFTs. 

\subsection{$U(1)$ Maxwell theory and higher condensation}\label{sec:Maxwell}

Our first example is the 2+1d free $U(1)$ Maxwell gauge theory:
\ie
{\cal L} = {1\over g^2} F\wedge \star F\, ,
\fe
where $F=dA$ is the field strength of the dynamical 1-form gauge field $A$.  The gauge transformation is $A\sim A+d\alpha$.  
The pure $U(1)$ gauge theory is exactly dual to a free compact, massless scalar theory.\footnote{The duality is given by $dA\sim \star d \phi$, where $\phi$ is a compact scalar. See, for instance,  \cite{tong2018gauge}.}

The pure $U(1)$ gauge theory has a  $U(1)$ electric 1-form global symmetry, which shifts the gauge field by  a flat connection, $A\to A+\lambda$. 
The $U(1)$ 1-form symmetry is free of 't Hooft anomaly.\footnote{On the other hand, it has  a mixed anomaly with the magnetic $U(1)$ 0-form symmetry.} 
We can therefore gauge any of its $\mathbb{Z}_N$ 1-form symmetry subgroup on a surface for any positive integer $N$. 
The resulting condensation surface defects $S_N$ are non-invertible and obey the fusion rule  \eqref{generalSSZN} with $k=0$:
\ie\label{Maxwellfusion}
S_N \times S_{N'} = \left(\mathcal{Z}_{\text{gcd}(N,N')}\right) \, S_{\text{lcm}(N,N')}\,.
\fe
This can be viewed as the categorical version of the elementary identity $N\times N' = \text{gcd}(N,N')$ $\times\text{lcm}(N,N')$.\footnote{Relatedly,  this algebra admits a 1-dimensional representation $\langle S_N \rangle = N$. }

The action for the non-invertible surface in the 2+1d free $U(1)$ gauge theory takes the following form \cite{Choi:2022zal}:\footnote{SHS would like to thank Yichul Choi and Ho Tat Lam for enlightening discussions on this point and collaborations on a related project \cite{Choi:2022zal} in 3+1d.}
\ie\label{MaxwellLag}
S &= {1\over g^2} \int_{x<0} F_L\wedge \star F_L  +{1\over g^2} \int_{x>0} F_R\wedge \star F_R\\
&+ {iN\over 2\pi} \int _{x=0}  \phi (dA_L -dA_R)\,,
\fe
where $A_L,A_R$ are the bulk 1-form gauge fields on the two sides of the surface defect, which for notational convenience is placed at $x=0$.  
Here $\phi$ is a compact scalar field (i.e., $\phi\sim \phi+2\pi$) that only lives on the defect at $x=0$. 
When we integrate out $\phi$, it gauges the $\mathbb{Z}_N$ 1-form symmetry along $x=0$. 

Following similar steps as in \cite{Banks:2010zn,Kapustin:2014gua}, we can dualize the worldsheet action for the defect to 
\ie
i \int_{x=0} b \left[ d\bar\phi  - N (A_L-A_R) \right]\,,
\fe
where $b=d\phi$ is a 1-form gauge field  and $\bar\phi$ is a compact scalar field, both living only on the defect. 
In this presentation, $\bar\phi$ is a  Stueckelberg field that Higgses the $U(1)$ gauge field $A_L-A_R$ to $\mathbb{Z}_N$ only along the surface.  
More explicitly, this worldsheet action for the defect can be viewed as the low-energy limit of  a Higgs Lagrangian:
\ie
\int_{x=0} d\tau dy \left[ \, \left| \partial \Phi - i N (A_L-A_R)\Phi  \right|^2 +V(\Phi)\right] \,,
\fe
where  the potential $V(\Phi)$ is chosen so that the complex scalar field $\Phi = \rho e^{i\bar\phi}$ condenses. 
Importantly, the Higgs field $\Phi$ only lives on the 2-dimensional defect. 
Therefore, the Higgs mechanism only takes place on the surface $x=0$, but not in the bulk of the spacetime. 
This higher Higgs mechanism is the reason why we call these surfaces ``condensation" defects.

 Finally, let us reproduce the non-invertible fusion rule \eqref{Maxwellfusion} using the explicit defect action \eqref{MaxwellLag}.  
We place two parallel surfaces labeled by $N$ and $N'$ at $x=0$ and $x=\epsilon$, respectively, and bring them close to each other.  
The total action is
\ie
S &= {1\over g^2} \int_{x<0} F_L\wedge \star F_L  +{1\over g^2} \int_{0<x<\epsilon } F_I\wedge \star F_I
+{1\over g^2} \int_{x>0} F_R\wedge \star F_R\\
&+ {iN\over 2\pi} \int _{x=0}  \phi_L (dA_L -dA_I)+ {iN'\over 2\pi} \int _{x=\epsilon}  \phi_R (dA_I -dA_R)\,,
\fe
 where $A_I$ is the dynamical 1-form gauge field that lives in the intermediate region between the two surfaces. 
 In the limit $\epsilon\to0$ when the two surfaces $S_N$ and $S_{N'}$ are on top of each other, $A_I$ only lives on the $x=0$ surface.  The worldsheet action for the surface defect then becomes
 \ie
\int _{x=0}\left[ {iN\over 2\pi}    \phi_L (dA_L -dA_I)+ {iN'\over 2\pi}   \phi_R (dA_I -dA_R)\right]\,.
 \fe
 Let us perform a change of variables on the two compact scalars $\phi_L,\phi_R$ to another pair $\phi, \varphi$:
 \ie
& \phi_L = { \text{lcm}(N,N')\over N} \phi + \alpha \varphi\,,\\
& \phi_R = { \text{lcm}(N,N')\over N'} \phi + \beta \varphi\,,
 \fe
 where $\alpha,\beta$ are two integers obeying $-\alpha N + \beta N' =  \text{gcd}(N,N')$. Note that the new scalars $\phi,\varphi$ have proper $2\pi$ periodicities. 
 In terms of the new variables, the worldsheet  action for the defect becomes
 \ie
 \int_{x=0} \left[\,
 {i\text{lcm}(N,N') \over 2\pi} \phi (dA_L-dA_R)
 +{i\text{gcd}(N,N')\over 2\pi} \varphi da
 \,\right]
 \fe
 where $a\equiv A_I + {N\over \text{gcd}(N,N')}\alpha A_L-{N'\over \text{gcd}(N,N')}\beta A_R$. 
 The first term above is the condensation surface $S_{\text{lcm}(N,N')}$, and the second term is a decoupled 1+1d $\mathbb{Z}_{\text{gcd}(N,N')}$ gauge theory.  Hence we have reproduced the non-invertible fusion rule \eqref{Maxwellfusion} in the 2+1d $U(1)$ gauge theory from the explicit Lagrangians for these condensation defects. 

\subsection{$U(1)_{2N}$ Chern-Simons theory}\label{sec:U1CS}

$U(1)_{2N}$ Chern-Simons theory is arguably the simplest 2+1d bosonic QFT with a 1-form global symmetry. 
In this subsection, we will realize  the topological surfaces of this theory, studied in \cite{Kapustin:2010if} (see also \cite{Fuchs:2002cm}), from higher gauging. 
In particular, we will show that the fusion rule of these topological surfaces is a special case of our general formula \eqref{generalSSZN}.
Furthermore, we will give explicit Lagrangian descriptions for these condensation surfaces and reproduce their fusion rule.

\subsubsection{Condensation surfaces and the charge conjugation symmetry}

The action of the  $U(1)_{2N}$ Chern-Simons theory is:
\ie
S= {2iN\over 4\pi } \int AdA\, ,
\fe
where $A$ is a 1-form dynamical gauge field. 
The Wilson lines and their spins are
\ie
e^{i a \oint A}\,,~~~\theta(a) =\exp\left(2\pi i { a^2\over 4N} \right) \,,~~~a=0,1,\cdots, 2N-1\,.
\fe
 These Wilson lines generate a $\mathbb{Z}_{2N}$ 1-form symmetry \cite{Gaiotto:2014kfa}.  
 Since the spin of the generator is not a $2N$-th root of unity, we see that the $\mathbb{Z}_{2N}$ 1-form symmetry is not 1-gaugeable (see \eqref{ZN1cond}).

While the whole $\mathbb{Z}_{2N}$ 1-form symmetry group is 1-anomalous, its $\mathbb{Z}_N$ subgroup generated by 
\ie
\eta= \exp\left(2 i \oint A\right)
\fe
 is 1-gaugeable, and so is its  $\mathbb{Z}_{n}$  subgroup   with $n$ a divisor of $N$, i.e., $N=mn$ for some $m$. 
Indeed, the spin of the generator $\exp(2i m  \oint A)$ of $\mathbb{Z}_n$ is $\theta = \exp(2\pi i m^2/N)$, which is an $n$-th root of unity (see \eqref{ZN1cond}).

Therefore, the condensation surfaces in the $U(1)_{2N}$ Chern-Simons theory are $S_n$ with $n$ a divisor of $N$. 
These are exactly the surfaces studied in  \cite{Kapustin:2010if}, which was based on the earlier work of \cite{Fuchs:2002cm}. 
The fusion rule derived in the above reference is a special case of \eqref{generalSSZN} with $k=1$:\footnote{By noting $N = \ell nn' / \mathrm{gcd}(n,n')$, we see that $\mathrm{gcd}(N/n,N/n') = \ell$, $\mathrm{gcd}(n,N/n') = {\mathrm{gcd}(n,n',\ell) n \over \mathrm{gcd}(n,n')}$ and $\mathrm{gcd}(n',N/n) = {\mathrm{gcd}(n,n',\ell) n' \over \mathrm{gcd}(n,n')}$. Using these identities we find \eqref{KS} is equivalent to the fusion rule presented in \cite{Kapustin:2010if}.}
 \begin{equation}\label{KS}
	S_n \times S_{n'} = \left( \mathcal{Z}_{\mathrm{gcd}(n,n',\ell)}  \right) S_{\frac{\mathrm{gcd}(n,n',\ell)nn'}{\mathrm{gcd}(n,n')^2}} \,,
\end{equation}
where $\ell = {N\over \text{lcm}(n,n')}$.
In particular, we have
\ie
&S_n \times S_1 = S_1\times S_n  =S_n\,,~~~\forall ~n|N\,,\\
&S_N\times S_N=1\,.
\fe
Hence, $S_1=1$ is the trivial defect and $S_N$ is the charge conjugation $\mathbb{Z}_2$ 0-form symmetry that maps $A\to -A$. To see the action $A\to -A$, we need to understand how $S_N$ acts on gauge invariant operators, which include the  Wilson lines $W^a=\exp(i a \oint A)$. Indeed, we will verify in \eqref{SnactionCS} that $S_N$ maps $W^a \to W^{-a}$.

Let us go through a few examples.  $U(1)_2$ does not have any 1-gaugeable 1-form symmetry. 
Consequently, $U(1)_2$ does not have a charge conjugation symmetry. 
$U(1)_4$ has a 1-gaugeable $\mathbb{Z}_2$ 1-form symmetry, which is a fermion. Its condensation defect is the charge conjugation symmetry, which is a (non-anomalous) $\mathbb{Z}_2$ 0-form symmetry. 

The simplest example of a non-invertible surface is in $U(1)_8$. 
There are two nontrivial condensation surfaces, $S_2$ and $S_4$, arising from the higher gauging of the 1-gaugeable $\mathbb{Z}_2$ and $\mathbb{Z}_4$ subgroups, respectively.  
Their fusion rules are
\ie
&S_2\times S_2 = (\mathcal{Z}_2) \, S_2\,,~~~S_4\times S_4=1\,,~~~S_2\times S_4 =S_4\times S_2 =S_2\,.
\fe
Here $S_4$ is the charge conjugation symmetry and $S_2$ is a non-invertible surface.

Let us make a comment on  anomalies.  
Since there is no 't Hooft anomaly for any $\mathbb{Z}_2$ 0-form symmetry in 2+1d bosonic QFT, the charge conjugation symmetry generated by $S_N$ in the $U(1)_{2N}$ Chern-Simons theory is always 0-gaugeable. 
However, $S_N$ arises from the higher gauging of a 0-anomalous 1-form  $\mathbb{Z}_N$  symmetry. 
It is curious that an anomalous 1-form symmetry gives rise to a non-anomalous 0-form symmetry via higher gauging. 
See also the discussion at the end of Section \ref{sec:z2fusion}.

The action of the condensation surface $S_n$   on the lines is given by the general formula \eqref{Snaction} with $k=1$.  
Let $W=\exp(i\oint A)$ be the minimal Wilson line, then the  generator of the 1-gaugeable $\mathbb{Z}_N$ 1-form symmetry subgroup is  $\eta= W^2$. 
We have $B(\eta, W )  =  \theta(W)^4 =  \exp(2\pi i /N)$, hence the $\mathbb{Z}_N$ 1-form symmetry charge of $W$ is $Q=1$. 
Applying \eqref{Snaction}, we obtain the general action of the condensation surface on the Wilson lines in the $U(1)_{2N}$ Chern-Simons theory: 
\ie\label{SnactionCS}
&S_n \cdot W^b = \sum_{\substack{a=0 \\ ma +b=0~\text{mod}~n }}^{n-1} W^{2ma+b} =  \begin{cases}
\displaystyle{\sum_{c=0}^{\text{gcd}(m,n) -1}} W^{-b +  {2N\over\text{gcd}(m,n)} c} \,, &\text{if}~~\text{gcd}(m,n)| b\\
\, 0\,,  &\text{otherwise}
\end{cases}\,,\\
&m=N/n\,.
\fe
In particular, this implies that $W^n$ has a junction with itself  on $S_n$. 
On the other hand, $W^m$ and $W^{-m}$ can meet at a junction on $S_n$.  
(See the general discussion on junctions around Figure \ref{fig:cylinder S}.)  
Finally, when $n=N$, we find the charge conjugation symmetry action, $S_N \cdot W^b =W^{-b}$.

\subsubsection{Lagrangian for the defects}

We now discuss the explicit worldsheet Lagrangian presentation for the (generally non-invertible) condensation surfaces $S_n$ in the $U(1)_{2N}$ Chern-Simons theory.  
Our presentation differs slightly from that in \cite{Kapustin:2010if}.

A topological surface in $U(1)_{2N}$ Chern-Simons theory is described by imposing topological  sewing conditions on the gauge field on both sides of the surface. 
For the surface defect $S_n$ with $N=nm$, the appropriate sewing conditions compatible with \eqref{SnactionCS} is
\ie
  &n(A_L -A_R)|=\text{pure gauge}, \\
   &m(A_L +A_R)|=\text{pure gauge}~, \label{bc}
\fe
where $A_L$ ($A_R$) is the gauge field on the left (right) on the surface. In other words, we impose that the above combinations of the gauge fields on the two sides of the surface are pure gauge. We will shortly see that this is equivalent to the higher gauging of a $\mathbb{Z}_{n}$ 1-form symmetry subgroup   on the surface.

For notational simplicity, we place the topological surface $S_n$ on the  2-manifold  $x=0$ in 2+1d.  
The action for the 2+1d $U(1)_{2N}$ Chern-Simons theory with $S_n$ inserted is given by\footnote{Our action differs from \cite{Kapustin:2010if} by the last term. As a result, their action does not have the full left and right gauge symmetries implemented by $\alpha_L,\alpha_R$, while ours does. }
\ie\label{L1}
&{2iN\over 4\pi } \int_{x<0} A_LdA_L 
+{2iN\over 4\pi} \int_{x>0} A_R dA_R\\
&+{i  n\over 2\pi} \int_{x=0} d\phi (A_L- A_R) +{iN\over 2\pi} \int_{x=0} A_L A_R\,.
\fe
The gauge transformations are
\ie
&A_L\sim A_L+d\alpha_L\,,~~~A_R\sim A_R+d\alpha_R\,,\\
&\phi \sim  \phi +  m(\alpha_L+\alpha_R)\,.
\fe
The equations of motion of $A_L,A_R$ on the boundary give
\ie\label{mbc}
m( A_L+A_R)|_{x=0}= d\phi\,.
\fe
This realizes one of the conditions in \eqref{bc}. 
The other one can be obtained as follows. 
Starting from \eqref{L1}, we can dualize the scalar $\phi$ on the defect. 
Let us define a 1-form field $b\equiv d\phi$. Then the defect Lagrangian becomes
\ie
& \int_{x=0}\left[
 {i n\over 2\pi} b(A_L-A_R) -{ i\over 2\pi} \bar\phi db +{iN\over 2\pi} A_LA_R\right]\, ,\\
\fe
where $\bar\phi$ is a Lagrange multiplier enforcing the condition $db=0$.  
We therefore obtain the following dual action:
\ie\label{L2}
&{2iN\over 4\pi } \int_{x<0} A_LdA_L 
+{2iN\over 4\pi} \int_{x>0} A_R dA_R\\
&+\int_{x=0}\left[
 -{i\over 2\pi} b ( d\bar\phi -  n A_L+ n A_R)  +{iN\over 2\pi} A_LA_R\right]\,.
\fe
The gauge transformations are
\ie
&A_L\sim A_L+d\alpha_L\,,~~~A_R\sim A_R+d\alpha_R\,,\\
&b \sim  b + m(d\alpha_L+d\alpha_R)\,,\\
&\bar\phi \sim \bar\phi+  n (\alpha_L-\alpha_R)\,.
\fe
In the Higgs presentation \eqref{L2}, $\bar\phi$ is a Stueckelberg field that Higgses the gauge field $A_L-A_R$ to $\mathbb{Z}_n$ on the defect. As noted before, this justifies why the surface is called a ``condensation" defect.

In the dual action \eqref{L2}, the equation of motion for $b$ then gives the second condition in \eqref{bc}:
\ie
n (A_L -A_R) |_{x=0}= d\bar\phi\,.
\fe
We conclude that the action \eqref{L1} (or its dual \eqref{L2}) gives the correct sewing conditions \eqref{bc} for the condensation defect $S_n$.

\subsubsection{Non-invertible fusion rules from Lagrangians}\label{sec:U1Lagfusion}

Similar to Section \ref{sec:Maxwell}, the fusion of two parallel defects $S_{n_1}$ and $S_{n_2}$ is described by the Lagrangian 
\ie
& \frac{i N}{2\pi} \int_{x<0} A_L d A_L+\frac{i N}{2\pi} \int_{x=0}^{x=\epsilon} A_I d A_I+\frac{i N}{2\pi} \int_{x>\epsilon} A_R d A_R\\
&  +\frac{i n_1}{2\pi} \int_{x=0}  d\f_L(A_L-A_I)+ \frac{i N}{2 \pi} \int_{x=0} A_L A_I \\
 &+\frac{i n_2}{2\pi} \int_{x=\epsilon}  d\f_R(A_I-A_R)+ \frac{i N}{2 \pi} \int_{x=\epsilon} A_I A_R ~,
\fe
where for simplicity we consider flat defects along the $x=0$ and $x=\epsilon$ planes, respectively. Also for simplicity in what follows we focus on the case of $U(1)_{2n^2}$ theories and further simplify to the case with $n_1=n_2=n$. To compute the fusion $S_n \times S_n$ we just need to take the limit $\epsilon\rightarrow 0$ in the above action. 
The worldsheet action for the defect at $x=0$  then becomes 
\begin{align}
\int_{x=0}  \left[ {in\over2\pi} d\f_L(A_L-A_I)+{i N\over2\pi} A_L A_I +{i n\over2\pi} d\f_R(A_I-A_R)+ {iN\over2\pi} A_I A_R\right]  ~.
\end{align}
The gauge transformations are
\ie
&A_L\sim A_L+d\alpha_L\,,~~~A_R\sim A_R+d\alpha_R\,,~~~A_I\sim A_I+d\alpha_I\,,\\
&\phi_L \sim  \phi _L+  n(\alpha_L+\alpha_I)\,,~~~\phi_R \sim  \phi _R+  n(\alpha_I+\alpha_R)\,.
\fe
We can rewrite the worldsheet  action for the fused surface defect as
\ie
& \int_{x=0}
\left[{in\over 2\pi}  d\bar\phi (A_L+A_R) -{in^2\over 2\pi}A_LA_R
+{i\over 2\pi}(d\bar\phi - n  A_L +n A_R)( d\varphi -na)
\right]
\fe
where  
\ie
&\bar\phi\equiv \phi_L-\phi_R\,,~~~\varphi \equiv \phi_L\,,~~~a\equiv A_I +A_L\,.
\fe
In particular, $a$ and $\varphi$ have the independent gauge transformations $a\sim a+d\lambda$ and $ \varphi\sim \varphi+ n\lambda$.

Next, we dualize the scalar field $\bar\phi$ as before using a 1-form field $\bar b=d\bar\phi$ and introduce a Lagrange multiplier $\phi$ to enforce $d\bar b=0$:
\ie
 \int_{x=0}
\left[-{i\over 2\pi} \bar b( d\phi - n A_L-n A_R) -{in^2\over 2\pi}A_LA_R
+{i\over 2\pi}(\bar b- n  A_L +n A_R)( d\varphi -na)
\right]
\fe
The gauge transformation of $\phi$ is $\phi\sim \phi+n(\alpha_L+\alpha_R)$. 

Finally, we define a gauge invariant 1-form field $c\equiv\bar b-nA_L+nA_R$. 
Integrating out $\varphi$ then gives $dc=0$. 
Locally, we can solve this equation by $c=d\tilde\varphi$. 
After this dualization, the worldsheet action for the surface now takes a factorized form:
\ie
\int_{x=0}
\left[{in\over 2\pi}  d\phi( A_L -A_R)+{in^2\over 2\pi}A_LA_R
+{in\over 2\pi} \tilde ad\tilde\varphi \, ,
\right]
\fe
where $\tilde a \equiv a-A_L-A_R=  A_I-A_R$.  
The first two terms are the worldsheet action for $S_n$, while the last term is a decoupled 1+1d $\mathbb{Z}_n$ gauge theory. This can be seen by doing the change of variable $A_I \mapsto A_I + A_R$

 Hence, we conclude that 
\begin{equation}
S_n \times S_n = (\mathcal{Z}_n  )\, S_n~.
\end{equation}
This agrees with the fusion rule \eqref{generalSSZN} with $n=n'$ and $k=1$.

\subsection{$\mathbb{Z}_2$ gauge theory} \label{sec:z2gauge}

In this subsection, we study the surface defects and their fusion rules in the 2+1d $\mathbb{Z}_2$ gauge theory, i.e., the low energy limit of the $\mathbb{Z}_2$ toric code \cite{Kitaev:1997wr}.  
We find six topological surfaces, which consists of one trivial surface, a $\mathbb{Z}_2$ surface for the electromagnetic symmetry exchanging the $\ee$ and $\mm$ anyons, and four non-invertible surfaces. 
Two of the non-invertible topological surfaces turn out to be the Cheshire strings in the 2+1d version of the models in  \cite{Else:2017yqj,Johnson-Freyd:2020twl}, which were based on the earlier work in \cite{Alford:1990mk,Preskill:1990bm,Bucher:1991bc,Alford:1992yx}.

The 2+1d $\mathbb{Z}_2$ gauge theory has a $\mathbb{Z}_2^\ee$ electric and a  $\mathbb{Z}_2^\mm$ magnetic 1-form symmetry. 
The full $\mathbb{Z}_2^\ee \times \mathbb{Z}_2^\mm$  is not 0-gaugeable (i.e., cannot be gauged in the whole spacetime). 
Its 0-anomaly is parameterized by  $k_\ee = k_\mm = 0$ and $k_{\ee\mm} = 1$ (see Section \ref{sec:ZN1ZN2}).  
However, it is 1-gaugeable, i.e., the full $\mathbb{Z}_2^\ee \times \mathbb{Z}_2^\mm$ can be gauged on a codimension-1 surface in spacetime. 
We denote the generators by $\eta_\ee$ and $\eta_\mm$, respectively. Their spins and braiding are given by
\begin{equation}
	\theta(\eta_\ee) =\theta(\eta_\mm) = 1\,, \quad B(\eta_\ee,\eta_\mm) = -1\,.
\end{equation}
These 1-form symmetry lines obey the non-commutative algebra given in \eqref{commutation.relations}.

\subsubsection{Six condensation surfaces}

The condensation surfaces correspond to different ways of gauging subgroups of the 1-gaugeable $\mathbb{Z}_2^\ee \times \mathbb{Z}_2^\mm$. In total we have six  surfaces, including the trivial surface. There are two $\mathbb{Z}_2$ subgroups generated by the electric and magnetic boson lines, and another one generated by the fermion line $\eta_\psi \equiv \eta_\ee \eta_\mm$. Thus, we get three  condensation surfaces from the higher gauging of these three $\mathbb{Z}_2$ 1-form symmetry subgroups:
\ie \label{toric.code.surface}
	S_\ee(\Sigma) &= \frac{1}{\sqrt{|H_1(\Sigma,\mathbb{Z}_2)|}} \sum_{\gamma \in H_1(\Sigma,\mathbb{Z}_2)} \eta_\ee(\gamma)  \, , \\
	S_\mm(\Sigma) &= \frac{1}{\sqrt{|H_1(\Sigma,\mathbb{Z}_2)|}} \sum_{\gamma \in H_1(\Sigma,\mathbb{Z}_2)} \eta_\mm(\gamma)  \, , \\
	S_\psi(\Sigma) &= \frac{1}{\sqrt{|H_1(\Sigma,\mathbb{Z}_2)|}} \sum_{\gamma \in H_1(\Sigma,\mathbb{Z}_2)} \eta_\psi(\gamma)  \,.
\fe
From \eqref{genusgfusionZ2}, we learn that $S_\ee$ and $S_\mm$ are non-invertible because they arise from the higher gauging of a boson line, while $S_\psi$ is an invertible $\mathbb{Z}_2$ surface because it is the condensation defect of a fermion line. 

There are two more  condensation surfaces corresponding to two possible discrete torsions $H^2(B(\mathbb{Z}_2^\ee \times \mathbb{Z}_2^\mm),U(1))=\mathbb{Z}_2$ for the higher gauging of $\mathbb{Z}_2^\ee \times \mathbb{Z}_2^\mm$ on a surface:
\ie
	S_{\ee\mm}(\Sigma) &= \frac{1}{\sqrt{|H_1(\Sigma,\mathbb{Z}_2 \times \mathbb{Z}_2)|}} \sum_{\gamma,\gamma' \in H_1(\Sigma,\mathbb{Z}_2)} \eta_\ee(\gamma)\eta_\mm(\gamma')  \, , \\
	S_{\mm\ee}(\Sigma) &= \frac{1}{\sqrt{|H_1(\Sigma,\mathbb{Z}_2 \times \mathbb{Z}_2)|}} \sum_{\gamma,\gamma' \in H_1(\Sigma,\mathbb{Z}_2)} (-1)^{ \langle \gamma, \gamma' \rangle }\eta_\ee(\gamma)\eta_\mm(\gamma')  \, .
\fe
Note that these two surfaces are orientation reversal of each other, while the  other three condensation surfaces $S_\ee, S_\mm,S_\psi$  are invariant under orientation reversal:
\ie
 \overline{S_{\ee\mm}}= S_{\mm\ee} \,,~~~\overline S_\ee = S_\ee\,,~~~\overline S_\mm = S_\mm \,,~~~\overline S_\psi =S_\psi\,.
\fe

The fusion rules are computed in Appendix \ref{app.zp.gauge.theory}.\footnote{In  Appendix \ref{app.zp.gauge.theory} we derive the fusion rule for the surfaces in the more general $\mathbb{Z}_p$ gauge theory with prime $p$.  The translation between the notation here and there is given by: $$S_\ee = S_{\mathbb{Z}_{2}^{(\infty)}} \,, \quad S_\mm = S_{\mathbb{Z}_{2}^{(0)}} \,, \quad S_\psi =  S_{\mathbb{Z}_{2}^{(1)}} \,, \quad S_{\ee\mm} = S_{(\mathbb{Z}_2 \times \mathbb{Z}_2)^{(\infty)}} \,, \quad S_{\mm\ee}= S_{(\mathbb{Z}_2 \times \mathbb{Z}_2)^{(0)}}\,.$$
} The minimal list of fusion rules are: 
\begin{equation} \label{z2.gauge.theory.fusion}
\begin{array}{lllll}
	S_\ee \times S_\ee = (\mathcal{Z}_2) S_\ee \,, &~~~~& S_\ee \times S_\mm = S_{\ee\mm} \,, &~~~~& S_{\ee\mm} \times S_{\ee\mm} = S_{\ee\mm} \,,\\
	S_\mm \times S_\mm = (\mathcal{Z}_2) \, S_\mm \,, &~~~~& S_\ee \times S_\psi = S_{\ee\mm} \,, &~~~~& S_{\ee\mm} \times S_{\mm\ee} = (\mathcal{Z}_2) \,  S_{\ee} \,, \\
	S_\psi \times S_\psi = 1 \,, &~~~~& S_\psi \times S_\mm = S_{\ee\mm} \,, &~~~~& S_{\mm\ee} \times S_{\ee\mm} = (\mathcal{Z}_2) \,  S_{\mm} \,,\\
	S_\ee \times S_{\ee\mm} = (\mathcal{Z}_2) \, S_{\ee\mm} \,, &~~~~& S_\mm \times S_{\ee\mm} =  S_{\mm} \,, &~~~~& S_\psi \times S_{\ee\mm} = S_\mm \,, \\
	S_\mm \times S_{\mm\ee} = (\mathcal{Z}_2) \, S_{\mm\ee} \,, &~~~~& S_\ee \times S_{\mm\ee} =  S_{\ee} \,, &~~~~& S_\psi \times S_{\mm\ee} = S_\ee \,.
\end{array}
\end{equation}
The other ten fusion rules can be obtained by taking the orientation reversal of the above fusion rules.\footnote{Suppose ${\cal N}_1\times {\cal N}_2 = {\cal N}_3$, then using the definition of the fusion product in Figure \ref{fig:fusion} and the orientation reversal \eqref{orientationrev}, we have $\overline {\cal N}_2 \times \overline {\cal N}_1 = \overline {\cal N}_3$. }
 To summarize, in addition to the trivial surface, we have three non-invertible surfaces $S_\ee, S_\mm, S_{\ee\mm}$, and one invertible  $\mathbb{Z}_2$ surface $S_\psi$.
These six topological surfaces in the 2+1d $\mathbb{Z}_2$ gauge theory have been previously studied in \cite{Lan:2014uaa}.

Using \eqref{action.of.fermioncondensation} and \eqref{action.of.znxzn.condensation}, we find the action of these surfaces on the anyon lines:
\ie
	S_\ee \cdot 1 &= 1+\eta_\ee \,, \quad &S_\ee \cdot \eta_\ee &= 1+\eta_\ee \,, \quad & S_\ee \cdot \eta_\mm &= 0 \,, \quad & S_\ee \cdot \eta_\psi &= 0 \,, \\
	S_\mm \cdot 1 &= 1+\eta_\mm \,, \quad &S_\mm \cdot \eta_\ee &= 0 \,, \quad & S_\mm \cdot \eta_\mm &= 1+\eta_\mm \,, \quad & S_\mm \cdot \eta_\psi &= 0 \,, \\
	S_\psi \cdot 1 &= 1 \,, \quad&	S_\psi \cdot \eta_\ee & = \eta_\mm \,, \quad & S_\psi \cdot \eta_\mm &= \eta_\ee \,, \quad & S_\psi \cdot \eta_\psi  &= \eta_\psi \,,\\
	S_{\ee\mm} \cdot 1 &= 1+\eta_\ee \,, \quad &S_{\ee\mm} \cdot \eta_\ee &= 0 \,, \quad & S_{\ee\mm} \cdot \eta_\mm &= 1+\eta_\ee \,, \quad & S_{\ee\mm} \cdot \eta_\psi &= 0 \,,\\
	S_{\mm\ee} \cdot 1 &= 1+\eta_\mm \,, \quad &S_{\mm\ee} \cdot \eta_\ee &= 1+\eta_\mm \,, \quad & S_{\mm\ee} \cdot \eta_\mm &= 0\,, \quad & S_{\mm\ee} \cdot \eta_\psi &= 0 \,.
\fe
In particular, we see that $S_\psi$ exchanges the two anyons $\eta_\ee,\eta_\mm$, i.e., it generates the  electromagnetic $\bZ_2$ symmetry.  
This $\mathbb{Z}_2$ symmetry has been realized on a lattice in \cite{Wen:2003yv}. 
See also \cite{Freed:2018cec,Ji:2019ugf} for the relation between this $\mathbb{Z}_2$ symmetry in 2+1d and the Kramers-Wannier duality of the Ising model.

Note that the electromagnetic $\mathbb{Z}_2$ 0-form symmetry is 0-gaugeable, even though it arises from an anomalous 1-form symmetry line, the fermion line $\eta_\psi$. 
Indeed, gauging the electromagnetic $\mathbb{Z}_2$ 0-form symmetry in the 2+1d $\mathbb{Z}_2$ gauge theory yields either the doubled Ising TQFT, or the $Spin(3)_1\times Spin(13)_{1}$ Chern-Simons theory, depending on the choice of the 2+1d SPT class in $H^3(B\mathbb{Z}_2,U(1))=\mathbb{Z}_2$ \cite{2013arXiv1302.2634L}.  

\subsubsection{Non-invertible surfaces as products of boundaries}

\begin{figure}
	\centering
	\begin{tikzpicture}[scale = 1]
		\node[cylinder, 
		draw = black, 
		text = purple,
		cylinder uses custom fill, 
		cylinder body fill =gray!50,
		cylinder end fill = magenta!40, 
		minimum width = 2cm,
		minimum height = 3.1cm,
		shape border rotate = 0] (c) at (-1.5,0) {}; 
		\node[color = violet] at (0,-1.5) {$\ket{B}$};
		
		\node[cylinder, 
		draw = black, 
		text = purple,
		cylinder uses custom fill, 
		cylinder body fill =gray!50,
		cylinder end fill = gray!70, 
		minimum width = 2cm,
		minimum height = 3.1cm,
		shape border rotate = 0] (c) at (2.3,0) {}; 
		
		\node[ellipse,
		draw = violet,
		text = violet,
		fill = magenta!40,
		minimum width = 0.2cm, 
		minimum height = 2cm] (e) at (1,0) {};
		
		\node[color = violet] at (1,-1.5) {$\bra{B'}$};     
		
		\node[] at (5,0) {$\rightarrow$};    
		
		\node[cylinder, 
		draw = black, 
		text = purple,
		cylinder uses custom fill, 
		cylinder body fill =gray!50,
		cylinder end fill = gray!70, 
		minimum width = 2cm,
		minimum height = 5cm,
		shape border rotate = 0] (c) at (9,0) {}; 
		
		\node[ellipse,
		draw = violet,
		text = violet,
		fill = magenta!40,
		minimum width = 0.2cm, 
		minimum height = 2cm] (e) at (9,0) {};
		
		\node[color = violet] at (9,-1.5) {$\ket{B}\bra{B'}$}; 
	\end{tikzpicture}
	\caption{Fusing a boundary condition $\ket{B}$ with the orientation reversal $\bra{B'}$ of another boundary condition we can construct a surface defect $S=\ket{B}\bra{B'}$.}
	\label{fig: Factorized surfaces}
\end{figure}
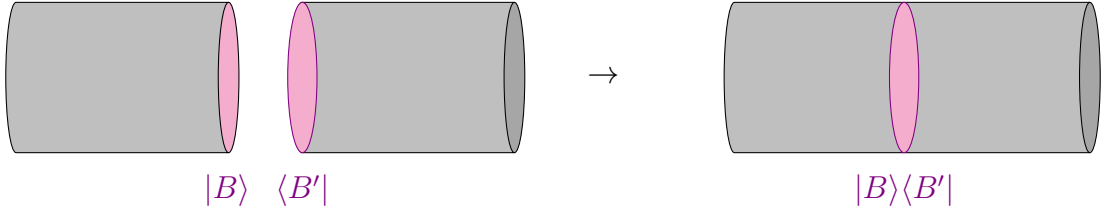

We now show that  the four non-invertible defects are factorized into the tensor products of two gapped boundary conditions (see Figure \ref{fig: Factorized surfaces}). This provides an intuitive understanding of the non-invertible symmetries in the 2+1d $\mathbb{Z}_2$ gauge theory.

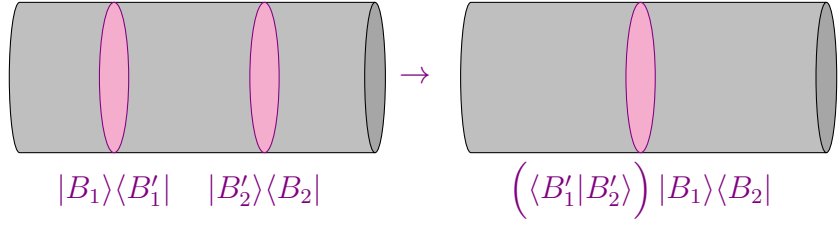
\begin{figure}
	\centering
	\begin{tikzpicture}[scale = 1]
		\node[cylinder, 
		draw = black, 
		text = purple,
		cylinder uses custom fill, 
		cylinder body fill =gray!50,
		cylinder end fill = gray!70, 
		minimum width = 2cm,
		minimum height = 5cm,
		shape border rotate = 0] (c) at (0,0) {}; 
		
		\node[ellipse,
		draw = violet,
		text = violet,
		fill = magenta!40,
		minimum width = 0.2cm, 
		minimum height = 2cm] (e) at (-1,0) {};
		
		\node[color = violet] at (-1,-1.5) {$\ket{B_1}\bra{B'_1}$}; 
		
		\node[ellipse,
		draw = violet,
		text = violet,
		fill = magenta!40,
		minimum width = 0.2cm, 
		minimum height = 2cm] (e) at (1,0) {};
		
		\node[color = violet] at (1,-1.5) {$\ket{B_2'}\bra{B_2}$};
		
		\node[color = violet] at (3,0) {$\rightarrow$}; 
		
		\node[cylinder, 
		draw = black, 
		text = purple,
		cylinder uses custom fill, 
		cylinder body fill =gray!50,
		cylinder end fill = gray!70, 
		minimum width = 2cm,
		minimum height = 5cm,
		shape border rotate = 0] (c) at (6,0) {}; 
		
		\node[ellipse,
		draw = violet,
		text = violet,
		fill = magenta!40,
		minimum width = 0.2cm, 
		minimum height = 2cm] (e) at (6,0) {};
		
		\node[color = violet] at (6,-1.5) {$ \Big( \langle B'_1 | B'_2 \rangle \Big) \, | B_1 \rangle \langle B_2 | $};
		
	\end{tikzpicture}
	\caption{The fusion of two surfaces can be understood from the fusion of the corresponding boundary conditions.}
	\label{fig: Fusion from boundary conditions}
\end{figure}

We start with a brief review of the topological/gapped boundary conditions of the 2+1d $\mathbb{Z}_2$ gauge theory. 
There are two gapped boundary conditions of the $\bZ_2$ gauge theory,  given by Dirichlet boundary conditions for one of the two gauge fields whose Wilson lines are $\eta_\ee$ and $\eta_\mm$  \cite{Bravyi:1998sy} (see also \cite{Kapustin:2010hk,Kitaev:2011dxc,2013PhRvX...3b1009L,Barkeshli:2013jaa}). 
In  condensed matter physics, the two topological boundaries correspond to condensing one of these two anyon lines. 
We denote these two boundary conditions, respectively, by $\ket{B_\ee}$ and $\ket{B_\mm}$. 
Because of the Dirichlet boundary conditions they satisfy\footnote{We slightly abuse the notation and use $\ket{B}$ for both the abstract boundary condition and also the boundary state on $\Sigma$, i.e.\, $\ket{B}\in \mathcal{H}(\Sigma)$.}
\ie
	\eta_\ee(\gamma) \ket{B_\ee} = \ket{B_\ee}\,, \qquad \eta_\mm(\gamma) \ket{B_\mm} = \ket{B_\mm}\,.
\fe
We claim that the non-invertible defects can be factorized into boundaries as
\begin{equation} \label{factorization}
	S_\ee = \ket{B_\ee}\bra{B_\ee}\,, \quad S_\mm = \ket{B_\mm}\bra{B_\mm}\,, \quad S_{\ee\mm} = \ket{B_\ee}\bra{B_\mm}\,, \quad S_{\mm\ee} = \ket{B_\mm}\bra{B_\ee}\,. 
\end{equation}
Before showing the factorization \eqref{factorization}, we note that the fusion rules of the non-invertible surfaces in \eqref{z2.gauge.theory.fusion} immediately follow from the fact that they are factorized gapped boundaries. 
This is because the fusion rule of general factorized defects is given by (see Figure \ref{fig: Fusion from boundary conditions})
\begin{equation} \label{fusion.of.factorized.defects}
	| B_1 \rangle \langle B'_1 | \times | B'_2 \rangle \langle B_2 | = \Big( \langle B'_1 | B'_2 \rangle \Big) \, | B_1 \rangle \langle B_2 | \,,
\end{equation}
where $\langle B | B' \rangle$ denotes the 1+1d TQFT obtained by reducing the bulk 2+1d TQFT on an interval with boundary conditions $B$ and $B'$. 
In the case of the 2+1d $\mathbb{Z}_2$ gauge theory, we have
\begin{equation}
	\langle B_\mm | B_\mm \rangle = \langle B_\ee | B_\ee \rangle = (\mathcal{Z}_2) \,, \qquad \langle B_\ee | B_\mm \rangle = \langle B_\mm | B_\ee \rangle = 1\,,
\end{equation}
where $(\mathcal{Z}_2)$ stands for the 1+1d $\mathbb{Z}_2$ gauge theory.

\begin{figure}
	\centering
\begin{tikzpicture}
\draw[smooth,fill = gray!50, opacity =0.2] (5.5,-0.85) to[out=180,in=30] (5,-1) to[out=210,in=-30] (3,-1) to[out=150,in=30] (2,-1) to[out=210,in=-30] (0,-1) to[out=150,in=-150] (0,1) to[out=30,in=150] (2,1) to[out=-30,in=210] (3,1) to[out=30,in=150] (5,1) to[out=-30,in=180] (5.5,0.85);
\draw[smooth] (0.4,0.1) .. controls (0.8,-0.25) and (1.2,-0.25) .. (1.6,0.1);
\draw[smooth,fill = white] (0.5,0) .. controls (0.8,0.2) and (1.2,0.2) .. (1.5,0);
\draw[smooth,fill = white] (0.52,0) .. controls (0.8,-0.25) and (1.2,-0.25) .. (1.48,0);
\draw[smooth] (3.4,0.1) .. controls (3.8,-0.25) and (4.2,-0.25) .. (4.6,0.1);
\draw[smooth,fill = white] (3.5,0) .. controls (3.8,0.2) and (4.2,0.2) .. (4.5,0);
\draw[smooth,fill = white] (3.52,0) .. controls (3.8,-0.25) and (4.2,-0.25) .. (4.48,0);
\node at (6.5,0) {$. \; \; . \; \; .$};
\node at (0.1,1.5) {\large $Y$};
\draw[smooth,fill = gray!50, opacity =0.2] (7.5,0.85) to[out=0,in=210] (8,1) to[out=30,in=150] (10,1) to[out=-30,in=210] (11,1) to[out=30,in=150] (13,1) to[out=-30,in=30] (13,-1) to[out=210,in=-30] (11,-1) to[out=150,in=30] (10,-1) to[out=210,in=-30] (8,-1) to[out=150,in=0] (7.5,-0.85);
\draw[smooth] (8.4,0.1) .. controls (8.8,-0.25) and (9.2,-0.25) .. (9.6,0.1);
\draw[smooth,fill = white] (8.5,0) .. controls (8.8,0.2) and (9.2,0.2) .. (9.5,0);
\draw[smooth,fill = white] (8.52,0) .. controls (8.8,-0.25) and (9.2,-0.25) .. (9.48,0);
\draw[smooth] (11.4,0.1) .. controls (11.8,-0.25) and (12.2,-0.25) .. (12.6,0.1);
\draw[smooth,fill = white] (11.5,0) .. controls (11.8,0.2) and (12.2,0.2) .. (12.5,0);
\draw[smooth,fill = white] (11.52,0) .. controls (11.8,-0.25) and (12.2,-0.25) .. (12.48,0);
\draw[color =gray!80,fill = gray!50] (5.5,-0.85) arc(270:85:0.3 and 0.85);
\draw[color =gray!80,fill = gray!50] (5.5,-0.85) arc(270:450:0.3 and 0.85);
\draw[color =gray!80,fill = gray!50] (7.5,-0.85) arc(270:85:0.3 and 0.85);
\draw[color =gray!80,dashed,fill = gray!50] (7.5,-0.85) arc(270:450:0.3 and 0.85);

\node[ellipse,
		draw = red,
		text = violet,
		minimum width =2 cm, 
		minimum height = 1cm] (e) at (1,0) {};
		
\draw[color = blue] (1,-1.3) arc(270:85:0.3 and 0.55);
\draw[dashed,color = blue] (1,-1.3) arc(270:450:0.3 and 0.55);
\node[color = red] at (2,0.5) {$\a_1$};
\node[color = blue] at (0.3,-0.8) {$\b_1$};

\node[ellipse,
		draw = red,
		text = violet,
		minimum width =2 cm, 
		minimum height = 1cm] (e) at (4,0) {};
		
\draw[color = blue] (4,-1.3) arc(270:85:0.3 and 0.55);
\draw[dashed,color = blue] (4,-1.3) arc(270:450:0.3 and 0.55);
\node[color = red] at (5,0.5) {$\a_2$};
\node[color = blue] at (3.3,-0.8) {$\b_2$};

\node[ellipse,
		draw = red,
		text = violet,
		minimum width =2 cm, 
		minimum height = 1cm] (e) at (9,0) {};
		
\draw[color = blue] (9,-1.3) arc(270:85:0.3 and 0.55);
\draw[dashed,color = blue] (9,-1.3) arc(270:450:0.3 and 0.55);
\node[color = red] at (10,0.5) {$\a_{g-1}$};
\node[color = blue] at (8.3,-0.8) {$\b_{g-1}$};

\node[ellipse,
		draw = red,
		text = violet,
		minimum width =2 cm, 
		minimum height = 1cm] (e) at (12,0) {};
		
\draw[color = blue] (12,-1.3) arc(270:85:0.3 and 0.55);
\draw[dashed,color = blue] (12,-1.3) arc(270:450:0.3 and 0.55);
\node[color = red] at (13,0.5) {$\a_{g}$};
\node[color = blue] at (11.3,-0.8) {$\b_{g}$};
\end{tikzpicture}
	\caption{The manifold $Y$ depicted above is a three-dimensional genus-$g$ handlebody such that $\partial Y = \Sigma$. The 1-cycles $\a_{i=1,\dots,g}$ and $\b_{i=1,\dots,g}$ are on the boundary and generate $H_1(\Sigma,\bZ)$. The $\a$-cycles are non-intersecting 1-cycles of $\Sigma$ generating $H_1(Y,\bZ_2)$ while the $\b$-cycles are contractible inside $Y$.}
	\label{fig:handlebody}
\end{figure}
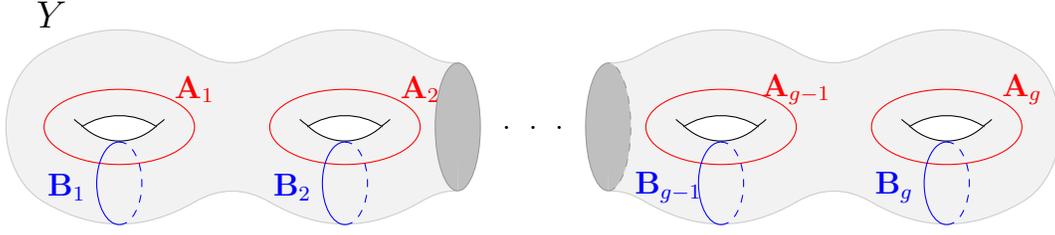

To show the factorization of defect operators, we need the boundary states on arbitrary two-dimensional surfaces $\Sigma$. We first specify a basis for the Hilbert space $\mathcal{H}(\Sigma)$ on $\Sigma$. Let $\Sigma$ be a genus-$g$ surface and choose $Y$ to be a three-dimensional genus-$g$ handlebody such that $\partial Y = \Sigma$ (see Figure \ref{fig:handlebody}). A basis for the Hilbert space on $\Sigma$ is prepared by the path integral on $Y$ with the insertion of different anyons of $\bZ_2$ gauge theory inside $Y$ \cite{turaev2016quantum}.

Different configurations of anyons inside $Y$ are labeled by 1-cycles inside $H_1(Y, \bZ_2^\ee \times \bZ_2^\mm)$. However this assignment is not canonical because of the non-trivial linking phase between $\eta_\ee$ and $\eta_\mm$ (i.e., the mixed anomaly between $\bZ_2^\ee$ and $\bZ_2^\mm$) which is not captured by the homology class of the curves. To address this issue we choose representative loops that link trivially with each other inside $Y$. We can further push such loops to the boundary of $Y$ to get a maximal set of non-intersecting 1-cycles on $\Sigma$. We refer to such 1-cycles on $\Sigma$ as ``$\a$-cycles" and denote the subgroup generated by them as $\Lambda_\a(\Sigma)$. Therefore, we get a decomposition of 1-cycles on $\Sigma$ into $\a$-cycles and $\b$-cycles. 
The $\a$-cycles are a maximal set of non-intersecting cycles generating $H_1(Y, \bZ_2)$, and $\b$-cycles, denoted by $\Lambda_\b(\Sigma)$, are those that are contractible inside $Y$ (see Figure \ref{fig:handlebody}) \footnote{$\Lambda_\a(\Sigma)$ and $\Lambda_\b(\Sigma)$ are maximal isotropic subgroups of $H_1(\Sigma,\bZ_2)$ that are called Lagrangian subgroups in the literature.}. Thus we get the decomposition
\begin{equation} \label{polarization}
	H_1(\Sigma,\bZ_2) = \Lambda_\a(\Sigma) \oplus \Lambda_\b(\Sigma) \,.
\end{equation}
Such a decomposition is known as a choice of polarization in the literature. Thus, for any 1-cycle $\gamma$ in $\Sigma$ we have the unique decomposition
\begin{equation} \label{decomposition.cycles}
	\gamma = \gamma_\a + \gamma_\b\,,
\end{equation}
for $\gamma_\a \in \Lambda_\a(\Sigma)$ and $\gamma_\b \in \Lambda_\b(\Sigma)$.

We denote the state prepared by inserting $\eta_\ee$ along $\alpha_\ee \in\Lambda_\a(\Sigma)$ and $\eta_\mm$ along $\alpha_\mm \in\Lambda_\a(\Sigma)$ by
\begin{equation} \label{hilbert.space.basis}
	|\alpha_\ee,\alpha_\mm\rangle \equiv \eta_\ee(\alpha_\ee) \eta_\mm(\alpha_\mm) |0,0\rangle \in \mathcal{H}(\Sigma) \,,
\end{equation}
where $|0,0\rangle$ is the state prepared by the path integral on the empty handlebody $Y$. Notice since $\alpha_\ee$ and $\alpha_\mm$ are non-intersecting cycles the order of $\eta_\ee(\alpha_\ee) \eta_\mm(\alpha_\mm)$ does not matter and can be reversed. The inner product of any two such states is given by the path integral on the three-manifold given by gluing $Y$ with its orientation reversal and with the insertion of anyons $\eta_\ee(\alpha_\ee-\alpha'_\ee)$ and $\eta_\mm(\alpha_\mm-\alpha'_\mm)$. Thus, we find $\langle \alpha'_\ee,\alpha'_\mm |\alpha_\ee,\alpha_\mm\rangle =2^{g-1} \, \delta_{\alpha_\ee,\alpha'_\ee} \delta_{\alpha_\mm,\alpha'_\mm}$,\footnote{The insertion of anyons gives a delta function, and the normalization is given by the partition function on $\overline{Y} \sqcup_\Sigma Y = (S^2 \times S^1)^{\# g}$ where $\#$ denotes connected sum. Note that $Z\left( (S^2 \times S^1)^{\# g} \right) = { Z( S^2 \times S^1 )^g \over Z( S^3)^{g-1}} = 2^{g-1}$, since $Z(S^3) = 1/2$.}
where $g$ is the genus of $\Sigma$. Since $2^{g-1}$ is just an Euler counterterm and is not physical, we can safely do the redefinition $|\alpha_\ee,\alpha_\mm\rangle \mapsto 2^{(1-g)/2}|\alpha_\ee,\alpha_\mm\rangle$ to make the basis \eqref{hilbert.space.basis} orthonormal.

Note that the $\ket{B_\ee}$ ($\ket{B_\mm}$) boundary condition on $\Sigma$ is given by gauging the $\bZ_2^\ee$ ($\bZ_2^\mm$) 1-form symmetry inside the handlebody $Y$. Since the gauging is done by summing over the insertion of 1-form symmetry lines inside $Y$~(see \cite{Kaidi:2021gbs, Choi:2021kmx}), we get
\begin{equation}
	\ket{B_\ee} = \sum_{\alpha_\ee \in\Lambda_\a(\Sigma)} |\alpha_\ee,0\rangle \,, \qquad \ket{B_\mm} = \sum_{\alpha_\mm \in\Lambda_\a(\Sigma)} |0,\alpha_\mm\rangle \,.
\end{equation}
Now to establish \eqref{factorization}, we need to write the line operators $\eta_\ee(\gamma)$ and $\eta_\ee(\gamma)$ in terms of the operators $ |\alpha'_\ee,\alpha'_\mm\rangle\langle \alpha_\ee,\alpha_\mm |$. By using the decomposing $\gamma=\gamma_\a+\gamma_\b$ (see equation \eqref{decomposition.cycles}) we find
\ie
	\eta_\ee(\gamma) |\alpha_\ee,\alpha_\mm\rangle &= (-1)^{\langle \gamma_\b, \alpha_\mm \rangle} |\alpha_\ee + \gamma_\a ,\alpha_\mm\rangle \,,\\
	\eta_\mm(\gamma) |\alpha_\ee,\alpha_\mm\rangle &= (-1)^{\langle \gamma_\b, \alpha_\ee \rangle} |\alpha_\ee ,\alpha_\mm + \gamma_\a\rangle \,,\\
	\eta_\psi(\gamma) |\alpha_\ee,\alpha_\mm\rangle &= (-1)^{\langle \gamma_\b, \alpha_\ee+\alpha_\mm+ \gamma_\a \rangle} |\alpha_\ee + \gamma_\a ,\alpha_\mm + \gamma_\a\rangle \,.
\fe
Therefore, we arrive at
\ie
	\eta_\ee(\gamma) &= \sum_{\alpha_\ee,\alpha_\mm \in\Lambda_\a(\Sigma)} (-1)^{\langle \gamma_\b, \alpha_\mm \rangle} |\alpha_\ee +\gamma_\a ,\alpha_\mm\rangle\langle \alpha_\ee,\alpha_\mm | \,, \\
	\eta_\mm(\gamma) &= \sum_{\alpha_\ee,\alpha_\mm \in\Lambda_\a(\Sigma)} (-1)^{\langle \gamma_\b, \alpha_\ee \rangle} |\alpha_\ee ,\alpha_\mm +\gamma_\a\rangle\langle \alpha_\ee,\alpha_\mm | \,.
\fe
Plugging these equations into \eqref{toric.code.surface} we verify the factorization \eqref{factorization}.

\subsubsection{Higher quantum symmetry lines on the surfaces}

Note that each of the two boundary conditions host a single $\bZ_2$ topological line that can be obtained by fusing the bulk lines with those boundaries. We denote them by\footnote{Note that the composite boundary states $\ket{B_\ee, \gamma}$ for different $\gamma$ span a basis for the Hilbert space on $\Sigma$ (and similarly for $\ket{B_\mm, \gamma}$). The basis $|\alpha_\ee,\alpha_\mm\rangle$ correspond to choosing a Lagrangian subgroup of $H_1(\Sigma,\bZ)$ (i.e.\ the $\mathbf{A}$-cycles) whereas the basis $\ket{B_\ee, \gamma}$ correspond to the Lagrangian subgroup of anyons generated by $\eta_\ee$.}
\ie \label{composite.boundaries}
	\ket{B_\ee, \gamma} &\equiv \eta_\mm(\gamma) \ket{B_\ee} = \sum_{\alpha_\ee \in\Lambda_\a(\Sigma)} (-1)^{ \langle \gamma_\b, \alpha_\ee \rangle} |\alpha_\ee ,  \gamma_\a\rangle \, ,\\
	\ket{B_\mm, \gamma} &\equiv \eta_\ee(\gamma) \ket{B_\mm} = \sum_{\alpha_\mm \in\Lambda_\a(\Sigma)} (-1)^{ \langle \gamma_\b, \alpha_\mm \rangle} | \gamma_\a, \alpha_\mm\rangle \,.
\fe
We find the fusion of bulk lines with boundaries is
\ie \label{fusion.with.boundaries}
	\eta_\mm \times \ket{B_\ee} &= \eta_\psi \times \ket{B_\ee} = \ket{B_\ee, \gamma}\,, \qquad &\eta_\ee \times \ket{B_\ee} &= \ket{B_\ee} \,,\\
	\eta_\ee \times \ket{B_\mm} &= \eta_\psi \times \ket{B_\mm} = \ket{B_\mm, \gamma} \,, \qquad &\eta_\mm \times \ket{B_\mm} &= \ket{B_\mm} \,.
\fe
In condensed matter physics, the two equations on the right are interpreted as  $\eta_\ee$ ``condensing" on $B_\ee$ and $\eta_\mm$ ``condensing" on $B_\mm$, respectively.

By fusing the composite boundaries \eqref{composite.boundaries} with their orientation reversal we get all the higher quantum symmetry lines on the non-invertible defects. Furthermore, from \eqref{hat.eta.tilde.eta}, there are four higher quantum symmetry lines on $S_\psi$ given by:
\ie
	\hat{\eta}_\psi(\gamma,\Sigma) &\equiv \frac{1}{\sqrt{|H_1(\Sigma,\mathbb{Z}_2)|}} \sum_{\gamma' \in H_1(\Sigma,\mathbb{Z}_2)} (-1)^{ \langle \gamma , \gamma' \rangle} \eta_\psi(\gamma')  \,, \\
	\tilde{\eta}_\psi(\gamma,\Sigma) &\equiv \frac{1}{\sqrt{|H_1(\Sigma,\mathbb{Z}_2)|}} \sum_{\gamma' \in H_1(\Sigma,\mathbb{Z}_2)} \eta_\ee(\gamma'+\gamma) \eta_\mm(\gamma')  \,, \\
	\hat{\eta}_\psi\tilde{\eta}_\psi(\gamma,\Sigma) &\equiv \frac{1}{\sqrt{|H_1(\Sigma,\mathbb{Z}_2)|}} \sum_{\gamma' \in H_1(\Sigma,\mathbb{Z}_2)} (-1)^{ \langle \gamma , \gamma' \rangle} \eta_\ee(\gamma'+\gamma) \eta_\mm(\gamma')  \,. \\
\fe
with the fusion rules:
\ie \label{fusion.with.Spsi}
	\hat{\eta}_\psi = \eta_\psi \times S_\psi = S_\psi \times \eta_\psi \,,\qquad \tilde{\eta}_\psi = \eta_\ee \times S_\psi = S_\psi \times \eta_\mm \,.
\fe

To find the rest of the fusion rules we need the fusions between $\ket{B_\ee, \gamma}$, $\ket{B_\mm, \gamma}$ and $S_\psi$. We note that
\begin{equation}
	S_\psi(\Sigma) |\alpha_\ee,\alpha_\mm\rangle = \sum_{\substack{ \alpha\in \Lambda_\a(\Sigma) \\ \beta \in \Lambda_\b(\Sigma) }} {(-1)^{\langle \beta, \alpha_\ee+\alpha_\mm+ \alpha \rangle} \over \sqrt{|H_1(\Sigma,\mathbb{Z}_2)|} } |\alpha_\ee + \alpha ,\alpha_\mm + \alpha\rangle = |\alpha_\mm,\alpha_\ee\rangle \,.
\end{equation}
From this we find the fusion of $S_\psi$ with boundaries as
\ie \label{action.of.z2.exchange}
S_\psi \times \ket{B_\ee} &= \ket{B_\mm}  \,, &\quad S_\psi \times \ket{B_\ee, \gamma} &= \ket{B_\mm, \gamma} \,, \\
S_\psi \times \ket{B_\mm} &= \ket{B_\ee} \,, &\quad S_\psi \times \ket{B_\mm, \gamma} &= \ket{B_\ee, \gamma}\,.
\fe
The remaining fusion rules that we need are
\begin{equation} \label{z2.reducing.on.interval}
	\langle B_\mm, \gamma| B_\mm \rangle = \langle B_\ee,\gamma | B_\ee \rangle = (\mathcal{Z}_2, W) \,, \qquad \langle B_\ee,\gamma | B_\mm \rangle = \langle B_\mm,\gamma | B_\ee \rangle = 1\,.
\end{equation}
Recall that $(\mathcal{Z}_2, W)$ denoted the 1+1d $\bZ_2$ gauge theory with the insertion of a Wilson line.

In total we find 24 defects since there are six surfaces (including the trivial surface) and four higher quantum symmetry lines on each of them. They are:
\begin{equation}
\begin{array} {lllllll}
	1 \,, &~~~~& \eta_\ee \,, &~~~~& \eta_\mm \,, &~~~~& \eta_\psi \,,\\
	S_\psi \,, &~~~~& \hat{\eta}_\psi \,, &~~~~& \tilde{\eta}_\psi \,, &~~~~& \hat{\eta}_\psi\tilde{\eta}_\psi \,,\\
	\ket{B_\ee}\bra{B_\ee}\,, &~~~~& \ket{B_\ee, \gamma}\bra{B_\ee}\,, &~~~~& \ket{B_\ee}\bra{B_\ee, \gamma}\,, &~~~~& \ket{B_\ee, \gamma}\bra{B_\ee, \gamma}\,, \\
	\ket{B_\mm}\bra{B_\mm}\,, &~~~~& \ket{B_\mm, \gamma}\bra{B_\mm}\,, &~~~~& \ket{B_\mm}\bra{B_\mm, \gamma}\,, &~~~~& \ket{B_\mm, \gamma}\bra{B_\mm, \gamma}\,, \\
	\ket{B_\ee}\bra{B_\mm}\,, &~~~~& \ket{B_\ee, \gamma}\bra{B_\mm}\,, &~~~~& \ket{B_\ee}\bra{B_\mm, \gamma}\,, &~~~~& \ket{B_\ee, \gamma}\bra{B_\mm, \gamma}\,, \\
	\ket{B_\mm}\bra{B_\ee}\,, &~~~~& \ket{B_\mm, \gamma}\bra{B_\ee}\,, &~~~~& \ket{B_\mm}\bra{B_\ee, \gamma}\,, &~~~~& \ket{B_\mm, \gamma}\bra{B_\ee, \gamma}\,.
\end{array}
\label{z2 gauge theory fusion}
\end{equation}
The fusion of lines with surfaces can be obtained from \eqref{fusion.with.boundaries} and \eqref{fusion.with.Spsi}. To find the rest of the fusions we need the fusion between $\ket{B_\ee, \gamma}$, $\ket{B_\mm, \gamma}$ and $S_\psi$. These are given in equations \eqref{action.of.z2.exchange} and \eqref{z2.reducing.on.interval}.

\subsubsection{Non-invertible fusion rules from Lagrangians}

Having discussed the general fusion rules of the surfaces in the 2+1d $\mathbb{Z}_2$ gauge theory, here we give the Lagrangian description of the non-invertible surface $S_\ee$, which is  the Cheshire string in the 2+1d version of the models of \cite{Else:2017yqj,Johnson-Freyd:2020twl}, and derive its non-invertible fusion rule.

The 2+1d $\mathbb{Z}_2$ gauge theory can be realized as a BF action using two $U(1)$ gauge fields $A,\tilde A$ \cite{Maldacena:2001ss,Banks:2010zn,Kapustin:2014gua}:
\ie
{2i\over 2\pi} \int Ad\tilde A\,.
\fe
The gauge transformations are $A\sim A+d\alpha,\, \tilde A\sim \tilde A + d\tilde \alpha$.  
The 1-form topological symmetry lines, i.e., the anyons, are the Wilson lines of these gauge fields, $\eta_\ee= \exp(i\oint A),\eta_\mm =\exp(i\oint \tilde A)$. 

We focus on the non-invertible surface $S_\ee = |B_\ee \rangle \langle B_\ee|$, which corresponds to the higher gauging of $\eta_\ee$ on a surface. 
Under the folding trick, it maps to the topological  boundary in the  2+1d $\mathbb{Z}_2\times \mathbb{Z}_2$ gauge theory where both copies of the e anyons condense. 
The bulk  Lagrangian of the  2+1d $\mathbb{Z}_2\times \mathbb{Z}_2$ gauge theory is
\ie
{2i\over 2\pi}\int A_1 d\tilde A_1- {2i\over 2\pi}\int  A_2 d\tilde A_2\,.
\fe
The sign is chosen for later convenience when we apply the folding trick. (It can be removed by a field redefinition of, say, $A_2$.)

In the 2+1d $\mathbb{Z}_2\times \mathbb{Z}_2$ gauge theory, the topological boundary where both e anyons condense correspond to the Dirichlet boundary condition  for $A_1,A_2$:\footnote{SHS thanks Pranay Gorantla, Ho Tat Lam, and Nati Seiberg for discussions on the Lagrangians of boundary conditions in discrete gauge theory.}
\ie
A_1 \Big|_{x=0} = A_2 \Big|_{x=0}=0 \,.
\fe
Following \cite{Kapustin:2014gua,Gaiotto:2014kfa}, we introduce a Stueckelberg field $\Phi$ to write the action for this boundary condition:
\ie
&{2i\over 2\pi} \int_{x>0}  A_1 d\tilde A_1- {2i\over 2\pi}\int_{x>0}  A_2 d\tilde A_2
- {2i\over 2\pi}\int_{x=0} \Phi d(\tilde A_1-\tilde A_2)
\,.\\
\fe
The equation of motion of $\tilde A_{1,2}$ implies  $ A\equiv  A_1 \Big|_{x=0} =  A_2 \Big|_{x=0} =d\Phi$. 
The gauge symmetries are $A_i \sim A_i +d\alpha_i\,,\tilde A_i \sim \tilde A_i +d\tilde \alpha_i\,,\Phi\sim \Phi+\alpha$, 
where $\alpha$ is the restriction of $\alpha_1 |_{x=0} = \alpha_2|_{x=0}$ on the boundary.

After we unfold, the above boundary condition becomes the topological surface defect $S_\ee$ in the 2+1d $\mathbb{Z}_2$ gauge theory, whose action is
\ie
S_\ee:~~~&{2i\over 2\pi} \int_{x<0}  A_L d\tilde A_L+ {2i\over 2\pi}\int_{x>0}  A_R d\tilde A_R
- {2i\over 2\pi}\int_{x=0} \Phi d(\tilde A_L-\tilde A_R)
\,,\\
&  A_L \Big|_{x=0} =  A_R \Big|_{x=0} =d\Phi\,,
\fe
where $A_L,\tilde A_L$ and $A_R,\tilde A_R$ are the bulk gauge fields on the left and the right of the surface defect at $x=0$, respectively.  
The gauge symmetries are
\ie
&A_L \sim A_L +d\alpha_L\,,~~~A_R\sim A_R+d\alpha_R,\\
&\tilde A_L \sim \tilde A_L +d\tilde \alpha_L\,,~~~\tilde A_R\sim \tilde A_R +d\tilde \alpha_R\,,\\
&\Phi\sim \Phi+\alpha\,,
\fe
where $\alpha$ is the restriction of $\alpha_L |_{x=0} = \alpha_R|_{x=0}$ on the defect.  
Similar to Section \ref{sec:Maxwell}, the defect action at $x=0$ can be dualized to a Higgs action, with the dual scalar field being the Stueckelberg field.

Following similar steps in Section \ref{sec:Maxwell}, we can bring two such parallel surfaces on top of each other to compute their fusion rule. 
The worldsheet action  for the fusion of two such surface defects is given by
\ie
S_\ee \times S_\ee:~~&- {2i\over 2\pi}\int_{x=0} \left[
\Phi_1 d(\tilde A_L-\tilde A_I)
+ \Phi_2 d(\tilde A_I-\tilde A_R)\right]\\
&=  -{2i\over 2\pi}\int_{x=0} 
\left[ -\Phi d\tilde a
+ \Phi_2 (d\tilde A_L - d\tilde A_R)
\right]\, , 
\fe
where $A_I,\tilde A_I$ used to be the bulk gauge fields in the intermediate region, which only live on the surface in the limit when the two surfaces collide. 
In the last step we have defined $\Phi=\Phi_1-\Phi_2$  (which is gauge invariant) and $\tilde a = \tilde A_I - \tilde A_L$. 
The first term is a decoupled 1+1d $\mathbb{Z}_2$ gauge theory, and the second term is another copy of the surface defect $S_\ee$. 
We therefore have reproduced the non-invertible fusion rule of the  surface $S_\ee$ \eqref{z2.gauge.theory.fusion} (see also \eqref{higherprojection}):
\ie
S_\ee \times S_\ee =(\mathcal{Z}_2) \,S_\ee\,.
\fe
This is the fusion rule of the Cheshire string \cite{Else:2017yqj,Johnson-Freyd:2020twl}.

\subsection{$\mathbb{Z}_p$ gauge theory} \label{sec:zpgauge}

We further generalize the discussion in Section \ref{sec:z2gauge} to $\mathbb{Z}_p$ gauge theory with prime $p$.  
The 1-form symmetry is $\mathbb{Z}_p \times \mathbb{Z}_p$ which is 1-gaugeable and parameterized by the 0-anomaly $k_1 = k_2 = 0$ and $k_{12} = 1$. 

The condensation surfaces correspond to the higher gauging  of  subgroups of the 1-gaugeable $\mathbb{Z}_p \times \mathbb{Z}_p$, possibly with discrete torsions.
There are $p+1$ $\mathbb{Z}_p$ subgroups and $p$ choices of discrete torsion for gauging the whole $\mathbb{Z}_p \times \mathbb{Z}_p$. 
Together with the trivial surface, these lead to $1+(p+1)+p = 2p+2$ different surface defects. 
As we will see, four of these surface defects are non-invertible and the rest form the Dihedral group $D_{2(p-1)}$ of order $2(p-1)$.

\bigskip\centerline{\it Higher gauging of $\mathbb{Z}_p$}\bigskip

Given any non-zero element $(m_1,m_2) \in \mathbb{Z}_p \times \mathbb{Z}_p$ associated with the topological line $\eta_1^{m_1} \eta_2^{m_2}$, we can take the $\mathbb{Z}_p$ subgroup generated by it and denote it by $\mathbb{Z}_p^{(m_1/m_2)}$. Note that $\mathbb{Z}_{p}^{(m)}$ only depend on $m=\frac{m_1}{m_2} \pmod{p}$, where $m=\infty$ and $m=0$ correspond to the first and second $\mathbb{Z}_p$ subgroups which are  $0$-gaugeable. The surface defect given by condensing $\mathbb{Z}_{p}^{(m)}$ is denoted as
\begin{equation}
	S_{\mathbb{Z}_{p}^{(m)}}(\Sigma) = \frac{1}{\sqrt{|H_1(\Sigma , \mathbb{Z}_p)|}} \sum_{ \gamma \in H_1(\Sigma,\mathbb{Z}_p) } \eta_1(m_1 \gamma)  \, \eta_2( m_2 \gamma) \,.
\end{equation}
From the results of Section \ref{sec:ZN},  we find that  $S_{\mathbb{Z}_{p}^{(\infty)}}$ and $S_{\mathbb{Z}_{p}^{(0)}}$ are non-invertible and the rest are all invertible order-$2$ defects.

\bigskip\centerline{\it Higher gauging of $\mathbb{Z}_p\times \mathbb{Z}_p$}\bigskip

For any choice of discrete torsion $f \in \mathbb{Z}_p$, there exist a surface defect $S_{\mathbb{Z}_{p} \times \mathbb{Z}_{p}, f}$ given by \eqref{znxzn.condensation}. 
To simplify the fusion rules, we perform the change of variable $f=\frac{1}{m-1}$ and introduce the notation
\begin{equation}
	S_{(\mathbb{Z}_p \times \mathbb{Z}_p)^{(m)}} = \begin{cases}
		1 & m = 1 \pmod{p} \\
		S_{\mathbb{Z}_p \times \mathbb{Z}_p, \frac{1}{m-1}} & m = 2,\dots,p-1 \pmod{p}  \\
		S_{\mathbb{Z}_p \times \mathbb{Z}_p, -1} & m = 0 \\
		S_{\mathbb{Z}_p \times \mathbb{Z}_p, 0} & m = \infty
	\end{cases}
	\,.
\end{equation}

\bigskip\centerline{\it Summary of the fusion rule for the surfaces}\bigskip

The fusion rules of all these surfaces is computed in Appendix \ref{app.zp.gauge.theory}. Here we summarize them. There are $2(p+1)$ defects 
\begin{equation} 
	S_{\mathbb{Z}_{p}^{(m)}} \quad \text{and} \quad S_{(\mathbb{Z}_p \times \mathbb{Z}_p)^{(m)}} ~, \qquad \text{for} \quad m \in \{1,2, \dots, p-1 \} \cup \{0,\infty \} ~, \label{zp.gauge.theory.defects}
\end{equation}
where the identity defect is given by $S_{(\mathbb{Z}_p \times \mathbb{Z}_p)^{(1)}} =1$. More precisely the index $m$ takes value in $(\mathbb{Z}_p)^\times \cup \{0,\infty\}$ where $(\mathbb{Z}_p)^\times$ is the multiplicative group of integers modulo $p$. The orientation-reversals of these defects are given by
\begin{equation}
	\overline{S_{\mathbb{Z}_{p}^{(m)}}} = S_{\mathbb{Z}_{p}^{(m)}} \qquad \text{and} \qquad \overline{S_{(\mathbb{Z}_p \times \mathbb{Z}_p)^{(m)}}}= S_{(\mathbb{Z}_p \times \mathbb{Z}_p)^{(1/m)}}~,
\end{equation}
where
\begin{equation}
	\frac{1}{m} = \begin{cases}
		m^{-1} \pmod{p} & m \in (\mathbb{Z}_p)^\times \\
		\infty & m=0 \\
		0 & m = \infty
	\end{cases}~.
\end{equation}
The fusion rules are
\ie\label{zpgaugefusion}
S_{\mathbb{Z}_{p}^{(m)}} \times	S_{\mathbb{Z}_{p}^{(m')}} &= \begin{cases}
	S_{(\mathbb{Z}_p \times \mathbb{Z}_p)^{(m/m')}} & m/m' \text{ is well-defined}\\
	\left( \mathcal{Z}_p  \right) S_{\mathbb{Z}_{p}^{(m)}} & m/m' \text{ is ill-defined}
\end{cases} ~, \\
S_{(\mathbb{Z}_p \times \mathbb{Z}_p)^{(m)}} \times S_{(\mathbb{Z}_p \times \mathbb{Z}_p)^{(m')}} &= \begin{cases}
	S_{(\mathbb{Z}_p \times \mathbb{Z}_p)^{(m \times m')}} & m \times m' \text{ is well-defined} \\ 
	\left( \mathcal{Z}_p \right) S_{\mathbb{Z}_{p}^{(m)}} & m \times m' \text{ is ill-defined}
\end{cases} ~, \\
S_{\mathbb{Z}_{p}^{(m)}} \times S_{(\mathbb{Z}_p \times \mathbb{Z}_p)^{(m')}} &= \begin{cases}
	S_{\mathbb{Z}_p^{(m/m')}} &  m/m' \text{ is well-defined} \\
	\left( \mathcal{Z}_p \right) S_{(\mathbb{Z}_p \times \mathbb{Z}_p)^{(m')}} &  m/m' \text{ is ill-defined}
\end{cases} ~,\\
S_{(\mathbb{Z}_p \times \mathbb{Z}_p)^{(m)}} \times S_{\mathbb{Z}_{p}^{(m')}} &= \begin{cases}
	S_{\mathbb{Z}_p^{(m \times m')}} & m \times m' \text{ is well-defined}  \\
	\left( \mathcal{Z}_p \right) S_{(\mathbb{Z}_p \times \mathbb{Z}_p)^{(m)}} & m \times m' \text{ is ill-defined}
\end{cases} ~,
\fe
where
\begin{equation}
	m \times m' = \begin{cases}
		 m m' \pmod{p} & m,m' \in (\mathbb{Z}_p)^\times \\
		 0 & m=0,m'\neq \infty ~\text{ or }~ m\neq \infty,m'=0 \\
		 \infty & m=\infty,m' \neq 0 ~\text{ or }~ m\neq 0,m'=\infty \\
		 \text{ill-defined} & m=0,m'=\infty ~\text{ or }~ m=\infty,m'=0
	\end{cases}~,
\end{equation}
and $m/m' = m \times (1/m')$ for $m,m' \in (\mathbb{Z}_p)^\times \cup \{0,\infty\}$.  
Recall that $(\mathcal{Z}_p)$ stands for the 1+1d $\mathbb{Z}_p$ gauge theory.

The $2(p-1)$ defects labeled by $m \in \{1,2, \dots, p-1 \}$ in \eqref{zp.gauge.theory.defects} are invertible.
The other four condensation defects, $S_{\mathbb{Z}_{p}^{(\infty)}} ,\, S_{\mathbb{Z}_{p}^{(0)}} , \,  S_{(\mathbb{Z}_p \times \mathbb{Z}_p)^{(\infty)}} ,\,  S_{(\mathbb{Z}_p \times \mathbb{Z}_p)^{(0)}} $,  are non-invertible.  
They are related as follows:
 \ie
  S_{(\mathbb{Z}_p \times \mathbb{Z}_p)^{(\infty)}} = 	S_{\mathbb{Z}_{p}^{(\infty)}} \times S_{\mathbb{Z}_{p}^{(0)}} ,~~  S_{(\mathbb{Z}_p \times \mathbb{Z}_p)^{(0)}} = S_{\mathbb{Z}_{p}^{(0)}} \times S_{\mathbb{Z}_{p}^{(\infty)}}\,.
  \fe
Two comments in order:
\paragraph{1. Non-invertible defects:} Every non-invertible defect can be factorized into the product of two topological boundary conditions. $\mathbb{Z}_p$ gauge theory has two topological boundary conditions corresponding to gauging the  $\mathbb{Z}_{p}^{(\infty)}$ and $\mathbb{Z}_{p}^{(0)}$ 1-form symmetry subgroups on half of the spacetime. Let us denote them by $| \infty \rangle$ and $| 0 \rangle$, respectively. 
The four non-invertible defects are factorized into the products of these gapped boundaries  (see Figure \ref{fig: Factorized surfaces}):
\begin{equation}
	| \infty \rangle \langle \infty | = S_{\mathbb{Z}_{p}^{(\infty)}} \,, \quad | 0 \rangle \langle 0 | = S_{\mathbb{Z}_{p}^{(0)}} \,, \quad  | \infty \rangle \langle 0 | = S_{(\mathbb{Z}_p \times \mathbb{Z}_p)^{(\infty)}}\,, \quad  | 0 \rangle \langle \infty | = S_{(\mathbb{Z}_p \times \mathbb{Z}_p)^{(0)}}~.
\end{equation}
We can verify the fusion rules between the non-invertible defects using the general formula \eqref{fusion.of.factorized.defects} (see Figure \ref{fig: Fusion from boundary conditions}) and the fact that
\begin{equation}
	\langle 0 | 0 \rangle = \langle \infty | \infty \rangle  = \left( \mathcal{Z}_p \right) \, , \qquad 	\langle 0 | \infty \rangle = \langle \infty | 0 \rangle = 1  \,.
\end{equation}
Recall  that $\langle B | B' \rangle$ denotes the 1+1d TQFT obtained by reducing the bulk 2+1d TQFT on an interval with boundary conditions $B$ and $B'$.

\paragraph{2. Invertible defects:} 
It is known that the invertible 0-form symmetry of the 2+1d $\mathbb{Z}_p$ gauge theory is the Dihedral group $D_{2(p-1)}$ of order $2(p-1)$ \cite{Delmastro:2019vnj,Fuchs:2014ema}.  
Indeed, from \eqref{zpgaugefusion}, the $2(p-1)$ invertible defects obey the fusion rules of  $D_{2(p-1)}$:
\begin{gather}
	S_{(\mathbb{Z}_p \times \mathbb{Z}_p)^{(m)}} \times S_{(\mathbb{Z}_p \times \mathbb{Z}_p)^{(m')}} = S_{(\mathbb{Z}_p \times \mathbb{Z}_p)^{(mm')}} ~, \qquad (S_{\mathbb{Z}_{p}^{(m)}})^{\times 2}= 1~, \notag\\
	S_{\mathbb{Z}_{p}^{(m)}} \times S_{(\mathbb{Z}_p \times \mathbb{Z}_p)^{(m')}} \times S_{\mathbb{Z}_{p}^{(m)}} = S_{(\mathbb{Z}_p \times \mathbb{Z}_p)^{(1/m')}} ~,
\end{gather}
where the reflections and rotations are respectively given by $S_{\mathbb{Z}_{p}^{(m)}}$ and $S_{(\mathbb{Z}_p \times \mathbb{Z}_p)^{(m)}}$ for $m \in (\mathbb{Z}_p)^\times$.

\subsection{$\mathcal{G}_k\times \mathcal{G}_{-k}$ Chern-Simons theory}

In this subsection we demonstrate examples of fusion rules where the fusion ``coefficients" are more general 1+1d TQFTs beyond the $\mathbb{Z}_n$ gauge theory. 

Let us consider the 2+1d $\mathcal{G}_k\times \mathcal{G}_{-k}$ Chern-Simons theory, where $\mathcal{G}$ is a  simply connected Lie group.  
This 2+1d theory admits a topological boundary condition $|B\rangle$, which after folding, becomes the trivial surface defect of $\mathcal{G}_k$ Chern-Simons theory (see, for example, \cite{Kaidi:2021gbs} for  a recent discussion). 

Using the topological boundary $|B\rangle$, we can construct a topological surface defect in the $\mathcal{G}_k\times \mathcal{G}_{-k}$ Chern-Simons theory that factorizes:
\ie
S \equiv |B\rangle \langle B|\,.
\fe
The fusion of this surface is (see figure \ref{fig: Fusion from boundary conditions}):
\ie
S\times S = \left( \mathcal{G}_k/\mathcal{G}_k\right) \,S\,,
\fe
where the fusion ``coefficient" is the 1+1d $\mathcal{G}_k/\mathcal{G}_k$ TQFT. The latter is obtained from compactifying the 2+1d $\mathcal{G}_k\times \mathcal{G}_{-k}$ Chern-Simons theory on an interval with the boundary condition $|B\rangle$ imposed on both ends.

\subsection{QED with an even charge $q$ fermion}\label{sec:QED}

Here we discuss condensation surfaces of an interacting 2+1d QFT.\footnote{Throughout the paper, we have assumed the 2+1d QFT to be a bosonic QFT, with the exception of  this subsection. Most of our previous results for the bosonic QFT still apply here.} 
We will consider the 2+1d $U(1)_0$ gauge theory (without any bare Chern-Simons term) coupled to a Dirac fermion $\psi$ of even charge $q \in 2\mathbb{Z}$ with the Lagrangian (in the Lorentzian signature) 
\begin{equation}
	\mathcal{L} = -\frac{1}{4g^2} F_{\mu\nu}F^{\mu\nu} + i\overline{\psi} (\slashed{\partial} +  iq \slashed{A}) \psi + m \overline{\psi} \psi~,
\end{equation} 
where $A$ is the dynamical $U(1)$ gauge field and $F=dA$ its field-strength. We treat this Lagrangian as the UV theory and considers its IR limit in the following.

When the mass is large and positive, the fermion can be integrated out, and we are left with the $U(1)_{q^2/2}$ Chern-Simons theory in the IR. On the other hand, when the mass is large and negative, the low energy phase is the $U(1)_{-q^2/2}$ Chern-Simons theory. See \cite{Cordova:2017kue} for  a more complete discussion of the proposed phase diagram.

The UV theory has a charge conjugation 0-form symmetry $C^{UV}$ that acts as $A\to -A$ and $\psi\to \psi^\dagger$. 
It acts on the gauge-invariant local and line operators as follows:
\ie\label{CUV}
C^{UV}:
~~F\to -F\,,~~~W^{UV}_n \to W^{UV}_{-n}\,,~~~\eta\to \eta^{-1}\,.
\fe
Here $W^{UV}_n\equiv \exp( in\oint A)$ is  the non-topological Wilson line of charge $n$. 

In addition, the UV theory has a $\mathbb{Z}_{q}$ 1-form symmetry, generated by the topological line $\eta$ with $\eta^{q}=1$. 
The 0-anomaly of the $\mathbb{Z}_q$ 1-form symmetry is  $k=q/4$ \cite{Cordova:2017kue}, which can be obtained by matching with the IR $U(1)$ Chern-Simons theory.  
We now discuss the condensation defects from this UV 1-form symmetry. 

When $q=2$, the UV $\mathbb{Z}_2$ 1-form symmetry is not 1-gaugeable because $k$ is not an integer, and we do not have any condensation defect in the UV. 

When $q=4$, the higher gauging of the  $\mathbb{Z}_4$ 1-form symmetry gives an invertible $\mathbb{Z}_2$ 0-form symmetry $S^{UV}$ in the UV that acts as (using \eqref{zn.action.on.lines} with $N=4, k=1$):
\ie\label{SUV}
S^{UV}:~~F\to F\,,~~~W^{UV}_n \to \eta^{-n} W^{UV}_{n}\,,~~~\eta\to \eta^{-1}\,.
\fe
Note that the $\mathbb{Z}_4$ 1-form symmetry charges of $W_n^{UV}$ and $\eta$ are $n$ and 2, respectively.  
We conclude that the $q=4$ UV theory has \textit{two} $\mathbb{Z}_2$ 0-form symmetries, $C^{UV}$ and $S^{UV}$.\footnote{In the UV, there are additional global symmetries including the magnetic $U(1)$ symmetry and the time-reversal symmetry. We will not discuss them here.} 
Let us compare them:
\begin{itemize}
\item The charge conjugation symmetry $C^{UV}$ acts nontrivially on the local operators, whereas the $\mathbb{Z}_2$ condensation defect $S^{UV}$ acts trivially on all  local operators because it is made out of lines.
\item  The actions of $C^{UV}$ and $S^{UV}$ on the Wilson lines are different,  i.e., $W_{-n}^{UV}$ and $\eta^{-n}W^{UV}_n$, but the latter have the same $\mathbb{Z}_4$ 1-form symmetry charge. 
\end{itemize}

For $q=4$, let us understand what happen to these two UV $\mathbb{Z}_2$ 0-form symmetries  as we turn on a large and positive mass $m$ and flow to the IR $U(1)_{8}$ Chern-Simons theory. 
Let $W^{IR} =\exp(i\oint A^{IR})$ be the minimal topological Wilson line in the IR, which obeys
\ie
(W^{IR})^8=1\,.
\fe  
The topological spin of this line is
\ie
\theta(W^{IR})  =\exp\left(  {2\pi i /16}\right) \,.
\fe

As discussed in Section \ref{sec:U1CS}, the IR 1-form symmetry $\mathbb{Z}_8$ is not gaugeable. 
However, its $\mathbb{Z}_4$ subgroup is 1-gaugeable with anomaly $k=1$. 
It gives rise to the condensation defect $C^{IR}$ (which was denoted as $S_4$ in Section \ref{sec:U1CS}) that generates the IR charge conjugation symmetry (see \eqref{SnactionCS}):
\ie\label{CIR}
C^{IR} :~~ W^{IR}\to (W^{IR})^{-1}\,.
\fe

The line operators in the UV and in the IR are related as follows. 
The non-topological  UV Wilson line $W_1^{UV}$ flows to the topological line $W^{IR}$ in the IR Chern-Simons theory. 
On the other hand, the UV 1-form symmetry line $\eta$ becomes $\eta = (W^{IR})^{2}$. 
Comparing  \eqref{CUV}, \eqref{SUV} with the IR charge conjugation \eqref{CIR}, we conclude that the two UV $\mathbb{Z}_2$ 0-form symmetries $C^{UV}, S^{UV}$ flow to the same IR charge conjugation symmetry $C^{IR}$.

\section{All 0-form symmetries in TQFT from higher gauging}\label{sec:symTQFT}

One salient feature of condensation surface defects is that they act trivially on all local operators.  
This is because they are made out of lines, and are therefore ``porous" to the local operators. 
As discussed earlier, the simplest example is the charge conjugation symmetry in Chern-Simons theory. 

It is natural to ask the converse question: Do all (invertible and non-invertible) 0-form symmetries that act trivially on local operators arise from higher gauging? 
  In the context of 2+1d TQFT with no local operator, we provide a positive answer to this question.   
In fact, all 0-form symmetries in a 2+1d TQFT act trivially on local operators, since there is not any.

Importantly, for this statement to be true, we need to extend the definition of the higher gauging to including gauging non-invertible 1-form symmetries, i.e., non-abelian anyons, on surfaces.

\subsection{Higher gauging of non-invertible 1-form symmetries}

We start by reviewing the 0-gauging of non-invertible 1-form symmetries in a 2+1d TQFT $\cal T$, which is described by the mathematical framework of unitary modular tensor category (UMTC)  $\cal C$ plus a chiral central charge. 
This is known as the non-abelian anyon condensation in the condensed matter physics literature \cite{Bais:2008ni,2014NuPhB.886..436K,Burnell:2017otf}. 
It was developed in the mathematical literature in \cite{Kirillov:2001ti,2010arXiv1009.2117D} (see also \cite{Carqueville:2018sld}).  
More recently, it has also been discussed in \cite{Kaidi:2021gbs,Buican:2021axn,Yu:2021zmu,Benini:2022hzx} from the point of view of generalized global symmetries.  

Our review below can be viewed as  a reinterpretation of the results in \cite{Fuchs:2002cm,Kapustin:2010if}  in terms of generalized global symmetries. 
We will be brief about the detailed mathematical definitions, focusing more on the translation to the field theory language. The interested readers are referred to the  references above for a more detailed discussion.

The 0-gauging of a non-invertible 1-form symmetry in the bulk is given by a \textit{commutative algebra object}  $\cal A$ \cite{Kirillov:2001ti,2010arXiv1009.2117D}.\footnote{More precisely, the algebra  needs to be  both commutative and separable with a unique unit. Such an algebra is the same as the \textit{connected $\acute{e}$tale algebra} of \cite[Definition 3.1]{2010arXiv1009.2117D} and \textit{rigid algebra with trivial spin} of \cite[Section 3]{Kirillov:2001ti}. Here we refer to them simply as commutative algebra objects.}
A commutative algebra object generalizes the notion of a non-anomalous (0-gaugeable) 1-form symmetry subgroup in the invertible/abelian case.  
It is a superposition of boson lines $a$ that are gauged, ${\cal A} = \bigoplus_a Z_a \, a$, where $Z_a\in \mathbb{Z}_{\ge 0}$ and $Z_0=1$.
In addition, there is a multiplication junction $\mu \in \text{Hom}({\cal A}\otimes {\cal A},{\cal A})$, which generalizes the notion of the choice of SPT, and the unit $u \in \text{Hom}(1,{\cal A})$. The defining properties of a commutative algebra object are summarized as follows:
\ie \label{commutative.algebra}
\raisebox{-2.2em}{\begin{tikzpicture}
		\draw [thick, decoration = {markings, mark=at position .7 with {\arrow[scale=1.2]{stealth}}}, postaction=decorate] (0,0) to (0,.5) node[above] {\small $\cal A$};
		\draw [thick] (0,0) arc [radius=.3, start angle=-240, end angle=-120];
		\draw [thick, decoration = {markings, mark=at position .75 with {\arrow[scale=1.2]{stealth reversed}}}, postaction=decorate] (0,-0.52) to +(-30:.5) node[right] {\small $\cal A$};
		\draw [thick] (0,0) arc [radius=.3, start angle=60, end angle=-42];
		\draw [thick, decoration = {markings, mark=at position .8 with {\arrow[scale=1.2]{stealth reversed}}}, postaction=decorate] (0,-0.52) ++(-150:.1) to +(-150:.4) node[left] {\small $\cal A$};
		\draw [fill=black](0,0) circle (0.05) node[left] {\small $\mu$};
\end{tikzpicture}}
 &= 
\raisebox{-2.2em}{\begin{tikzpicture}
		\draw [thick, decoration = {markings, mark=at position .5 with {\arrow[scale=1.2]{stealth reversed}}}, postaction=decorate] (0,.8) node[above] {\small $\cal A$} to (0,0);
		\draw [thick, decoration = {markings, mark=at position .7 with {\arrow[scale=1.2]{stealth reversed}}}, postaction=decorate] (0,0)  to (-.6,-.5) node[left] {\small $\cal A$};
		\draw [thick, decoration = {markings, mark=at position .7 with {\arrow[scale=1.2]{stealth reversed}}}, postaction=decorate] (0,0)--(.6,-.5) node[right] {\small $\cal A$};
		\draw [fill=black](0,0) circle (0.05) node[left] {\small $\mu$};
\end{tikzpicture}} ~ \text{(commutativity)}\,, \quad&
\raisebox{-2.2em}{\begin{tikzpicture}
		\draw [thick, decoration = {markings, mark=at position .55 with {\arrow[scale=1.2]{stealth}}}, postaction=decorate] (0,-.6) node[left] {\small $\cal A$} to (0,0);
		\draw [thick, decoration = {markings, mark=at position .55 with {\arrow[scale=1.2]{stealth}}}, postaction=decorate] (0,.6) to (0,1.2) node[left] {\small $\cal A$};
		\draw [thick] (.3,.3) arc [radius=.3, start angle=0, end angle=360];
		\draw [thick, decoration = {markings, mark=at position 1 with {\arrow[scale=1.2]{stealth}}}, postaction=decorate] (0.3,0.38) to (0.3,0.39);
		\draw [thick, decoration = {markings, mark=at position 1 with {\arrow[scale=1.2]{stealth}}}, postaction=decorate] (-0.3,0.38) to (-0.3,0.39);
		\draw [fill=black](0,0) circle (0.05) node[anchor = west] {\small $\mu^\dagger$};
		\draw [fill=black](0,.6) circle (0.05) node[anchor = south west] {\small $\mu$};
\end{tikzpicture}}
 &= 
\raisebox{-2.2em}{\begin{tikzpicture}
		\draw [thick, decoration = {markings, mark=at position .5 with {\arrow[scale=1.2]{stealth}}}, postaction=decorate] (0,-.6) node[left] {\small $\cal A$} to (0,1.2) node[left] {\small $\cal A$};
\end{tikzpicture}} ~~~ \text{(separability)}\,, \\
\raisebox{-2.2em}{\begin{tikzpicture}
		\draw [thick, decoration = {markings, mark=at position .7 with {\arrow[scale=1.2]{stealth}}}, postaction=decorate] (-.3,-.6) node[below] {\small $\cal A$} to (0,0);
		\draw [thick, decoration = {markings, mark=at position .7 with {\arrow[scale=1.2]{stealth}}}, postaction=decorate] (.3,-.6) node[below] {\small $\cal A$} to (0,0);
		\draw [fill=black](0,0) circle (0.05) node[left] {\small $\mu$};
		\draw [thick, decoration = {markings, mark=at position .3 with {\arrow[scale=1.2]{stealth}}, mark=at position .85 with {\arrow[scale=1.2]{stealth}}}, postaction=decorate] (0,0) to (.6,1.2) node[right] {\small $\cal A$};
		\draw [thick, decoration = {markings, mark=at position .5 with {\arrow[scale=1.2]{stealth}}}, postaction=decorate] (.9,-.6) node[below] {\small $\cal A$} to (.3,.6);
		\draw [fill=black](.3,.6) circle (0.05) node[left] {\small $\mu$};
\end{tikzpicture}}
~ &= ~
\raisebox{-2.2em}{\begin{tikzpicture}
		\draw [thick, decoration = {markings, mark=at position 1 with {\arrow[scale=1.2]{stealth}}}, postaction=decorate] (-.3,-.6) node[below] {\small $\cal A$} to (0,0);
		\draw [thick, decoration = {markings, mark=at position .7 with {\arrow[scale=1.2]{stealth}}}, postaction=decorate] (.3,-.6) node[below] {\small $\cal A$} to (.6,0);
		\draw [fill=black](.6,0) circle (0.05) node[right] {\small $\mu$};
		\draw [thick, decoration = {markings, mark=at position .85 with {\arrow[scale=1.2]{stealth}}}, postaction=decorate] (0,0) to (.6,1.2) node[right] {\small $\cal A$};
		\draw [thick, decoration = {markings, mark=at position .8 with {\arrow[scale=1.2]{stealth}}, mark=at position .3 with {\arrow[scale=1.2]{stealth}}}, postaction=decorate] (.9,-.6) node[below] {\small $\cal A$} to (.3,.6);
		\draw [fill=black](.3,.6) circle (0.05) node[left] {\small $\mu$};
\end{tikzpicture}} ~~~~~  \text{(associativity)}\,, \quad&
\raisebox{-2.2em}{\begin{tikzpicture}
		\draw [thick, decoration = {markings, mark=at position .7 with {\arrow[scale=1.2]{stealth}}}, postaction=decorate] (-.5,-.6) node[above = 8pt] {\small $\cal A$} to (0,0);
		\draw [fill=black](-.5,-.6) circle (0.05) node[left] {\small $u$};
		\draw [thick, decoration = {markings, mark=at position .3 with {\arrow[scale=1.2]{stealth}},mark=at position .85 with {\arrow[scale=1.2]{stealth}}}, postaction=decorate] (0,-.6) node[below] {\small $\cal A$} to (0,.6) node[above] {\small $\cal A$};
		\draw [fill=black](0,0) circle (0.05) node[right] {\small $\mu$};
\end{tikzpicture}}
~ &= ~
\raisebox{-2.2em}{\begin{tikzpicture}
		\draw [thick, decoration = {markings, mark=at position .5 with {\arrow[scale=1.2]{stealth}}}, postaction=decorate] (0,-.6) node[below] {\small $\cal A$} to (0,.6) node[above] {\small $\cal A$};
\end{tikzpicture}} ~ \text{(unit)}\,.
\fe
These properties guarantee that a mesh of ${\cal A}$ inserted along the dual triangulation of the spacetime 3-manifold is independent of the choice of the triangulation.  
This generalizes the notion of 0-gaugeability for non-invertible 1-form symmetries in 2+1d.
  
 Importantly, the set of lines in a (commutative) algebra object $\cal A$ generally does not form a subcategory, i.e., they are generally not closed under fusion.  
For example, the tensor product TQFT ${\cal T}\times \overline{\cal T}$ always admit a commutative algebra object ${\cal A} = \sum_{a\in {\cal T}} a\otimes \tilde a$ (which is furthermore Lagrangian). Here $\overline{\cal T}$ is the orientation-reversal of the TQFT $\cal T$, and $\tilde a$ is the line in $\overline{\cal T}$ corresponding to $a$ in $\cal T$. 
Clearly, the set of lines $a\otimes \tilde a$ is generally not closed under fusion. 
It is an interesting point that the non-invertible generalization of a gaugeable ``subgroup" need not be closed under fusion.

Next, we move on to the higher gauging of non-invertible 1-form symmetries on a surface. 
Since we now only gauge on a surface, the commutativity condition in \eqref{commutative.algebra}, which is a generalization of the trivial braiding, is no longer relevant.
This generalizes the discussion in Section \ref{sec:generalhighergauging}, where 1-gaugeability only demands the triviality of crossing, but not the braiding.   
We conclude that  every (not necessarily commutative) algebra object is 1-gaugeable.\footnote{This definition of the algebra object was originally introduced in \cite{ostrik2003module,muger2003subfactors}.  In the literature, it is also called the \textit{symmetric special Frobenius algebra} in \cite{Fuchs:2002cm,Fuchs:2012dt} and the \textit{$\Delta$-separable Frobenius algebra} in \cite{Carqueville:2017ono}. See \cite{Bhardwaj:2017xup,Huang:2021zvu} for recent discussions in the context of non-invertible 0-form symmetries in 1+1d.}

For example, in the 2+1d $\mathbb{Z}_2$ gauge theory, the algebra objects $1+\ee$ and $1+\mm$ are commutative, while $1+\psi$ and $1+\ee+\mm+\psi$ are not commutative.  In the case of ${\cal A}=1+\ee+\mm+\psi$, there are two inequivalent choices of the discrete torsion $\mu$, leading to two different condensation surfaces. 
In total, together with the trivial algebra object ${\cal A}=1$, these lead to the six condensation surfaces in Section \ref{sec:z2gauge}.

\subsection{Trivial condensation defects and Morita equivalence}\label{sec:morita}

Not every 1-gaugeable (invertible or non-invertible) 1-form symmetry, i.e., algebra object, gives rise to a non-trivial condensation defect.  
By a trivial condensation defect, we mean a surface that acts trivially on not just the local operators, but also the  lines.\footnote{There are nontrivial SETs associated with symmetries that do not permute the lines but have a nontrivial symmetry fractionalization class (see, for instance, \cite{2015PhRvX...5d1013C}). We do not discuss the freedom of choosing a symmetry fractionalization class in this paper.} 
Mathematically, an algebra object $\cal A$ is Morita trivial iff there exists another object ${\cal N}$ such that \cite[Example 7.8.18]{etingof2016tensor}
\ie
{\cal A} = {\cal N}\times \overline{\cal N}\,, \qquad \text{and} \qquad \raisebox{-1.8em}{\begin{tikzpicture}
		\draw [thick, decoration = {markings, mark=at position .7 with {\arrow[scale=1.2]{stealth}}}, postaction=decorate] (-.5,-.5) node[left] {\small $\cal A$} to (0,0);
		\draw [thick, decoration = {markings, mark=at position .7 with {\arrow[scale=1.2]{stealth}}}, postaction=decorate] (.5,-.5) node[right] {\small $\cal A$} to (0,0);
		\draw [thick, decoration = {markings, mark=at position .5 with {\arrow[scale=1.2]{stealth}}}, postaction=decorate] (0,0) to (0,.8) node[above] {\small $\cal A$};
		\draw [fill=black](0,0) circle (0.05) node[left] {\small $\mu$};
\end{tikzpicture}}
~ = ~
\raisebox{-2.2em}{\begin{tikzpicture}
		\draw [thick, decoration = {markings, mark=at position .7 with {\arrow[scale=1.2]{stealth}}}, postaction=decorate] (-.7,-.5) to (-.1,.1);
		\draw [thick, decoration = {markings, mark=at position .7 with {\arrow[scale=1.2]{stealth reversed}}}, postaction=decorate] (.7,-.5)  to (.1,.1);
		\draw [thick, decoration = {markings, mark=at position .7 with {\arrow[scale=1.2]{stealth reversed}}}, postaction=decorate] (-.5,-.5) to (0,0) node[below = 12pt] {\small $\cal N$};;
		\draw [thick, decoration = {markings, mark=at position .7 with {\arrow[scale=1.2]{stealth}}}, postaction=decorate] (.5,-.5) to (0,0);
		\draw [thick, decoration = {markings, mark=at position .5 with {\arrow[scale=1.2]{stealth}}}, postaction=decorate] (-.1,.1) to (-.1,.8) node[above left] {\small $\cal N$};
		\draw [thick, decoration = {markings, mark=at position .5 with {\arrow[scale=1.2]{stealth reversed}}}, postaction=decorate] (.1,.1) to (.1,.8) node[above right] {\small $\cal N$};
\end{tikzpicture}} \,.
\fe
Every Morita trivial algebra object gives rise to the trivial condensation surface. This is because a mesh of $\cal N \times \overline{\cal N}$ inserted on any closed 2-dimensional surface is topologically trivial since it can be shrunk to a single contractible loop of $\cal N$ (see, for instance, \cite{Chang:2018iay}). More generally two algebras $\cal A$ and $\cal A'$ lead to the same condensation surface (i.e., are Morita equivalent) if\footnote{Mathematically, two algebras $\cal A$ and $\cal A'$ are Mortia equivalent iff the category of (right) $\cal A$-modules and $\cal A'$-modules are the same (left) module categories over the modular tensor category of all topological lines \cite{ostrik2003module}. The condensation defects only depend on the Morita equivalent class of the algebra objects \cite{Fuchs:2012dt,Carqueville:2017ono}.} 
\ie
{\cal A}' = {\cal N}\times {\cal A} \times \overline{\cal N}\,, \qquad \text{and} \qquad \raisebox{-1.8em}{\begin{tikzpicture}
		\draw [thick, decoration = {markings, mark=at position .7 with {\arrow[scale=1]{stealth}}}, postaction=decorate] (-.7,-.7) node[below left] {\small $\cal A'$} to (0,0);
		\draw [thick, decoration = {markings, mark=at position .7 with {\arrow[scale=1]{stealth}}}, postaction=decorate] (.7,-.7) node[below right] {\small $\cal A'$} to (0,0);
		\draw [thick, decoration = {markings, mark=at position .5 with {\arrow[scale=1]{stealth}}}, postaction=decorate] (0,0) to (0,.8) node[above] {\small $\cal A'$};
		\draw [fill=black](0,0) circle (0.05) node[left] {\small $\mu'$};
\end{tikzpicture}}
~ = ~
\raisebox{-2.2em}{\begin{tikzpicture}
		\draw [color=purple, thick, decoration = {markings, mark=at position .7 with {\arrow[scale=1]{stealth}}}, postaction=decorate] (-.9,-.7) to (-.1,.1);
		\draw [color=purple, thick, decoration = {markings, mark=at position .7 with {\arrow[scale=1]{stealth reversed}}}, postaction=decorate] (.9,-.7)  to (.1,.1);
		\draw [thick, decoration = {markings, mark=at position .7 with {\arrow[scale=1]{stealth}}}, postaction=decorate] (-.7,-.7) node[below left] {\small $\cal A$} to (0,0);
		\draw [thick, decoration = {markings, mark=at position .7 with {\arrow[scale=1]{stealth}}}, postaction=decorate] (.7,-.7) node[below right] {\small $\cal A$} to (0,0);
		\draw [color=purple, thick, decoration = {markings, mark=at position .7 with {\arrow[scale=1]{stealth reversed}}}, postaction=decorate] (-.5,-.7) to (0,-.2);
		\draw [color=purple, thick, decoration = {markings, mark=at position .7 with {\arrow[scale=1]{stealth}}}, postaction=decorate] (.5,-.7) to (0,-.2) node[below = 10pt] {\small $\cal N$};
		\draw [color=purple, thick, decoration = {markings, mark=at position .5 with {\arrow[scale=1]{stealth}}}, postaction=decorate] (-.1,.1) to (-.1,.8) node[above left] {\small $\cal N$};
		\draw [color=purple, thick, decoration = {markings, mark=at position .5 with {\arrow[scale=1]{stealth reversed}}}, postaction=decorate] (.1,.1) to (.1,.8) node[above right] {\small $\cal N$};
		\draw [thick, decoration = {markings, mark=at position .7 with {\arrow[scale=1]{stealth}}}, postaction=decorate] (0,0) to (0,.8) node[above] {\small $\cal A$};
		\draw [fill=black](0,0) circle (0.05) node[left] {\small $\mu$};
\end{tikzpicture}} \,.
\fe
Similarly, this is because a mesh of ${\cal N}\times {\cal A} \times \overline{\cal N}$ inserted on any closed surface is topologically equivalent to a mesh of $\cal A$ and a single contractible loop of $\cal N$.

Let us demonstrate this in the example of the 2+1d Ising TQFT.  There are two nontrivial topological lines. There is a 0-anomalous $\mathbb{Z}_2$ fermion line $\psi$ with $\theta(\psi)=-1$. There is another non-invertible line (non-abelian anyon) $\sigma$. Together, they  obey the fusion rule:
\ie
\psi^2 =1\,,~~~\sigma \times \sigma =  1+\psi\,,~~~\psi\times \sigma=\sigma\times \psi=\sigma\,.
\fe
Furthermore, $\sigma$ braids nontrivially with $\psi$, i.e., $B(\psi, \sigma)=-1$. 
Following the discussion in Section \ref{sec:z2}, a fermion line is 1-gaugeable and leads to an invertible $\mathbb{Z}_2$ 0-form symmetry $S$.  
However, using \eqref{action.of.fermioncondensation}, we see that  $S$ acts trivially on all the lines:
\ie
S\cdot 1=1\,,~~~S\cdot \psi=\psi\,,~~~ S\cdot \sigma =  \sigma\,.
\fe
This is indeed consistent with the general mathematical statement that a Morita trivial algebra object, in this case ${\cal A}=1+\psi$ in the Ising TQFT, gives rise to a trivial surface defect.

\subsection{All surfaces are made out of lines}

In \cite{Fuchs:2012dt}, it is shown that in every 2+1d TQFT\footnote{Here we only consider unitary 2+1d TQFTs with no non-trivial local operator.}, there is a one-to-one correspondence between the topological surfaces and the Morita equivalence classes of algebra objects. 
Importantly, the latter do not need to be commutative. 
Translating into our formalism, this theorem shows that every (invertible and non-invertible)  0-form symmetry in a 2+1d TQFT arises from the higher gauging of  an (invertible or non-invertible) 1-form symmetry.  
At the level of the slogan, we have arrived at the conclusion that in 2+1d TQFT, \textit{all surfaces are made out of lines}.

We now relate this general conclusion on the surface defects to topological boundary conditions in a 2+1d TQFT. 
By the folding trick, a topological surface in a 2+1d TQFT  $\cal T$ is equivalent to a topological boundary condition of the doubled theory $\cal T \times \overline{\cal T}$. 
It is known that the topological boundary conditions of  a TQFT are in one-to-one correspondence with Lagrangian algebras of the underlying  UMTC $\cal C$ \cite{2010arXiv1009.2117D,Fuchs:2012dt,2014NuPhB.886..436K,Kaidi:2021gbs}.\footnote{In the context of fully extended TQFT, the necessary and sufficient condition for existence of  a topological boundary condition was proven rigorously in  \cite{Freed:2020qfy}.} 
Therefore, the folding trick implies that for every UMTC $\cal C$,  \textit{the Morita equivalence classes of algebras in  $\cal C$ is in one-to-one correspondence with the Lagrangian algebras in $\cal C \boxtimes \overline{\cal C}$.} 
(Note that if $\cal C$ describes the anyons of $\cal T$, then $\cal C \boxtimes \overline{\cal C}$ describes the anyons of $\cal T \times \overline{\cal T}$.)

This physics expectation follows from a more general mathematical theorem.  
It was proven in \cite[Proposition 4.8]{2010arXiv1009.2117D} that there is a one-to-one correspondence between the Morita equivalence classes of algebras in a unitary fusion category $\cal C$ and the Lagrangian algebras in its Drinfeld center ${\cal Z}({\cal C})$ (i.e., quantum double).  
In the case of a TQFT, the unitary fusion category ${\cal C}$ is a UMTC, and it follows that ${\cal Z}({\cal C}) = {\cal C}\boxtimes \overline{\cal C}$.  
We therefore have proved the above  physics expectation.

\subsection{RCFT interpretation}

Now we comment on the RCFT interpretation of the condensation surfaces. 
As shown in \cite{Fuchs:2002cm, Kapustin:2010if}, given a chiral algebra, the possible (non-diagonal) modular invariant 1+1d RCFTs   correspond to different condensation  surface defects in the corresponding 2+1d TQFT. 
This is obtained by compactifying the 2+1d TQFT on an interval with holomorphic and anti-holomorphic boundary conditions given by the chiral algebra and an insertion of the condensation surface defect in the middle \cite{Kapustin:2010if,Komargodski:2020mxz}. In particular, there is a correspondence between non-diagonal modular invariant $\mathcal{G}_K$ WZW models and condensation surfaces in $\mathcal{G}_K$ Chern-Simons theory.

An interesting example is $SU(2)_K$, where the modular invariants, and equivalently the condensation surfaces, have an ADE classification. The multiplication of the corresponding torus partition function matrices gives the fusion rules of the condensation defects of $SU(2)_K$ Chern-Simons theory on a torus. At every level $K$, the $SU(2)_K$ Chern-Simons theory has a $\bZ_2$ 1-form symmetry line with spin $e^{2\pi i K/4}$. Thus, the 1-form symmetry is 1-gaugeable for even $K$, which corresponds to the D-series modular invariants. Let us denote the corresponding condensation defects by $S_2$. Moreover, at $K=10,16,28$ there exist a non-invertible 1-gaugeable algebra leading to the E-series modular invariants. We denote these condensation defects by $S_\mathrm{E}$. The fusion of these condensation defects on the torus are \cite{Fuchs:2002cm}:
\ie
	SU(2)_{4l}: \quad & S_2 \times S_2 = 2 S_2\,,\\
	SU(2)_{4l+2}: \quad & S_2 \times S_2 = 1\,,\\
	SU(2)_{10}: \quad &  S_2 \times S_\mathrm{E} = S_\mathrm{E} \times S_2 = S_\mathrm{E},  \quad& S_\mathrm{E} \times S_\mathrm{E} &= 2 S_\mathrm{E}\,,\\
	SU(2)_{16}:  \quad & S_2 \times S_\mathrm{E} = S_\mathrm{E} \times S_2 = 2S_\mathrm{E}, \quad& S_\mathrm{E} \times S_\mathrm{E} &=S_2 + S_\mathrm{E}\,,\\
	SU(2)_{28}:  \quad & S_2 \times S_\mathrm{E} = S_\mathrm{E} \times S_2 = 2S_\mathrm{E}, \quad& S_\mathrm{E} \times S_\mathrm{E} &= 4S_\mathrm{E}\,.
\fe
Since this is just the fusion algebra on the torus, the fusion coefficients above are the torus partition function of the actual fusion coefficients. It would be interesting to find the fusion rules on higher genus surfaces and thus find the actual 1+1d TQFT fusion coefficients. In particular, consider the fusion
\begin{equation}
	S_\mathrm{E}(\Sigma) \times S_\mathrm{E}(\Sigma) = c_1(\Sigma) S_2(\Sigma) + c_2(\Sigma) S_\mathrm{E}(\Sigma)
\end{equation}
in $SU(2)_{16}$ Chern-Simons theory. The 1+1d TQFTs $c_1$ and $c_2$ are almost trivial whose partition functions have the form
$c_1(\Sigma) = \lambda_1^{\chi(\Sigma)}$ and $c_2(\Sigma) = \lambda_2^{\chi(\Sigma)}$. The relative Euler counter-term $\lambda_1/\lambda_2$ is an interesting data of the fusion rule to be computed from the fusion on higher genus surfaces. We leave this computation for future works.

\section*{Acknowledgements}

We  thank M.\ Cheng, Y.\ Choi, C.\ Cordova, T.\ D.\ D\'ecoppet, T.\ Dumitrescu, P.\ Gorantla,  D.\ Harlow, P.-S.\ Hsin, J.\ Kaidi, R.\ Kobayashi, Z.\ Komargodski, H.\ T.\ Lam, M.\ Metlitski, R.\ Radhakrishnan, N.\ Seiberg, and Y.\ Wang  for interesting discussions.  
We are grateful to Y.\ Choi, D.\ Delmastro, A.\ Kapustin,  R.\ Kobayashi, and K.\ Ohmori for helpful comments on a draft. 
SHS would  particularly like to thank Y.\ Choi and H.\ T.\ Lam for numerous enlightening discussions on a related project  \cite{Choi:2022zal} on higher gauging in 3+1d. 
SHS would also like to thank P.\ Gorantla, H.\ T.\ Lam, and N.\ Seiberg for discussions on anyon condensations. KR would like to thank T.\ Dumitrescu for many useful discussions on this topic. The work of KR is supported by the Mani L. Bhaumik Institute for Theoretical Physics. SS is supported in part by the Simons Foundation grant 488657 (Simons Collaboration on the Non-Perturbative Bootstrap).
The authors of this paper were ordered alphabetically.

\appendix

\section{More on the fusion rules}

\subsection{Fusion rules from the higher gauging of $\mathbb{Z}_N$ 1-form symmetries}\label{app:ZN}

In this appendix we will derive the fusion rule \eqref{zn.general.fusion} between the higher quantum symmetry lines on the condensation surfaces that arise from a 1-gaugeable $\mathbb{Z}_N$ 1-form symmetry. The 0-anomaly of the 1-gaugeable  $\mathbb{Z}_N$ 1-form symmetry is labeled by an integer $k$ (see Section \ref{sec:ZN}). 

\bigskip\centerline{\it $S_n \times S_{n'}$ fusion}\bigskip

As a warm up, we fist derive the fusion rule \eqref{generalSSZN} between the condensation surfaces  $S_n$, which come from the higher gauging of the $\mathbb{Z}_n$ 1-form symmetry subgroup with $n|N$. We will place the condensation surfaces on  a genus-$g$ surface $\Sigma_g$ with generating 1-cycles $\a_{i=1,\dots,g},\b_{i=1,\dots,g} \in H_1(\Sigma_g , \mathbb{Z}_N)$.

Let $n,n'$ be two divisors of $N$. The fusion of the corresponding condensation surfaces $S_n$ and $S_{n'}$ is
\ie
&	S_{n}(\Sigma_g) \times S_{n'}(\Sigma_g) \\
&= \frac{1}{n^g {n'}^g} \sum_{\mu_i ,\nu_i \in \mathbb{Z}_n} \sum_{\mu'_i,\nu'_i \in \mathbb{Z}_{n'}} e^{2\pi i \frac{kN}{n n'} (\mu_i \nu'_i - \nu_i \mu'_i )} \, \eta\left( \sum_i N(\frac{\mu_i}{n}+\frac{\mu'_i}{n'}) \a_i + N(\frac{\nu_i}{n}+\frac{\nu'_i}{n'}) \b_i \right) \, .
\fe
Below we will do  a change of variables on $\mu_i ,\nu_i,\mu'_i,\nu'_i$ in order to simplify the above expression.

To proceed, we first note that for any pair of numbers $(\mu,\mu') \in \mathbb{Z}_n \times \mathbb{Z}_{n'}$ we can do the change of variable
\ie \label{app.change.of.variable}
\mu &= p \alpha + \frac{n}{\mathrm{gcd}(n,n')}\lambda ~,\\
\mu' &= p' \alpha - \frac{n'}{\mathrm{gcd}(n,n')}\lambda ~,
\fe
for $\alpha \in \left\{ 1, \dots, \mathrm{lcm}(n,n') \right\}$ and $\lambda \in \mathbb{Z}_{\mathrm{gcd}(n,n')}$, where
\begin{equation}\label{app.bezout}
	p \frac{n'}{\mathrm{gcd}(n,n')} + p' \frac{n}{\mathrm{gcd}(n,n')} = 1
\end{equation}
for some fixed integers $p$ and $p'$. Note that  \eqref{app.change.of.variable} defines a map from $\left\{ 1, \dots, \mathrm{lcm}(n,n') \right\} \times \mathbb{Z}_{\mathrm{gcd}(n,n')}$ to $\mathbb{Z}_n \times \mathbb{Z}_{n'}$ whose inverse is given by
\ie 
\alpha &= \frac{n'}{\mathrm{gcd}(n,n')}\mu + \frac{n}{\mathrm{gcd}(n,n')} \mu' \in  \mathbb{Z}_{\mathrm{lcm}(n,n')} ~,\\
\lambda &= p' \mu - p\mu' \in  \mathbb{Z}_{\mathrm{gcd}(n,n')}~.
\fe
Therefore we conclude that the map is bijective and hence the change of variable is legitimate.

Now let us apply the change of variable \eqref{app.change.of.variable} to $\mu_i,\mu_i',\nu_i,\nu_i'$ and write
\ie
	\mu_i &= p \alpha_i + \frac{n}{\mathrm{gcd}(n,n')}\lambda_i  ~,\\
	\mu'_i &= p' \alpha_i - \frac{n'}{\mathrm{gcd}(n,n')}\lambda_i~,\\
	\nu_i &= p \beta_i + \frac{n}{\mathrm{gcd}(n,n')}\rho_i  ~,\\
	\nu'_i &= p' \beta_i - \frac{n'}{\mathrm{gcd}(n,n')}\rho_i  ~,
\fe
where $\alpha_i,\beta_i \in \left\{ 1, \dots, \mathrm{lcm}(n,n') \right\}$, $\lambda_i,\rho_i \in \mathbb{Z}_{\mathrm{gcd}(n,n')}$, and $pn'+p'n=\mathrm{gcd}(n,n')$ for some fixed integers $p$ and $p'$. Using these new set of  variables we get
\begin{equation}
	S_{n}(\Sigma_g)\times S_{n'}(\Sigma_g) = \frac{1}{n^g {n'}^g} \sum_{\substack{
			1 \leq \alpha_i,\beta_i \leq \mathrm{lcm}(n,n') \\
			\lambda_i,\rho_i \in \mathbb{Z}_{\mathrm{gcd}(n,n')}}} e^{2\pi i \frac{k\ell}{\mathrm{gcd}(n,n')} (\beta_i \lambda_i - \alpha_i \rho_i )} \, \eta\bigg( \ell (\alpha_i \, \a_i + \beta_i \, \b_i ) \bigg) \, , 
\end{equation}
where $\ell = \frac{N}{\mathrm{lcm}(n,n')}$. Summing over  $\lambda_i,\rho_i \in \mathbb{Z}_{\mathrm{gcd}(n,n')}$ gives the Kronecker deltas 
\ie
\mathrm{gcd}(n,n')^{2g} \,  \delta_{\, \left(\frac{k\ell}{\mathrm{gcd}(n,n')} \alpha_i \mod{\mathbb{Z}}\right),0} \, \delta_{\,\left(\frac{k\ell}{\mathrm{gcd}(n,n')} \beta_i \mod{\mathbb{Z}}\right),0}
\fe
 which forces $\alpha_i$ and $\beta_i$ to be divisible by $\frac{\mathrm{gcd}(n,n')}{\mathrm{gcd}(n,n',k\ell)}$. Therefore we do the replacement $\alpha_i,\beta_i \mapsto \frac{\mathrm{gcd}(n,n')}{\mathrm{gcd}(n,n',k\ell)} \bar\alpha_i, \frac{\mathrm{gcd}(n,n')}{\mathrm{gcd}(n,n',k\ell)}\bar\beta_i$ for $1 \leq \bar\alpha_i,\bar\beta_i \leq \frac{\mathrm{gcd}(n,n',k\ell)nn'}{\mathrm{gcd}(n,n')^2}$ and get
\begin{align}
	S_n(\Sigma_g)\times S_{n'}(\Sigma_g) &= \frac{\mathrm{gcd}(n,n')^{2g}}{n^g{n'}^g} \sum_{ 1 \leq \bar\alpha_i,\bar\beta_i \leq \frac{\mathrm{gcd}(n,n',k\ell)nn'}{\mathrm{gcd}(n,n')^2} } \eta^{\frac{\ell \, \mathrm{gcd}(n,n')}{\mathrm{gcd}(n,n',k\ell)}} \Bigg(\bar\alpha_i \, \a_i+ \bar\beta_i \, \b_i \Bigg) \, , \notag \\
	&= \left( \mathrm{gcd}(n,n',k\ell) \right)^g S_{\frac{\mathrm{gcd}(n,n',k\ell)nn'}{\mathrm{gcd}(n,n')^2}}(\Sigma_g)	\, .		
\end{align}
We conclude that the fusion algebra is
\begin{equation}
	S_n \times S_{n'} =
\left(	 \mathcal{Z}_{\mathrm{gcd}(n,n',k\ell)}\right)\, S_{\frac{\mathrm{gcd}(n,n',k\ell)nn'}{\mathrm{gcd}(n,n')^2}} \,,
\end{equation}
where $\left(	 \mathcal{Z}_{n}\right)$ stands for the 1+1d $\mathbb{Z}_n$ gauge theory. 

\bigskip\centerline{\it$\hat{\eta}_{n}^a \tilde{\eta}_{n}^b \times \hat{\eta}_{n'}^{a'} \tilde{\eta}_{n'}^{b'} $ fusion}\bigskip

Now we will derive the fusion between the higher quantum symmetry lines living on the surfaces \eqref{zn.general.fusion}. We put the composite defects $\hat{\eta}_{n}^a \tilde{\eta}_{n}^b$ of \eqref{hat.eta.tilde.eta} on $(\bar\gamma, \Sigma)$ and perform the computations covariantly without specifying a basis for $H_1(\Sigma, \mathbb{Z})$:
\ie
	&\hat{\eta}_{n}^a \tilde{\eta}_{n}^b (\bar{\gamma}, \Sigma)\times \hat{\eta}_{n'}^{a'} \tilde{\eta}_{n'}^{b'} (\bar{\gamma}, \Sigma) \\
	&= \sum_{\substack{\gamma \in H_1(\Sigma,\mathbb{Z}_{n}) \\
			\gamma' \in H_1(\Sigma,\mathbb{Z}_{n'})   }}  \frac{e^{ 2\pi i \langle \bar{\gamma}, {a \over n}\gamma + {a' \over n'}\gamma' \rangle + {2\pi ik \over N}\langle \frac{N}{n}\gamma + b\bar{\gamma}, \frac{N}{n'}\gamma' + b'\bar{\gamma} \rangle }}{\sqrt{|H_1(\Sigma ,\mathbb{Z}_n)||H_1(\Sigma ,\mathbb{Z}_{n'})|}} \, \eta\left( \frac{N}{n}\gamma+\frac{N}{n'}\gamma' + (b+b')\bar{\gamma} \right) \,.
\fe

Next, we apply the same change of variable as in \eqref{app.change.of.variable} to $\gamma \in H_1(\Sigma,\mathbb{Z}_{n})$ and $\gamma' \in H_1(\Sigma,\mathbb{Z}_{n'})$. We write
\ie 
\gamma &= p \alpha + \frac{n}{\mathrm{gcd}(n,n')}\lambda  ~,\\
\gamma' &= p' \alpha - \frac{n'}{\mathrm{gcd}(n,n')}\lambda  ~,
\fe
for $\alpha \in \left\{ \sum_{i,j} \alpha_i \a_i + \beta_j \b_j \in H_1(\Sigma,\mathbb{Z}) \mid \alpha_i,\beta_j \in \{ 1, \dots, \mathrm{lcm}(n,n')  \}  \right\}$ and $\lambda \in H_1(\Sigma,\mathbb{Z}_{\mathrm{gcd}(n,n')})$, where the integers $p$ and $p'$ satisfy $p n' + p'n = \mathrm{gcd}(n,n')$. Here $\a_{i},\b_{j}$ are a fixed set of generators for $H_1(\Sigma,\mathbb{Z})$.

In terms of these new variables we get:
\ie
&\hat{\eta}_{n}^a \tilde{\eta}_{n}^b(\bar{\gamma}, \Sigma) \times \hat{\eta}_{n'}^{a'} \tilde{\eta}_{n'}^{b'} (\bar{\gamma}, \Sigma) \\
&=  \sum_{\substack{\alpha \in H_1(\Sigma,\mathbb{Z}_{\mathrm{lcm}(n,n')}) \\
		\lambda \in H_1(\Sigma,\mathbb{Z}_{\mathrm{gcd}(n,n')}) }} {e^{ {2\pi i \over \mathrm{gcd}(n,n')} \langle (c-c')\bar{\gamma} - k\ell \alpha, \lambda \rangle + {2\pi i}\langle \bar{\gamma}, ( \frac{pc}{n} + \frac{p'c'}{n'} ) \alpha \rangle } \over \sqrt{|H_1(\Sigma ,\mathbb{Z}_n)||H_1(\Sigma ,\mathbb{Z}_{n'})|}}  \; \eta\big( \ell \alpha + (b+b')\bar{\gamma} \big) \,,
\fe
where $c=a-kb'$, $c'=a'+kb$, and $\ell = {N \over \mathrm{lcm}(n,n')}$. The sum over $\lambda$ gives a Kronecker delta, multiplied by $|H_1(\Sigma,\mathbb{Z}_{\mathrm{gcd}(n,n')})|$, that enforces $\alpha$ to satisfy
\ie \label{constraint}
(c-c')\bar{\gamma} - k\ell \alpha = 0 \in H_1(\Sigma,\mathbb{Z}_{\mathrm{gcd}(n,n')}) \,.
\fe
Thus, we can  rewrite the fusion as
\ie \label{app:general.fusion.expression}
& \frac{1}{|H_1(\Sigma ,\mathbb{Z}_{\mathrm{gcd}(n,n')})|} \; \hat{\eta}_{n}^a \tilde{\eta}_{n}^b (\bar{\gamma}, \Sigma)\times \hat{\eta}_{n'}^{a'} \tilde{\eta}_{n'}^{b'} (\bar{\gamma}, \Sigma) \\
&=  \sum_{\substack{\alpha \in H_1(\Sigma,\mathbb{Z}_{\mathrm{lcm}(n,n')}) \\
		(c-c')\bar{\gamma} - k\ell \alpha = 0 \text{ mod }\mathrm{gcd}(n,n') }} {e^{ {2\pi i}\langle \bar{\gamma}, ( \frac{pc}{n} + \frac{p'c'}{n'} ) \alpha \rangle } \over \sqrt{|H_1(\Sigma ,\mathbb{Z}_n)||H_1(\Sigma ,\mathbb{Z}_{n'})|}}  \; \eta\big( \ell \alpha + (b+b')\bar{\gamma} \big) \,.
\fe

If there is no solution to the constraint \eqref{constraint} we get zero on the righthand side. Otherwise, any solution can be written as
\begin{equation} \label{solution.to.constraint}
	\alpha = {c-c' \over \mathrm{gcd}(n,n',k\ell) } \left({k\ell \over \mathrm{gcd}(n,n',k\ell)} \right)^{-1}_{N \over x\ell} \bar{\gamma} + {\mathrm{gcd}(n,n') \over \mathrm{gcd}(n,n',k\ell) } \bar{\alpha} \,, \qquad \bar{\alpha} \in H_1(\Sigma,\mathbb{Z}_x)\,,
\end{equation}
where \ie x \equiv \frac{\mathrm{gcd}(n,n',k\ell)nn'}{\mathrm{gcd}(n,n')^2}\,, \fe and $\big ({k\ell \over \mathrm{gcd}(n,n',k\ell)}\big)^{-1}_{N \over x\ell} \in (\mathbb{Z}_{N \over x\ell})^\times$ is the multiplicative inverse of ${k\ell \over \mathrm{gcd}(n,n',k\ell)}$ modulo ${N \over x\ell} = {\mathrm{gcd}(n,n') \over \mathrm{gcd}(n,n',k\ell)}$. The multiplicative inverse always exists since ${k\ell \over \mathrm{gcd}(n,n',k\ell)}$ is coprime with respect to ${\mathrm{gcd}(n,n') \over \mathrm{gcd}(n,n',k\ell)}$.

We substitute \eqref{solution.to.constraint} into \eqref{app:general.fusion.expression} to get
\ie
& \frac{1}{|H_1(\Sigma ,\mathbb{Z}_{\mathrm{gcd}(n,n')})|} \; \hat{\eta}_{n}^a \tilde{\eta}_{n}^b(\bar{\gamma}, \Sigma) \times \hat{\eta}_{n'}^{a'} \tilde{\eta}_{n'}^{b'} (\bar{\gamma}, \Sigma) \\
&= \sum_{ \substack{ \bar\alpha \in H_1(\Sigma,\mathbb{Z}_x) \\ {(c-c') \bar{\gamma} \over \mathrm{gcd}(n,n',k\ell) } \in H_1(\Sigma,\mathbb{Z}) }    }  \frac{e^{ {2\pi i \over x}  \frac{pcn'+p'c'n}{\mathrm{gcd}(n,n')} \langle \bar{\gamma}, \bar{\alpha} \rangle }}{\sqrt{|H_1(\Sigma ,\mathbb{Z}_n)||H_1(\Sigma ,\mathbb{Z}_{n'})|}}\,  \, \eta \Bigg( \begin{array}{c}
	(b+b')\bar{\gamma} + {N \over x} \bar \alpha \\
	+{\ell(c-c') \over \mathrm{gcd}(n,n',k\ell) } \big({k\ell \over \mathrm{gcd}(n,n',k\ell)} \big)^{-1}_{N \over x\ell} \bar{\gamma}
\end{array} \Bigg) \,.
\fe
Comparing this with \eqref{hat.eta.tilde.eta} we find
\ie
& \hat{\eta}_{n}^a \tilde{\eta}_{n}^b(\bar{\gamma}, \Sigma) \times \hat{\eta}_{n'}^{a'} \tilde{\eta}_{n'}^{b'} (\bar{\gamma}, \Sigma) \\
& =  \delta_{(c-c')\bar\gamma \,\text{ mod gcd}(n,n',k\ell) , 0} \, \sqrt{|H_1(\Sigma,\mathbb{Z}_{\mathrm{gcd}(n,n',k\ell)})|} \; \hat{\eta}_{x}^{\frac{pcn'+p'c'n}{\mathrm{gcd}(n,n')}} \tilde{\eta}_{x}^{{\ell(c-c') \over \mathrm{gcd}(n,n',k\ell)}(\frac{k\ell}{\mathrm{gcd}(n,n',k\ell)})^{-1}_{N \over x\ell}+b+b'}(\bar\gamma, \Sigma) \,. \notag
\fe

The fusion coefficient $\delta_{(c-c')\bar\gamma \,\text{ mod gcd}(n,n',k\ell) , 0} \, \sqrt{|H_1(\Sigma,\mathbb{Z}_{\mathrm{gcd}(n,n',k\ell)})|}$ is the partition function of the 1+1d $\mathbb{Z}_{\mathrm{gcd}(n,n',k\ell)}$ gauge theory on $\Sigma$ with the Wilson line $W^{c-c'}$ on $\bar\gamma$ that we denote by $(\mathcal{Z}_{\mathrm{gcd}(n,n',k\ell)}, W^{c-c'})(\bar\gamma,\Sigma)$. This can be seen by
\ie
(\mathcal{Z}_N, W^{c}) (\bar\gamma,\Sigma) = {1\over \sqrt{|H_1(\Sigma, \mathbb{Z}_N)|}} \sum_{\gamma\in H_1(\Sigma,\mathbb{Z}_N)} e^{\frac{2\pi i}{N} c \langle \bar\gamma,\gamma \rangle } = \delta_{c\bar\gamma \,\text{ mod }N , 0} \, \sqrt{|H_1(\Sigma,\mathbb{Z}_N)|}  \,.
\fe
We therefore conclude that
\begin{equation}
	\hat{\eta}_{n}^a \tilde{\eta}_{n}^b \times \hat{\eta}_{n'}^{a'} \tilde{\eta}_{n'}^{b'} = \left( \mathcal{Z}_{\mathrm{gcd}(n,n',k\ell)}, W^{c-c'} \right) \, \hat{\eta}_{x}^{\frac{pcn'+p'c'n}{\mathrm{gcd}(n,n')}} \tilde{\eta}_{x}^{{\ell(c-c') \over \mathrm{gcd}(n,n',k\ell)}\left(\frac{k\ell}{\mathrm{gcd}(n,n',k\ell)}\right)^{-1}_{N \over x\ell}+b+b'} \,.
\end{equation}

\subsection{Fusion rules in the $\mathbb{Z}_p$ gauge theory} \label{app.zp.gauge.theory}

In this appendix we give the detail of the computation for the condensation surfaces in the $\mathbb{Z}_p$ gauge theory in Section \ref{sec:zpgauge}.

\bigskip\centerline{\it Fusion of condensation defects from higher gauging $\mathbb{Z}_p \times \mathbb{Z}_p$}\bigskip

Let us choose $f,f' \in \mathbb{Z}_p$ and compute the fusion between $S_{\mathbb{Z}_p \times \mathbb{Z}_p, f}$ and $S_{\mathbb{Z}_p \times \mathbb{Z}_p, f'}$
\ie
&S_{\mathbb{Z}_p \times \mathbb{Z}_p, f}(\Sigma) \times S_{\mathbb{Z}_p \times \mathbb{Z}_p, f'}(\Sigma) \\ &= {1 \over {|H_1(\Sigma, \mathbb{Z}_p)|}^2 }\sum_{ \gamma_i, \gamma'_i} {e^{{2\pi i \over p} \left({f \langle \gamma_1, \gamma_2\rangle + f' \langle \gamma'_1, \gamma'_2\rangle}\right) } } \, \eta_1(\gamma_1) \eta_2(\gamma_2) \, \eta_1(\gamma'_1)\eta_2(\gamma'_2)  \\
&= {1 \over {|H_1(\Sigma, \mathbb{Z}_p)|}^2} \sum_{ \gamma_i, \gamma'_i} {e^{{2\pi i \over p} \left({f \langle \gamma_1, \gamma_2\rangle + (f+f'+1) \langle \gamma'_1, \gamma'_2\rangle + (f+1) \langle \gamma_2, \gamma'_1 \rangle + f \langle \gamma'_2, \gamma_1 \rangle }\right) } } \, \eta_1(\gamma_1)  \, \eta_2(\gamma_2) \,.
\fe
To simplifying the RHS, there are qualitatively two different cases depending on whether $f+f'+1 = 0 \pmod{p}$ or not:
\paragraph{Case 1 ($f+f'+1\neq0$):} The sum over $\gamma'_2$ results in the Kronecker delta function factor ${|H_1(\Sigma, \mathbb{Z}_p)|} \, \delta_{f\gamma_1 - [f+f'+1]\gamma'_1, 0} $ which sets $\gamma'_1 = \frac{f}{f+f'+1}\gamma_1$ and we get
\ie
S_{\mathbb{Z}_p \times \mathbb{Z}_p, f}(\Sigma) \times S_{\mathbb{Z}_p \times \mathbb{Z}_p, f'}(\Sigma) = \frac{1}{{|H_1(\Sigma, \mathbb{Z}_p)|}} \sum_{ \gamma_1, \gamma_2 }  e^{\frac{2\pi i}{p} \frac{ff' }{f+f'+1} \langle \gamma_1, \gamma_2\rangle } \, \eta_1(\gamma_1)  \, \eta_2(\gamma_2) \,.
\fe
This translates to the invertible fusion rule $	S_{\mathbb{Z}_p \times \mathbb{Z}_p, f} \times S_{\mathbb{Z}_p \times \mathbb{Z}_p, f'} = S_{\mathbb{Z}_p \times \mathbb{Z}_p, \frac{ff'}{f+f'+1}}$. If we do the crucial change of variable $f=\frac{1}{m-1}$, the fusion rule simplifies to
\begin{equation}
	S_{\mathbb{Z}_p \times \mathbb{Z}_p, \frac{1}{m-1}} \times S_{\mathbb{Z}_p \times \mathbb{Z}_p, \frac{1}{m'-1}} = S_{\mathbb{Z}_p \times \mathbb{Z}_p, \frac{1}{mm'-1}} ~.
\end{equation}

\paragraph{Case 2 ($f+f'+1=0$):} In this case we can perform the sum over $\gamma'_1$ and $\gamma'_2$ to get
\ie
S_{\mathbb{Z}_p \times \mathbb{Z}_p, f}(\Sigma) \times S_{\mathbb{Z}_p \times \mathbb{Z}_p, f'}(\Sigma) &= \sum_{ \gamma_1, \gamma_2 } \delta_{(f+1)\gamma_2,0} \, \delta_{f\gamma_1,0} \, e^{{2\pi i \over p} {f \langle \gamma_1, \gamma_2\rangle } } \, \eta_1(\gamma_1)  \, \eta_2(\gamma_2) \\
&= \begin{cases}
	\sqrt{|H_1(\Sigma, \mathbb{Z}_p)|} \, S_{\mathbb{Z}_{p}^{(\infty)}}(\Sigma)\,, & f=0 \\
	\sqrt{|H_1(\Sigma, \mathbb{Z}_p)|} \, S_{\mathbb{Z}_{p}^{(0)}}(\Sigma)\,, & f'=0 \\
	1\,, & f,f' \neq 0
\end{cases}
\fe
For reasons that become clear later, we have introduced a notation where $S_{\mathbb{Z}_{p}^{(\infty)}}$ and $S_{\mathbb{Z}_{p}^{(0)}}$ correspond to the higher gauging of the first and second $\mathbb{Z}_p$ subgroups (generated by $\eta_1$ and $\eta_2$), respectively.

Putting everything together the fusion rules can be summarized as:
\begin{equation}
	S_{\mathbb{Z}_p \times \mathbb{Z}_p, \frac{1}{m-1}} \times S_{\mathbb{Z}_p \times \mathbb{Z}_p, \frac{1}{m'-1}} = \begin{cases}
			\left(  \mathcal{Z}_p \right) S_{\mathbb{Z}_{p}^{(m)}} & m=\frac{1}{m'} =0 \text{ or }\infty \\ 
			S_{\mathbb{Z}_p \times \mathbb{Z}_p, \frac{1}{mm'-1}} & \text{otherwise}
		\end{cases} ~,
\end{equation}
where we have introduced another notation of $S_{\mathbb{Z}_p \times \mathbb{Z}_p, \infty} \equiv 1$. Therefore, we see that the surfaces  $	S_{\mathbb{Z}_p \times \mathbb{Z}_p, f} $ with $f\neq 0,-1$ form a cyclic group $\mathbb{Z}_{p-1}$ under fusion. 
In contrast,  the surface defects $	S_{\mathbb{Z}_p \times \mathbb{Z}_p, 0} $ and $	S_{\mathbb{Z}_p \times \mathbb{Z}_p, -1} $ are non-invertible.

\bigskip\centerline{\it Fusion of condensation defects from higher gauging $\mathbb{Z}_p$}\bigskip

The $\mathbb{Z}_p \times \mathbb{Z}_p$ 1-form symmetry has $p+1$ $\mathbb{Z}_p$ subgroups. 
Given any non-zero element $(m_1,m_2) \in \mathbb{Z}_p \times \mathbb{Z}_p$ associated with the topological line $\eta_1^{m_1} \eta_2^{m_2}$, we can take the $\mathbb{Z}_p$ subgroup generated by it and denote it by $\mathbb{Z}_p^{(m_1/m_2)}$. Note that $\mathbb{Z}_{p}^{(m)}$ only depend on $m=\frac{m_1}{m_2} \pmod{p}$, where $m=\infty$ and $m=0$ corresponds to the first and second $\mathbb{Z}_p$ subgroups which are  $0$-gaugeable. 
Therefore, from the results of Section \ref{sec:ZN}, $S_{\mathbb{Z}_{p}^{(\infty)}}$ and $S_{\mathbb{Z}_{p}^{(0)}}$ are non-invertible and the rest are all invertible order $2$ defects.

Denote the surface defect from the higher gauging of  $\mathbb{Z}_{p}^{(m)}$ by
\begin{equation}
	S_{\mathbb{Z}_{p}^{(m)}}(\Sigma) = \frac{1}{\sqrt{|H_1(\Sigma, \mathbb{Z}_p)|}} \sum_{ \gamma \in H_1(\Sigma,\mathbb{Z}_p) } \eta_1(m_1 \gamma)  \, \eta_2( m_2 \gamma) \,.
\end{equation}
The fusion of this surface with itself has been computed in Section \ref{sec:znn}. Thus let us take $m \neq m'$ and compute the fusion of $S_{\mathbb{Z}_{p}^{(m)}}$ with $S_{\mathbb{Z}_{p}^{(m')}}$:
\begin{align}
	S_{\mathbb{Z}_{p}^{(m)}}(\Sigma)\times	S_{\mathbb{Z}_{p}^{(m')}}(\Sigma) &= \sum_{ \gamma, \gamma' } {e^{\frac{2\pi i}{p} m_2m'_1 \langle \gamma, \gamma'\rangle } \over {|H_1(\Sigma, \mathbb{Z}_p)|}} \, \eta_1(m_1 \gamma + m'_1\gamma') \eta_2( m_2 \gamma + m'_2 \gamma') \notag \\
	&= \sum_{ \gamma, \gamma' } {e^{\frac{2\pi i}{p} \frac{m'}{m-m'} \langle m_1 \gamma + m'_1\gamma', m_2 \gamma + m'_2 \gamma' \rangle } \over {|H_1(\Sigma, \mathbb{Z}_p)|}} \,  \eta_1(m_1 \gamma + m'_1\gamma') \eta_2( m_2 \gamma + m'_2 \gamma') \notag \\
	&= S_{\mathbb{Z}_p \times \mathbb{Z}_p, \frac{1}{\frac{m}{m'}-1}} (\Sigma) ~ .
\end{align}

\bigskip\centerline{\it Fusion between condensation defects from higher gauging $\mathbb{Z}_p$ and $\mathbb{Z}_p\times \mathbb{Z}_p$}\bigskip

 Take $m=m_1/m_2$ and $f= 1/(m'-1)$, we have
\ie
S_{\mathbb{Z}_{p}^{(m)}}(\Sigma)\times S_{\mathbb{Z}_{p} \times \mathbb{Z}_{p} , \frac{1}{m'-1}}(\Sigma) &= \frac{1}{\sqrt{|H_1(\Sigma, \mathbb{Z}_p)|}^3} \sum_{ \gamma, \gamma'_i } e^{\frac{2\pi i}{p} f \langle \gamma'_1, \gamma'_2 \rangle } \, \eta_1(m_1 \gamma) \eta_2( m_2 \gamma) \eta_1(\gamma_1') \eta_2( \gamma'_2)  \\
&= \frac{1}{\sqrt{|H_1(\Sigma, \mathbb{Z}_p)|}} \sum_{ \gamma'_1, \gamma'_2 } e^{\frac{2\pi i}{p} f \langle \gamma'_1, \gamma'_2 \rangle } \, \delta_{m_2(f+1) \gamma'_1, m_1f \gamma'_2} \; \eta_1(\gamma'_1) \eta_2(\gamma'_2)  \\
&= \begin{cases}
	\left( \mathcal{Z}_p \right) S_{\mathbb{Z}_{p} \times \mathbb{Z}_{p} , \frac{1}{m'-1}}(\Sigma)\,, &  m=m' \in \{0, \infty\} \\
	S_{\mathbb{Z}_p^{(m/m')}}(\Sigma)\,, & \text{otherwise}
\end{cases} \,.
\fe
and
\ie
S_{\mathbb{Z}_{p} \times \mathbb{Z}_{p} ,  \frac{1}{m'-1}}(\Sigma) \times S_{\mathbb{Z}_{p}^{(m)}}(\Sigma) &= \frac{1}{\sqrt{|H_1(\Sigma, \mathbb{Z}_p)|}^3} \sum_{ \gamma, \gamma'_i } e^{\frac{2\pi i}{p} f \langle \gamma'_1, \gamma'_2 \rangle } \, \eta_1(\gamma_1') \eta_2( \gamma'_2) \eta_1(m_1 \gamma) \eta_2( m_2 \gamma)  \\
&= \frac{1}{\sqrt{|H_1(\Sigma, \mathbb{Z}_p)|}} \sum_{ \gamma'_1, \gamma'_2 } e^{\frac{2\pi i}{p} f \langle \gamma'_1, \gamma'_2 \rangle } \, \delta_{m_2f \gamma'_1, m_1(f+1) \gamma'_2} \; \eta_1(\gamma'_1) \eta_2(\gamma'_2)  \\
&= \begin{cases}
	\left(  \mathcal{Z}_p \right) S_{\mathbb{Z}_{p} \times \mathbb{Z}_{p} ,  \frac{1}{m'-1}}(\Sigma) & m=1/m' \in \{0, \infty \}  \\
	S_{\mathbb{Z}_p^{(m'm)}}(\Sigma) & \text{otherwise}
\end{cases} \,.
\fe

Putting everything together, we have derived the fusion rule \eqref{zpgaugefusion} of the surfaces, which includes the invertible $D_{2(p-1)}$ 0-form symmetry, of the 2+1d $\mathbb{Z}_p$ gauge theory.

\bibliographystyle{JHEP}

\bibliography{condensation_draft}

\end{document}